\documentclass[letterpaper, 11pt]{article}

\usepackage[nocompress]{cite}
\usepackage{jheppub}

\usepackage{graphicx}
\usepackage{epstopdf}
\usepackage{amsmath, amssymb}

\usepackage{caption}
\usepackage{subcaption}

\usepackage{multirow} 
\usepackage{longtable} 

\usepackage{romannum}

\usepackage{harpoon}
\usepackage[vcentermath]{youngtab}

\usepackage[us,12hr]{datetime} 

\newcommand{\be}{\begin{eqnarray}}
\newcommand{\ee}{\end{eqnarray}}

\newcommand{\bn}{\begin{enumerate}}
\newcommand{\en}{\end{enumerate}}

\newcommand\Tstrut{\rule{0pt}{3.8ex}}         
\newcommand\Bstrut{\rule[-0.9ex]{0pt}{0pt}}   

\def\sfund{\scalebox{.3}{\yng(1)}}
\def\ssym{\scalebox{.3}{\yng(2)}}
\def\santi{\scalebox{.3}{\yng(1,1)}}
\def\bfund{\scalebox{.6}{\yng(1)}}
\def\bsym{\scalebox{.6}{\yng(2)}}
\def\banti{\scalebox{.6}{\yng(1,1)}}

\def\CA{{\cal A}}

\def\CN{{\cal N}}
\def\CO{{\cal O}}

\def\CQ{{\cal Q}}

\def\s{\sigma}

\def\half{\frac{1}{2}}

\def\Tr{{\rm Tr}}

\newcommand{\bea}{\begin{eqnarray}}
\newcommand{\eea}{\end{eqnarray}}

\def\CA{{\cal A}}

\def\CN{{\cal N}}
\def\CO{{\cal O}}

\def\CQ{{\cal Q}}

\def\s{\sigma}

\def\half{\frac{1}{2}}

\def\Tr{{\rm Tr}}

\allowdisplaybreaks

\title{Classification of large N superconformal gauge theories with a dense spectrum}

\affiliation[a]{CRST and School of Physics and Astronomy, Queen Mary University of London\\London E1 4NS, United Kingdom}
\affiliation[b]{Asia Pacific Center for Theoretical Physics, Pohang, Gyeongbuk 37673, Korea}
\affiliation[c]{Department of Physics, Korea Advanced Institute of Science and Technology\\
291 Daehak-ro, Yuseong-gu, Daejeon 34141, Korea}

\author[a]{Prarit Agarwal,}
\author[b]{Ki-Hong Lee,}
\author[b, c]{and Jaewon Song}

\emailAdd{p.agarwal@qmul.ac.uk}
\emailAdd{kihong.lee@apctp.org}
\emailAdd{jaewon.song@kaist.ac.kr}

\abstract
{
We classify the large $N$ limits of four-dimensional supersymmetric gauge theories with simple gauge groups that flow to superconformal fixed points. We restrict ourselves to the ones without a superpotential and with a fixed flavor symmetry. We find 35 classes in total, with 8 having a dense spectrum of chiral gauge-invariant operators.  
The central charges $a$ and $c$ for the dense theories grow linearly in $N$ in contrast to the $N^2$ growth for the theories with a sparse spectrum. The difference between the central charges $a-c$ can have both signs, and it does not vanish in the large $N$ limit for the dense theories. We find that there can be multiple bands separated by a gap, or a discrete spectrum above the band. We also find a criterion on the matter content for the fixed point theory to possess either a dense or sparse spectrum. We discover a few curious aspects regarding supersymmetric RG flows and $a$-maximization along the way. For all the theories with the dense spectrum, the AdS version of the Weak Gravity Conjecture (including the convex hull condition for the cases with multiple $U(1)$'s) holds for large enough $N$ even though they do not have weakly-coupled gravity duals. 
}

\preprint{APCTP-Pre2020-011, QMUL-PH-20-20}

\begin{document}
\maketitle

\section{Introduction}
One of the siginificant hurdles in understanding non-Abelian gauge theories arises from an apparent lack of a small dimensionless parameter for perturbative expansions. This problem gets greatly ameliorated in 't Hooft's large $N$ limit \cite{tHooft:1973alw}. Therefore understanding their large $N$ behavior is of paramount importance in our quest to understand gauge theories in general. 
An important aspect of large $N$ gauge theories is the AdS/CFT correspondence \cite{Maldacena:1997re, Gubser:1998bc, Witten:1998qj}, which is an exact equivalence between gauge theory and quantum gravity (or string theory). The most precise form of this correspondence has been worked out for highly supersymmetric setups usually `derived' in string/M-theory, such as the duality between 4d $\CN=4$ super Yang-Mills theory and type IIB superstring theory on $AdS_5 \times S^5$. 

However, the AdS/CFT correspondence is not restricted to these well-controlled cases only. Generally, one can regard CFT as a definition of some quantum gravity in anti-de Sitter space. Since conformal field theory is a completely well-defined quantum mechanical system, it should also be true in AdS quantum gravity. However, the gravitational theory living in anti-de Sitter space is not always guaranteed to be similar to the weakly-coupled Einstein gravity. A number of criterion for the CFT to be holographically dual to Einstein-like gravity has been put forward by \cite{Heemskerk:2009pn, ElShowk:2011ag} for example. 

Since the spectrum of gauge theories is given in terms of gauge-invariant operators made out of elementary fields, it is natural to expect the scaling dimensions of the low-lying operators to be sparse. This is certainly true for all the theories with large amounts of supersymmetry and also for the theories with weakly-coupled gravity duals. In fact, it has been argued in \cite{ElShowk:2011ag}, that this is one of the necessary conditions for a CFT to be holographic. 
In \cite{Agarwal:2019crm}, two of the authors of the current paper showed that there exists a rather unconventional large $N$ gauge theory. It is a simple 4d $\CN=1$ supersymmetric $SU(N)$ gauge theory with an adjoint and a pair of fundamental and anti-fundamental chiral multiplets. 
It was shown that the mesonic chiral operators of the theory form a dense spectrum such that their scaling dimensions lie within a band with the gap between the dimension of the $i$-th and the $(i+1)$-th operator being $\mathcal{O}(1/N)$. It is possibly the closest example of the Liouville field theory or non-compact CFTs in two-dimensions with a continuum spectrum. 
Also, the central charges $a$ and $c$ grow linearly in $N$ instead of $N^2$. The ratio between the central charges $a/c$ can be either greater or smaller than 1 and it does not go to 1 in the large $N$ limit.\footnote{This ratio $a/c$ (or the difference $c-a$) appears in multiple contexts, see for example \cite{Buchel:2008vz, DiPietro:2014bca, Ardehali:2015bla, Perlmutter:2015vma}.} 
Such a theory is not expected to be dual to a weakly coupled gravity in AdS. Nevertheless, it was found in \cite{Agarwal:2019crm} that a version of Weak Gravity Conjecture (WGC) \cite{ArkaniHamed:2006dz} in AdS \cite{Nakayama:2015hga} still holds. Though WGC was originally proposed through a semi-classical analysis based on the evaporating black holes, the findings of \cite{Agarwal:2019crm} support the possibility that WGC is valid beyond the semi-classical regime and may be understandable from a more general principle. 

Given its `exotic' behavior, it is natural to ask whether such large $N$ gauge theories with dense spectrum occur frequently or are rather rare. 
To this end, we classify large $N$ limits of four-dimensional supersymmetric gauge theories with a simple gauge group. 
There are several recent attempts to classify superconformal gauge theories in 4d, for example with $\CN=2$ supersymmetry  \cite{Bhardwaj:2013qia}, all possible IR fixed points obtained via superpotential deformations \cite{Maruyoshi:2018nod, MNS2}, and the ones connected to free theory via exactly marginal deformations \cite{Razamat:2020pra}. 
There are also other classification programs that do not rely on a Lagrangian description. Especially, there has been progress in classifying 4d $\CN=2$ SCFTs via their Coulomb branch geometries \cite{Argyres:2015ffa,Argyres:2015gha,Argyres:2016xua,Argyres:2016xmc,Caorsi:2018zsq, Caorsi:2019vex, Argyres:2020nrr}. 
In our classification program, we turn off the superpotential and consider the large $N$ limit of gauge theories having a \emph{fixed global symmetry}. Our motivation comes from holography. In holography, we need a family of CFTs with a fixed global symmetry, which in turn becomes a gauge symmetry in the bulk. Fixing the global symmetry is, therefore, necessary in order to have a proper holographic interpretation, if it ever exists. This contrasts with the `Veneziano limit' considered in some of the literature, where one fixes the ratio of the number of flavors to the number of colors \emph{i.e.} $N_f/N_c$. For the latter case, flavor symmetry grows with $N_c$ and does not have a proper holographic interpretation. This excludes SQCD, where the IR theory is non-trivial only if $N_f/N_c$ is within a particular range called the conformal window. What we do is quite close in spirit with the paper \cite{Buchel:2008vz}, where they classify the large $N$ gauge theories with exactly marginal couplings. 

Upon classification, we find that many large $N$ superconformal theories have dense spectrum and linear growth of the central charges. More precisely, we find the following result: 
\begin{align}
	\begin{split}
		\textrm{Dense spectrum} \textrm{ if } & \sum_i T(\mathbf{R}_i) \sim h^\vee + \CO(1)  \\
		\textrm{Sparse spectrum}  \textrm{ if } & \sum_i T(\mathbf{R}_i) \ge 2h^\vee \pm \CO(1) 
	\end{split}
	\label{eq:dense_conj}
\end{align}
Here the index $i$ runs over all possible matter chiral multiplets in representation $R_i$, $T(R_i)$ denotes the Dynkin index of the representation $\mathbf{R}_i$, and $h^\vee$ denotes the dual Coxeter number of the gauge group which scales linearly with $N$. We also find that theories with dense spectrum always exhibit \emph{linear growth of central charges} $a, c$. The central charges for the theories with sparse spectrum grow quadratically in $N$ as usual. 

Roughly speaking, when the `amount of matter' is large enough, the spectrum becomes sparse and behaves according to conventional wisdom. On the other hand, when the amount of matter is `small', we find the CFT to be of the `exotic' type with a dense spectrum in large $N$. The examples we study are believed to flow to an interacting superconformal fixed point. When the matter representations are large enough, the IR fixed point is similar to the case of SQCD in the conformal window. In terms of 1-loop beta function, the second case has $b_0 \sim 2 h^\vee$, which is the value at the middle of the conformal window of SQCD ($N_f = 2N_c$ for $SU(N_c)$ with $N_f$ flavors). In this case, we expect the theory to be in the non-Abelian Coulomb phase. On the other-hand, when the matter representation is `small,' even though the theory flows to a non-trivial fixed point in the IR, it exhibits the exotic behavior of $\CO(N)$ growth of the degrees of freedom. 
This is rather similar to the case of Argyres-Douglas theories \cite{Argyres:1995jj, Argyres:1995xn}, where $\mathcal{O}(N)$ mutually non-local charged particles simultaneously become massless. Indeed, once deformed by a superpotential and appropriately coupled to gauge-singlets, some of the theories flow to $\CN=2$ supersymmetric Argyres-Douglas theories \cite{Maruyoshi:2016tqk, Maruyoshi:2016aim, Agarwal:2016pjo, Agarwal:2017roi, Benvenuti:2017bpg}. 
We find in total 35 classes of large $N$ gauge theories (20 $SU(N)$, 6 $SO(N)$, 9 $Sp(N)$ theories). There are 8 classes of theories with dense spectrum (4 $SU(N)$, 2 $SO(N)$, 2 $Sp(N)$). 
It is interesting to notice that the dense theories we obtain are exactly identical to the ones (with simple gauge group) studied in \cite{Intriligator:1995ax} some time ago, where dual descriptions were given. However, unlike \cite{Intriligator:1995ax}, we do not deform the theory with a superpotential.

The rest of this paper is organized as follows: In section \ref{sec:main}, we discuss the main idea of our classification and summarize the results. We spell out the criterion for the large $N$ gauge theories to have dense or sparse spectrum. We encounter interesting aspects of the RG flow exhibiting accidental symmetries arising from decoupled sectors and discuss how to treat them properly. We also briefly summarize the Weak Gravity Conjecture in AdS. In sections \ref{sec:SU}, \ref{sec:SO}, and \ref{sec:Sp}, we spell out the details of the classification, focusing on the theories exhibiting a dense spectrum. Finally, we conclude with possible future directions in section \ref{sec:conclusion}. 

\section{Main idea and results} \label{sec:main}

\subsection{Classification of large $N$ supersymmetric gauge theories} \label{subsec:classification}

Let us discuss our scheme of classification. We will be considering four-dimensional $\CN=1$ supersymmetric gauge theories that flow to interacting superconformal fixed points (without a superpotential) in the infrared with the following assumptions:
\begin{itemize}
	\item The large $N$ limit exists. 
	\item The gauge group is simple. 
	\item The flavor symmetry is fixed as we vary $N$. 
\end{itemize}
The first two conditions restrict the gauge group to be classical $SU(N)$, $SO(N)$, $Sp(N)$.

Imposing various consistency conditions on the gauge theory further constraints the representations $\{ \mathbf{R}_i \}$ of the chiral multiplets that can be incorporated into our theory. 
Firstly, the theory must be free of any gauge anomalies. This implies
\begin{align}
	\sum_{i} \CA(\mathbf{R}_i) = 0 \ , 
\end{align}
where $\CA(R_i)$ is the cubic Casimir of the massless fermions lying in representation $R_i$ of th gauge group. 
$\CA(R_i)$ is non-zero only for chiral representations of $SU(N)$ gauge theories. 
For the $Sp(N)$ gauge theories, we have to ensure that the Witten anomaly \cite{Witten:1982fp} vanishes. It implies that there has to be an even number of fundamental representations in $Sp(N)$ gauge theories.\footnote{More generally, if the $Sp(N)$ generators are normalized such that ${\rm Tr} T^a T^b = \half \delta^{ab}$ for the fundamental representation, then the consistency requires that the matter content be such that $T(\mathbf{R}): {\rm Tr}_{\mathbf{R}} T^a T^b = T(\mathbf{R}) \delta^{ab}$ is an integer \cite{Witten:1982fp}.}
Meanwhile, $SO(N)$ gauge theories are anomaly-free. Hence these considerations do not restrict their matter content.

Secondly, asymptotic freedom requires that the $\beta$-function must be negative. This implies
\begin{align}
	b_0 = \left(3h^\vee-\sum_i T(\mathbf{R}_i)\right) \geq 0 \ , 
\end{align}
where the sum is over all charged matter multiplets, and $T(\mathbf{R})$ is the Dynkin index. We use the  normalization given by $T(\square)=\half$.
Notice that the dual-Coxeter number $h^\vee$ of any classical simple Lie group grows linearly in $N$.\footnote{$h^\vee_{SU(N)} = N, h^\vee_{SO(N)} = N-2, h^\vee_{Sp(N)}=N+2.$} Thus the matter representations should have a Dynkin index $T(\mathbf{R}_i)\leq \mathcal{O}(N)$. Therefore, the allowed representations are given as follows:
\begin{itemize}
	\item fundamental and its complex conjugate: $Q_i$, $\widetilde{Q}_j$
	\item rank-2 anti-symmetric and its complex conjugate: $A_i$, $\widetilde{A}_j$
	\item rank-2 symmetric traceless and its complex conjugate: $S_i$, $\widetilde{S}_j$
	\item adjoint: $\Phi_i$
\end{itemize}
Notice that for the $SO(N)$ and $Sp(N)$ groups, the adjoint is given by rank-2 anti-symmetric and symmetric tensors, respectively. Also, for the $SO(N)$ and $Sp(N)$ groups, the representations are (pseudo-)real, so we do not have their complex conjugates. For lower-rank cases, one can also have other representations such as the rank-3 symmetric/anti-symmetric tensors, but they cannot be present in the large $N$ limit since they cause the beta function to become positive rather quickly. 

The numbers of rank-2 tensor matters and the adjoints are restricted to be of order $\CO(1)$ by the asymptotic freedom bound. However, the number of fundamental matters can grow with $N$ without violating the asymptotic freedom bound. In order to have a fixed flavor symmetry group, we need to fix the number of matter multiplets as we dial the value of $N$. This requirement forbids the usual SQCD with just the fundamental and anti-fundamental matter chiral multiplets. This is because, to have a superconformal fixed point in the IR (to be within the conformal window), the number of flavors should be of order $\CO(N)$. Therefore, in all of our examples, we will have $\CO(1)$ number of matter multiplets in each representation. 

We find our classification program to be tangible and exhibits various interesting phenomena. 
In the current paper, we will mainly focus on the spectrum of gauge-invariant operators. This boils down to performing $a$-maximization \cite{Intriligator:2003jj} for our set of gauge theories. We discuss technical but interesting issues involving the decoupling of operators under the renormalization group flow in section \ref{subsec:RGflow}. 

We find that all the superconformal fixed points we study in the current paper fall into two categories: 
\begin{itemize}
	\item Sparse spectrum of chiral operators in large $N$: There exists a set of operators for which the gap in the dimensions of the $i$-th lightest and the $(i+1)$-th lightest operators stays constant in large $N$ \emph{i.e.} $\Delta_{i+1} - \Delta_i \sim O(1)$. The central charges grow quadratically in $N$ \emph{i.e.} $a \sim c \sim O(N^2)$.
	\item Dense spectrum of chiral operators in large $N$:  There exists a band of operator dimensions $1 < \Delta < \bar{\Delta}$ such that the gap in the dimensions of the $i$-th lightest and the $(i+1)$-th lightest operators scales as $1/N$ \emph{i.e.} $\Delta_{i+1} - \Delta_i \sim O(\frac{1}{N})$. The central charges grow linearly in $N$ \emph{i.e.} $a \sim c \sim O(N)$.
\end{itemize}
The first one is the more familiar type of CFT one usually considers. For example, in 4d $\CN=4$ SYM theory and the $\CN=2$ superconformal QCD, the chiral operators are given by gauge-invariant combinations of the adjoint and the fundamental chiral multiplets of the form $\Tr \Phi^i$ and $Q\Phi^{j} \widetilde{Q}$. These have an $\CO(1)$ gap (actually exactly 1) in the large $N$ limit. The central charges grow like $N^2$ as can be easily expected from the matrix degrees of freedom of the gauge theory. 

\begin{figure}
	\centering
	\includegraphics[width=7cm]{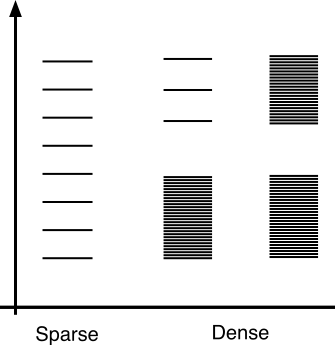}
	\caption{Illustration of the sparse and dense spectrum of large $N$ theories. Here we show 3 possible scenarios. The left one depicts the scaling dimension of the single trace gauge-invariant operators for the sparse case. The spacing between the operator dimensions scales as $\CO(1)$ at large $N$. We find two distinct cases for the dense theory. One can have a dense band of low-lying operators and discrete spectrum of heavy operators. The other case comes with multiple bands with an $\CO(N)$ gap between the bands. 
		For the theories with a dense spectrum, the spacing between the operator dimensions in a band scales as $\CO(1/N)$. }
	\label{fig:denseSparse}
\end{figure}

On the other hand, the second case is exotic. Once we list the (single-trace) gauge-invariant chiral operators according to their scaling dimensions, we find that there exists a band 
of fixed width, containing $O(N)$ number of operators. The spacing inside the band scales as $1/N$ so that the spectrum becomes effectively continuous at large $N$. See figure \ref{fig:denseSparse} for illustration.
As shown in the figure, it is possible to have one band of low-lying single-trace operators or more than one band. Within a band, the gap in the scaling dimensions of the operators goes like $1/N$. 
This is due to very large quantum corrections to the scaling dimension of the matter fields (adjoint or rank-2 tensors), which makes it nearly zero in large $N$. For example, in the adjoint SQCD, the gauge-invariant operators of the form $Q \Phi^j \tilde{Q}$ or $\Phi^j$ with adjoint $\Phi$ give a dense spectrum since the dimension of $\Phi$ is of $\CO(1/N)$. 

It is possible to understand the appearance of the dense spectrum by looking at the anomaly constraint. The anomaly-free condition for the $U(1)_R$ symmetry requires
\begin{align}
	T (\textbf{adj}) + \sum_i T (\mathbf{R}_i) (r_i - 1) = 0 \ , 
\end{align}
where $i$ runs over all chiral multiplets with representation $\mathbf{R}_i$ and the superconformal $R$-charge is given by $r_i$. In order to cancel the anomaly caused by the gaugino ($T(\textbf{adj}) = \CO(N)$), we need to have the second term of order $\CO(N)$. Under our assumption that the number of fundamentals is of $\CO(1)$, the dominant contribution for the second term should, therefore, come from rank-2 tensors, which have $T(\mathbf{R}) = \CO(N)$.\footnote{It may happen that the $R$-charge of the fundamentals is $O(N)$, hence making it possible for the fundamentals to contribute. However, it is highly unlikely since charged chiral multiplet have negative anomalous dimensions in perturbation theory.} Requiring the $R$-charge to be non-anomalous thus implies that the $R$-charge of the rank-2 tensors must be
\be
r_{\textbf{rank 2}} \sim 1 - \frac{T (\textbf{adj})}{\sum T (\mathbf{R}_\textbf{rank 2})} \ . 
\label{eq:RchargeRank2}
\ee
Now recall that in our scheme of classification, theories with a dense spectrum include at most a single adjoint chiral multiplet, in which case the RHS of \eqref{eq:RchargeRank2} becomes vanishingly small. On the other hand, for $SU(N)$ theories, instead of an adjoint chiral, we can also include a pair of chiral multiplets transforming in the rank-2 (anti-)symmetric representation and its conjugate respectively. In the large-$N$ limit for $SU(N)$ gauge groups,  the ratio $T (\textbf{adj})/ 2 T (\mathbf{R}_\textbf{rank 2}) \rightarrow 1$, which implies in this case too, the RHS of \eqref{eq:RchargeRank2} becomes vanishingly small. For $SO(N)$ and $Sp(N)$ theories, we will include a single rank-2 tensor. However, for these gauge groups, the ratio $T (\textbf{adj})/T (\mathbf{R}_\textbf{rank 2}) \rightarrow 1$ in the large-$N$ limit, once again ensuring that the RHS of \eqref{eq:RchargeRank2} becomes vanishingly small.
We thus see that if we have a single rank 2-tensor\footnote{For the $SU(N)$ theory, we need two of rank 2 tensors to cancel gauge anomaly. The $SU(N)$ adjoint can be thought of as a sum of symmetric and anti-symmetric.}, we need to have $r_{\textbf{rank 2}} \to 0$ to cancel the anomaly, hence we get a dense spectrum. If we have more than one rank 2-tensors, the $R$-charges can be of order one (eg. two adjoints will give $r_{\textbf{adj}} \to \half$) so that we get a sparse spectrum. 

Also, it turns out whenever the theory possesses a dense spectrum, the central charges grow linearly in $N$. To see this, recall that the central charges can be written in terms of trace anomalies of $R$-symmetry \cite{Anselmi:1997am}:
\begin{align} \label{eq:acTrR}
	\begin{split}
		a &= \frac{3}{32}(3 {\rm Tr} R^3 - {\rm Tr}R) \ ,  \\ 
		c &= \frac{1}{32}(9 {\rm Tr} R^3 - 5{\rm Tr}R) \ .
	\end{split}
\end{align}
For the gauge theories at our hand, we can write the trace anomaly explicitly as
\begin{align}
	\Tr R^{3} = |G| + \sum_i (r_i - 1)^{3} |\mathbf{R}_i| \ , \quad  \Tr R = |G| + \sum_i (r_i - 1) |\mathbf{R}_i| \ , 
\end{align}
where $|G|$ and $|\mathbf{R}|$ denotes the dimensions of the gauge group $G$ and the representation $\mathbf{R}$, respectively. We will now argue that, for the theories of our concern, $\mathcal{O}(N^2)$ contributions to both $\Tr R^{3}$ and $\Tr R$ vanish independently. To see this, notice that the contributions to either of the two anomalies can be seen as coming from two pieces: The first of these comes from the gauginos and is given by the dimension of the gauge group which is $\CO(N^2)$. The second piece consists of the contribution of the matter multiplets. The dimensions of fundamental representation is $\CO(N)$ and the $R$-charges of the corresponding chiral multiplets are at best $\CO(1)$ so their contributions to the $R$-anomalies is subleading in $N$. On the other-hand, the rank-2 tensors or adjoints contribute $\CO(N^2)$ and their $R$-charges scale as $\CO(1/N)$. Therefore, at large $N$, the $\CO(N^2)$ term in both $\Tr R^3$ and $\Tr R$ is captured by $|G| - \sum |\mathbf{R}_{\textbf{rank-2}}|$. It is now easy to see that for the theories having a dense spectrum, the $\CO(N^2)$ term in $|G| - \sum |\mathbf{R}_{\textbf{rank-2}}|$ vanishes exactly. It therefore follows, that both $\Tr R^{3}$ and $\Tr R$ as well as the central charges, $a$ and $c$, are at most $\CO(N)$.

We also notice that $16(a-c) = \Tr R = \CO(N)$ for all gauge theories that we consider in the current paper. In the sparse theories $a-c$ is subleading in $N$ so that $a/c \to 1$ in large $N$. On the other hand, for the dense theories, $a-c$ is of the same order as $a$ and $c$, therefore we do not expect $a/c$ to approach 1.\footnote{The value of $c-a$ controls the sub-leading correction to the celebrated holographic entropy-viscosity ratio bound $\frac{\eta}{s} \ge \frac{1}{4\pi}$ of Kovtun-Son-Starinets \cite{Kovtun:2004de}. The authors of \cite{Buchel:2008vz} pointed out if $c-a>0$ or equivalently $a/c<1$, the bound can be lower, potentially violating the KSS bound. We find both signs of $c-a$ in this paper. For the dense theories, $c-a$ is of the same order of magnitude as the central charges so that it is no longer a sub-leading correction.} 

\subsection{Renormalization group flow and the ordering of decoupling} \label{subsec:RGflow}

We study renormalization group fixed points of the supersymmetric gauge theories, as outlined in section \ref{subsec:classification}. The main tool we use is the procedure of $a$-maximization \cite{Intriligator:2003jj}, which we briefly review. 
As is well known, in $\CN=1$ supersymmetric theories in four-dimensions, the scaling dimensions of gauge-invariant chiral operators at the IR-fixed point is proportional to their IR $R$-charge:
\begin{eqnarray}
	\Delta_{\CO} = \frac{3}{2}R_\CO \ .
\end{eqnarray}
However, the IR $R$-symmetry is given by a non-trivial mixture of the UV $R$-symmetry with flavor currents, which is a priori unknown in general. 
Assuming that there is a superconformal fixed point, the exact $R$-symmetry at the IR fixed point can be obtained by maximizing the trial $a$-function with respect to the putative IR $R$-symmetry. Since the central charges $a$ and $c$ for the $\CN=1$ SCFT are determined via trace anomalies $\Tr R$ and $\Tr R^3$ as in \eqref{eq:acTrR}, it is a simple algebraic operation to investigate the IR fixed point. Once we obtain a non-trivial answer upon $a$-maximization that does not violate unitarity constraint, we can justify the assumption that there is a non-trivial SCFT at the IR fixed point. 

However, while $a$-maximization itself is a straight-forward procedure, one has to pay attention to the fact that quite often, a naive application gives rise to operators whose scaling dimensions are less than 1. In 4d CFTs, unitarity requires gauge-invariant scalar operators to have a scaling dimension $\Delta_{\mathcal{O}}$ such that:
\begin{align}
	\Delta_{\mathcal{O}} \geq 1, \ \forall \mathcal{O} \ .
\end{align} 
Moreover, operators that saturate the above inequality decouple from the interacting sector as free fields. The appearance of operators with a scaling dimension below 1 indicates that they must have decoupled from the interacting theory at some point during the RG flow. Therefore, one needs to redo $a$-maximization by removing the contribution of these decoupled operators from the interacting theory \cite{Kutasov:2003iy}. An efficient way to account for the decoupled operators is to introduce a `flip field' $X_\mathcal{O}$ for each decoupled operator $\mathcal{O}$ along with a superpotential term \cite{Barnes:2004jj}:\footnote{The role of these flip fields has been emphasized in recent studies of supersymmetric dualities and supersymmetry enhancing RG flows. See for example \cite{Benvenuti:2017lle, Fluder:2017oxm, Maruyoshi:2018nod, Agarwal:2018oxb}.}
\begin{align}
	W = X_\mathcal{O} \mathcal{O} \ .
\end{align} 
The $F$-term constraint for the chiral multiplet $X$ will set the operator $\CO$ to zero in the chiral ring, or equivalently this quadratic coupling renders the would-be free operator to be massive. 
Now, a second iteration of $a$-maximization will possibly generate a new set of operators with unitary-violating scaling dimensions. We will now have to decouple this new set of operators.
The above steps of $a$-maximization followed by removing decoupled operators will need to be iterated over repeatedly until there are no more operators to decouple. 

An interesting aspect to notice about the above procedure is that during any single iteration, it often happens that there is not just one but a number of operators to decouple from the theory. It is therefore natural to ask if we should remove all such operators from the theory simultaneously, or should we remove only a subset of these first? Generically, removing different subsets will produce a different IR theory. 
For example, in the $SU(7)$ gauge theory with one chiral multiplet transforming in the adjoint representation of the gauge group along with a single flavor of fundamental + anti-fundamental quarks (the theory considered in section \ref{sec:SUnAdj} with $N=7, N_f=1$), 12419 different possible decoupling sequences arise from considering all possible choices of subsets to decouple at each iteration. It turns out all of these decoupling sequences ultimately flow to two distinct fixed points:
\begin{align}
\begin{split}
	&\text{fixed point A:}\  a \simeq 2.83675, \ c \simeq 2.88125 \ ,  \\
	&\text{fixed point B:}\ a \simeq 2.85665, \ c \simeq 2.92113 \ .
\end{split}
\end{align}  
Moreover, if we should only remove a subset of these, how do we determine the correct choice of the subset to remove? These are precisely the questions we wish to answer in this section. We propose to consider decoupling all possible different subsets and then choose the one with the largest value of the central charge $a$. This can be thought of as an extension of a similar diagnostic of the IR phase of gauge theories that was first conjectured in \cite{Intriligator_2005}. The idea being that $\Delta a = a_{\text{UV}} - a_{\text{IR}}$ gives us a measure of the RG-distance between the UV and the IR theory and the correct IR fixed point is the one that lies closest to its UV parent. Let us also note that experimentally we found that in cases when there are multiple different decoupling sequences that correspond to the same value of the central charges $a$ and $c$,  it turns out that the IR spectrum of their chiral operators is also identical and hence either of those flows represents an equivalent choice.  

However, scanning over the set of all possibilities can be somewhat challenging in practice, given that the number of possibilities increases quite rapidly with the rank of the gauge group. We investigated the implications of our proposal for certain low-rank gauge groups and found that the proposal can be greatly simplified: it turns out that at any stage, it always suffices to only decouple the operator with the lowest scaling dimension (out of the ones that need to be decoupled). Therefore, we propose that during any single iteration, the correct $a$-maximization procedure involves removing only that unitarity-violating operator, which has the lowest scaling dimension at that point. Of course, this cycle has to be repeated over and over until there are no more operators to decouple.

We find another interesting issue that arises from the decoupling of operators: The `baryonic flavor symmetry' in the UV \emph{can mix} with the $R$-symmetry in the IR, contrary to the usual expectation. Let us clarify what we mean by this. 
Let $F$ denote a $U(1)$ flavor symmetry. In \cite{Intriligator:2003jj}, it was shown that if $F$ is such that ${\rm Tr} F =0$ (often such a flavor symmetry is said to be `baryonic'), then it can not mix with the $R$-symmetry of the theory. The appearance of unitary-violating operators that eventually decouple from the interacting sector, presents an interesting caveat to a naive application of this rule. 
The point is that the above rule applies to only those flavor symmetries whose trace is zero in the interacting IR-CFT. In principle one can construct theories where $\Tr F = 0$ in the UV but becomes non-zero in the IR. In particular, some of the operators that decouple from the interacting sector might carry a non-trivial charge with respect to $F$, in which case, the tracelessness of $F$ is only restored once the charges of these decoupled operators are also taken into account. Let us give an explicit example of such theories. Consider the $SU(N)$ gauge theory with the matter as follows: a chiral multiplet $S$ transforming in the rank-2 symmetric traceless tensor, a chiral multiplet $\widetilde{A}$ transforming in the conjugate of the rank-2 anti-symmetric tensor, $N_f$ fundamental chiral multiplets $Q$ and $N_f+8$ anti-fundamental chiral multiplets $\widetilde{Q}$. Notice that this theory is chiral. We study this theory in detail in section \ref{sec:TrFviolating}. In particular, this theory has a traceless $U(1)$ flavor symmetry with the following charge assignments: 
\begin{align}
	\begin{array}{c|c|c|c|c}
		& S & \widetilde{A} & Q & \widetilde{Q}\\ \hline
		U(1) & 0 & 0 & N_f+8 & -N_f 
	\end{array}
\end{align}
We find that this $U(1)$ flavor symmetry is no longer traceless in the IR, and hence, the $R$-symmetry mixes with it in a non-trivial fashion. This is due to the fact that some of the operators that get decoupled along the RG flow (such as $Q (S  \tilde{A})^n \tilde{Q}$) are charged under $F$. On the other hand, in non-chiral theories, it is always possible to parameterize the trial $R$-charge such that it commutes with the traceless $U(1)$ (as was done in \cite{Intriligator:2003jj}) thus ensuring that the above mentioned caveat due to decoupling of charged operators does not arise, since in non-chiral theories decoupled operators charged under $U(1)_F$ will always appear in pairs of opposite charges. 

Thus we see that the appearance of operators that decouple as free fields from the interacting theory has a rather important and non-trivial effect on the nature of the RG-flow of the theory. One should, therefore, carefully account for them when applying $a$-maximization.

\subsection{Weak Gravity Conjecture in AdS}

The `exotic' CFTs with a dense spectrum that we describe above do not have a weakly coupled gravity dual. Nevertheless, AdS/CFT correspondence implies that any CFT should be interpreted as a quantum theory of gravity in anti-de Sitter space, albeit it may be nothing like the weakly-coupled Einstein-like gravity. Generic CFTs might be dual to bulk theory in AdS with a curvature radius of Planck scale, light string states, and light Kaluza-Klien modes, strong quantum effects, and so on. Therefore, it would be interesting to ask: what are the possible generic features of this `exotic' quantum gravity that one can extract?

In view of this, we test the Weak Gravity Conjecture (WGC) \cite{ArkaniHamed:2006dz}. Simply put, WGC states that in any consistent theory with gravity, ``gravity is the weakest force.'' More precisely, WGC demands that there should be at least one state whose mass is smaller than its charge with respect to any other gauge interaction, such that no absolutely stable remnants may exist after black holes evaporate \cite{ArkaniHamed:2006dz}. These results were extended to the case of the AdS background in \cite{Nakayama:2015hga}. There the authors argued that the equivalent statement is that non-BPS Reissner-Nordstr\"om black holes are unstable and examined the implications of the black hole decay on the boundary CFT. Depending upon the mode of black hole decay, they arrive at a set of bounds for the spectrum of the boundary CFT. The simplest of these bounds states that the spectrum should contain an operator satisfying
\begin{align}
	\frac{q^2}{\Delta^2}\geq\frac{40}{9}\frac{C_F}{C_T}.
\end{align}
Here $C_F$ and $C_T$'s are coefficients of the two-point functions of the conserved flavor currents and the stress-energy tensor. In 4d $\mathcal{N}=1$ SCFT one may compute the coefficients as \cite{Nakayama:2015hga}
\begin{align}
	C_F=-\frac{9}{4\pi^4}\,\Tr (RF^2),\qquad C_T=\frac{40}{\pi^4}c = \frac{5}{4\pi^4}\left(9\,\text{Tr}R^3-5\,\text{Tr}R\right) \ , 
\end{align}
where $F$ and $R$ are flavor and R-symmetry generators. 

When the flavor symmetry of the boundary CFT is a product over multiple $U(1)$s, one needs to consider the decay of arbitrarily charged black holes in the dual theory. Requiring that black holes with an arbitrary charge vector should still not leave any stable remnants, leads to the so-called ``convex-hull condition" \cite{Cheung:2014vva}.
The convex-hull condition can be obtained as follows. Consider an extremal black hole of mass $M$ and charge $\vec{Q}$, where each component of $\vec{Q}$ denotes the charge with respect to the corresponding $U(1)$, with proper normalization.
The WGC demands that there are no absolutely stable remnants with any combination of charges. Suppose we have a charge-to-mass ratio vector $\vec{Z} = \vec{Q}/M$ that decays into a set of particles with mass and charges $(m_i, \vec{q}_i)$ with multiplicity $n_i$. Then we have
\begin{align}
	M >  \sum_i n_i m_i \ , \quad \vec{Q} = \sum_i n_i \vec{q}_i \ , 
\end{align} 
from the charge and energy conservation. We have 
\begin{align}
	\vec{Z} = \frac{\vec{Q}}{M} = \sum_i \frac{n_i \vec{q}_i }{M} = \sum_i \frac{n_i \vec{q}_i}{m_i} \frac{m_i}{M} = \sum_i \s_i \vec{z}_i  \ , 
\end{align}
with $\vec{z}_i = \vec{q}_i / m_i $ and $\s_i = n_i m_i /M$. We have $\sum_i \s_i < 1$ so that it defines a convex-hull generated by the charge-to-mass ratio vectors $\vec{z}_i$ that comes from the decay channel. On the other-hand, the extremal black hole satisfies $|\vec{Z}|=1$. This region is not necessarily inside the convex-hull region even if we have $|\vec{z}_i| >1 $ for each $i$. Therefore we have a more stringent condition than just requiring the individual $U(1)$'s to satisfy the WGC. 


We will show that all of our `exotic' theories with dense spectrum satisfy (the convex-hull version of) the Weak Gravity Conjecture for sufficiently large $N$, even though we do not have a clear interpretation in terms of black hole decay in the (highly quantum and stringy) AdS dual in the bulk. Sometimes we find the WGC to hold even for a small value of $N$. However, we find that for small values of $N$, the WGC is not always satisfied by our theories.  
\footnote{It was also the case for the SQCD in conformal window \cite{Nakayama:2015hga}.} It is possible that we did not correctly identify the charged operators in the non-BPS sector which makes the WGC to be valid eventually. If the version of WGC we used is strictly true in any SCFT, it means that there must be a light charged operator in the non-BPS sector. 

Another possibility is that at least the version of WGC that we use in this paper is not a generic property of 4d superconformal theory. It could either mean that the WGC is not satisfied in highly quantum/stringy setup, or there may be a weaker version of the WGC that is generic enough so that it holds for an arbitrary 4d SCFT. Indeed a version of WGC in AdS$_3$ was studied using the modular bootstrap \cite{Benjamin:2016fhe, Montero:2016tif, Bae:2018qym}, rigorously showing that the WGC holds for any unitary 2d CFT. 
We do not know whether such a generic constraint exists in higher dimensions, but see, for example \cite{Heidenreich:2016aqi, Montero:2018fns}. 

In summary, we demonstrate that the WGC to be valid for a large class of SCFTs in 4d even for a small value of $N$, which in turn indicates that the WGC goes beyond the semi-classical reasoning based on black hole physics. It does hold in AdS$_5$ `stringy' quantum gravity as well. However, our computation also suggests that this may not be a property of arbitrary SCFT. 

\section{SU(N) theories} \label{sec:SU}

In this section, we classify $SU(N)$ gauge theories in the large $N$ limit. As we discussed in section \ref{subsec:classification}, the only possible matter representations are fundamental ($\bfund$), adjoint ($\textbf{Adj}$), rank-2 symmetric ($\ssym$), rank-2 anti-symmetric ($\santi$) 
and their respective complex conjugates. Let us denote the multiplicities of each representation as $N_{\sfund}$, $N_{\textbf{Adj}}$, $N_{\ssym}$ and $N_{\santi}$ respectively. 
\begin{table}
	\centering
	\begin{tabular}{c|c|c|c}
		Irrep & dim & $T(\mathbf{R})$ & $\CA(\mathbf{R})$ \\
		\hline
		\bfund & $N$ & $\half$ & $1$ \\
		\textbf{Adj} & $N^2 - 1$ & $N$ & $0$ \\ 
		\bsym & $\frac{N(N+1)}{2}$ & $\half(N+2)$ & $N+4$ \\
		\banti & $\frac{N(N-1)}{2}$ & $\half(N-2)$ & $N-4$
	\end{tabular}
	\caption{Relevant representations of $SU(N)$ and their quadratic indices and anomalies.}
\end{table}
Then requiring the absence of gauge-anomalies along with a negative $\beta$-function implies
\begin{align}
\begin{split}
 (N+4)(N_{\ssym} -N_{\overline{\ssym}})+(N-4)(N_{\santi}-N_{\overline{\santi}})+(N_{\sfund}-N_{\overline{\sfund}}) \,=0 \ , \qquad \quad \\
 N\times N_{\text{\textbf{Adj}}} +\frac{N+2}{2}(N_{\ssym}+N_{\overline{\ssym}})+\frac{N-2}{2}(N_{\santi}+N_{\overline{\santi}})+\frac{1}{2}(N_{\sfund}+N_{\overline{\sfund}})\,\leq 3N \ .
\end{split}
\label{eq:Acond}
\end{align}
Henceforth, we assume $N_{\mathbf{R}}\sim \mathcal{O}(1)$ and do not scale with $N$ since we are interested in theories with a fixed global symmetry. When the equality is saturated, the one-loop beta function for the gauge coupling vanishes, and typically the theory does not flow to a superconformal theory. But sometimes the theory possesses a non-trivial conformal manifold \cite{Razamat:2020pra, Bhardwaj:2013qia} so that it can be continuously deformed to an interacting SCFT (sometimes with extended supersymmetry). 
The result is summarized in Table \ref{tab:SUlist}.\footnote{Some of the theories listed here also appear in \cite{Buchel:2008vz}. They classified the ones with vanishing one-loop beta function (not necessarily interacting). One of the entry (1 adjoint + 1 (\bsym + $\overline{\bsym}$) + 1  (\banti + $\overline{\banti}$)) was missing in v1. We thank Eric Perlmutter for informing us on the classification carried out in \cite{Buchel:2008vz}. } We find conditions for the matter multiplets (conformal window) so that the gauge theory flow to a superconformal fixed point. 
We find that there are 4 set of theories with a dense spectrum and 16 set of theories with a sparse spectrum under the assumption that the number of fundamentals $N_f$ much smaller than $N$. Those with a sparse spectrum have relatively familiar properties. Their central charges grow quadratically in $N$, and the dimensions of low-lying spectrum are of $\CO(1)$. 

\begin{table}[h]
	\centering
	\begin{tabular}{|c|ccc|c|}
	\hline
		Theory & $\beta_{\textrm{matter}}$ & chiral & dense & conformal window  \\\hline \hline 
	1 \,\textbf{Adj}\, + $N_f$  ( {\bfund} + {$\overline{\bfund}$} )& $\sim N$ & N & Y& $1 \leq N_f< 2N$\Tstrut\\ 
	1 \bsym\, + 1 $\overline{\bsym}$\, + $N_f$  ( {\bfund} + {$\overline{\bfund}$} )& $\sim N$ & N & Y & $0\le N_f < 2N-2$\Tstrut\\
	1 \banti\, + 1 $\overline{\banti}$\, + $N_f$   ( {\bfund} + {$\overline{\bfund}$} )& $\sim N$ & N & Y & $4\le N_f< 2N+2$\Tstrut \\
	1 \bsym\, + 1 $\overline{\banti}$\, + 8 $\overline{\bfund}$\, + $N_f$  ( {\bfund} + {$\overline{\bfund}$} )& $\sim N$ & Y & Y & $0\le N_f \leq 2N-4^*$\Tstrut  {\rule[-2ex]{0pt}{-3.0ex}} \\ \hline
	2 \bsym\, + 2 $\overline{\bsym}$\, + $N_f$  ( {\bfund} + {$\overline{\bfund}$} )& $\sim 2 \,N$ & N & N & $0\le N_f< N-4$\Tstrut  \\ 
	1 \bsym\, + 2 $\overline{\bsym}$\, + 1 \banti\, + 8 \bfund\, + $N_f$  ( {\bfund} + {$\overline{\bfund}$} ) & $\sim 2\,N$ & Y & N & $0\le N_f< N-6$\Tstrut \\
	1 \bsym\, + 1 $\overline{\bsym}$\, + 1 \banti\, + 1 $\overline{\banti}$\, + $N_f$  ( {\bfund} + {$\overline{\bfund}$} )& $\sim 2\,N$ & N & N & $0\le N_f \leq N^*$\Tstrut \\
	1 \bsym\, + 1 \banti\, + 2 $\overline{\banti}$\, + 8 $\overline{\bfund}$\, + $N_f$  ( {\bfund} + {$\overline{\bfund}$} )& $\sim 2\,N$ &Y & N & $0\le N_f<N-2$\Tstrut \\
		2 \bsym\, + 2 $\overline{\banti}$\, + 16 $\overline{\bfund}$\, + $N_f$  ( {\bfund} + {$\overline{\bfund}$} ) & $\sim 2\,N$ & Y & N & $0\leq N_f<N-8$\Tstrut \\
		1 \,\textbf{Adj}\, + 1 \bsym\, + 1 $\overline{\bsym}$\, + $N_f$  ( {\bfund} + {$\overline{\bfund}$} )& $\sim 2\,N$ & N & N & $0\le N_f< N-2$\Tstrut \\
		2 \banti\, + 2 $\overline{\banti}$\, + $N_f$  ( {\bfund} + {$\overline{\bfund}$} )& $\sim 2\,N$ & N & N & $0\le N_f< N+4$\Tstrut \\
		1 \,\textbf{Adj}\, + 1 \bsym\, + 1 $\overline{\banti}$\, + 8 $\overline{\bfund}$\, + $N_f$  ( {\bfund} + {$\overline{\bfund}$} )& $\sim 2\,N$ & Y & N & $0\le N_f \leq N-4^*$\Tstrut \\
		1 \,\textbf{Adj}\, + 1 \banti\, + 1 $\overline{\banti}$\, + $N_f$  ( {\bfund} + {$\overline{\bfund}$} )& $\sim 2\,N$ & N & N & $0< N_f< N+2$\Tstrut \\
		2 \,\textbf{Adj}\, + $N_f$  ( {\bfund} + {$\overline{\bfund}$} )& $\sim 2\,N$ & N & N & $0\le N_f \leq N^*$\Tstrut \\
		1 ( \bsym \,+ $\overline{\bsym}$ )\, + 2 ( \banti \,+ $\overline{\banti}$ )\, + $N_f$  ( {\bfund} + {$\overline{\bfund}$} )& $\sim 3\,N$ & N & N & $0 \le N_f < 2$\Tstrut \\
		3 \banti\, + 3 $\overline{\banti}$\, + $N_f$  ( {\bfund} + {$\overline{\bfund}$} )& $\sim 3\,N$ & N & N & $0 \le N_f < 6$\Tstrut \\
		1 \,\textbf{Adj}\, + 2 \banti\, + 2 $\overline{\banti}$\, + $N_f$  ( {\bfund} + {$\overline{\bfund}$} )& $\sim 3\,N$ & N & N & $0\leq N_f<4$\Tstrut \\
		1 \,\textbf{Adj}\, + ( \banti\, + $\overline{\banti}$ ) \, + ( {\bsym} + {$\overline{\bsym}$} )& $\sim 3\,N$ & N & N & $*$\Tstrut \\
		2 \,\textbf{Adj}\, + 1 \banti\, + 1 $\overline{\banti}$\, + $N_f$  ( {\bfund} + {$\overline{\bfund}$} )& $\sim 3\,N$ & N & N & $0\leq N_f \leq 2^*$\Tstrut  \\
		3 \textbf{Adj} & $\sim 3 \,N$ & N & N & $*$\Tstrut  \\ \hline
	\end{tabular}
	\caption{List of all possible superconformal $SU(N)$ theories with large $N$ limit and fixed global symmetry. $\beta_{\textrm{matter}}$ denotes the contribution to the 1-loop beta function from the matter multiplets and the fourth column denotes whether the theory possess a dense spectrum or not when $N_f \ll N$. 
	The last column denotes the condition for the theory to flow to a superconformal fixed point. 
The entries with $*$ (if $N_f$ saturates the upper bound) do not flow, but have non-trivial conformal manifolds \cite{Razamat:2020pra, Bhardwaj:2013qia}. We omit the theories that can be obtained via complex conjugation of the matter representations listed here.}
	\label{tab:SUlist}
\end{table}

In the rest of this section, we focus on the theories with a dense spectrum in detail. We compute the central charges at their fixed points and list the chiral operators in the theory and also identify their scaling dimensions. Moreover, we test the AdS version of the Weak Gravity Conjecture for each model. 

\subsection{One adjoint, $N_f$ fundamentals} \label{sec:SUnAdj}
1 \,\textbf{Adj}\, + $N_f$  ( {\bfund} + {$\overline{\bfund}$} ): 
There are two anomaly-free $U(1)$ symmetries that we call $U(1)_A$ and $U(1)_B$. Each field is charged under the symmetries as follows:
\begin{align}
		\begin{array}{c|c|c|c|c}
			& SU(N) & U(1)_B & U(1)_A & R \\ \hline
			Q & \bfund & 1 & N & 1-\frac{N}{N_f}R_\Phi \Tstrut \\
			\widetilde{Q}& \overline\bfund & -1 & N & 1-\frac{N}{N_f}R_\Phi \Tstrut\\
			\Phi & \textbf{Adj} & 0 & -N_f & R_\Phi \Tstrut
		\end{array}
\end{align}
The form of the single-trace gauge-invariant operators is given by 
\begin{itemize}
	\item $\text{Tr}\,\Phi^{n},\quad n=2,\dots,N$ (Coulomb branch operators)
	\item $Q_I(\Phi)^n\widetilde{Q}_J,\quad n=0,\dots,N-1$ (adjoint mesons)
	\item $ \mathcal{Q}_{I_1}^{n_1}\mathcal{Q}_{I_2}^{n_2}\cdots\mathcal{Q}_{I_N}^{n_N}$, (adjoint baryons)
	\item $\widetilde{\mathcal{Q}}_{I_1}^{n_1}\widetilde{\mathcal{Q}}_{I_2}^{n_2}\cdots\widetilde{\mathcal{Q}}_{I_N}^{n_N}$, (adjoint anti-baryons)
\end{itemize}
where $I, J, I_a, J_b$ indices run from $1, \ldots, N_f$ and we suppressed the gauge indices in this paper. Here we define the dressed quarks ($\mathcal{Q}_I$) and antiquarks ($\widetilde{\mathcal{Q}}_J$) as
\begin{align}
\mathcal{Q}_{I}^n = Q_I(\Phi)^n,\quad
\widetilde{\mathcal{Q}}_{J}^n = (\Phi)^n\widetilde{Q}_J \quad (n=0,\dots,N-1 \textrm{ and } I, J=1,\dots,N_f).
\end{align}
Thus we have in total of $N N_f$ dressed quarks and antiquarks. By combining $N$ different types of dressed (anti-)quarks we can form dressed (anti-)baryons. Notice that there can be non-trivial relation between the operators of the form $\CQ^N$ so one should be careful not to overcount them.\footnote{We have used the Hilbert series techniques to verify the gauge-invariant operator content throughout this paper. The adjoint SQCD case has been worked out in \cite{Hanany:2008sb}.}

Now we iterate the $a$-maximization until no gauge-invariant operators violate the unitarity bound $\Delta \geq 1$. We find that many, though not all,
Coulomb branch operators $\Phi^n$ and adjoint mesons $Q \Phi^n \widetilde{Q}$ decouple along the RG flow.\footnote{When $N_f=1$, one can flip all of these operators to obtain $(A_1, A_{2N-1})$ Argyres-Douglas theory \cite{Maruyoshi:2016aim, Fluder:2017oxm}.} In order to identify the interacting part at the fixed point, we introduce flip fields and appropriate superpotential for the decoupled operators as we discussed in Section \ref{subsec:RGflow}. 

\begin{figure}[h!]
	\centering
	\includegraphics[width=9cm]{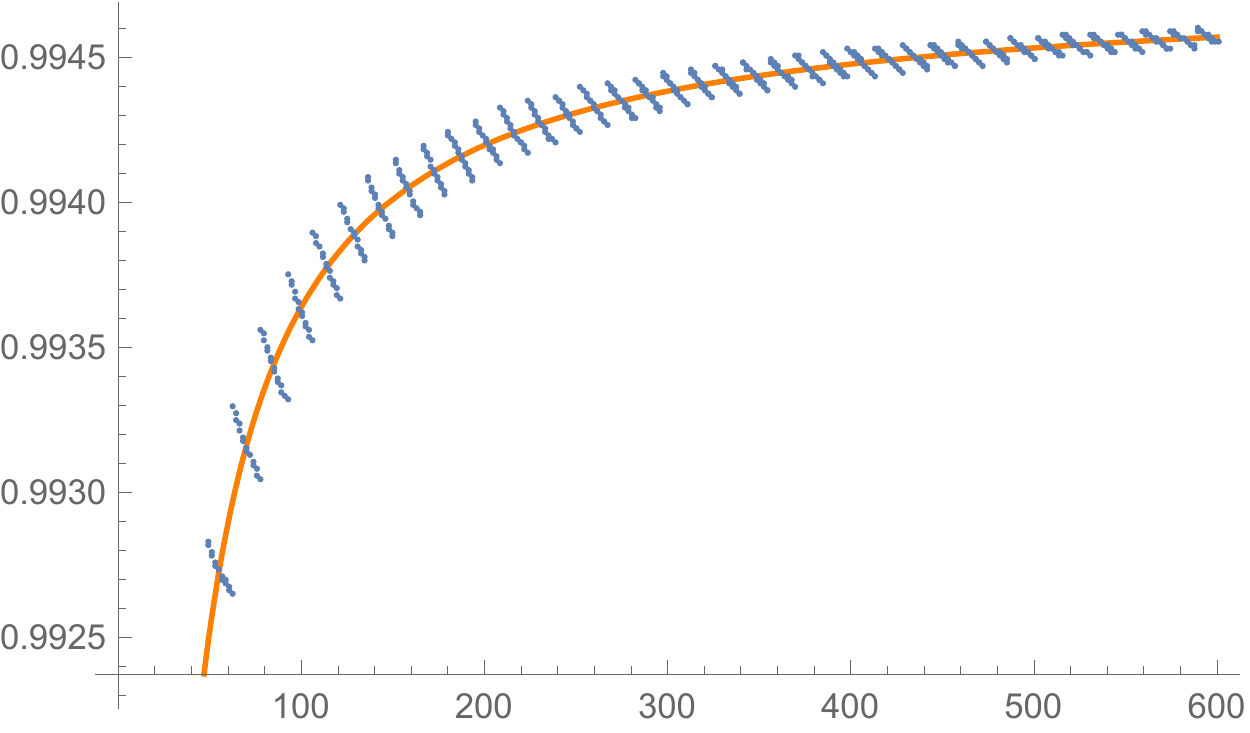}
	\caption{Plot of $a/c$ vs $N$ for the $SU(N)$ theory with 1 adjoint and $N_f=1$. The orange curve fits the plot with $a/c\sim0.994757\, -0.111888/N$.}
	\label{fig:ac_SU1adj1f1fb}
\end{figure}

\paragraph{$N_f=1$ theory} 
The $N_f=1$ case was already studied in \cite{Agarwal:2019crm}. We repeat the same analysis here with a slightly refined analysis of the Weak Gravity Conjecture. We obtain the central charges and $R$-charges to be
\begin{align}
\begin{split}
&a \simeq 0.500819 \,N-0.692539 \ ,\\
&c \simeq 0.503462 \,N-0.640935 \ , \\
&4\pi^4\,C_A\simeq 9.90492\,N^3+9.99795 \,N^2-180.279 \,N+7523.16 \ ,\\
&4\pi^4\,C_B\simeq 12.8808\, N-10.7703\ , \\
&R_{\Phi} \simeq0.712086/N \ ,\\
&R_{Q} \simeq0.284372+0.609971/N \ , 
\end{split}
\end{align}
where we fit the result for $N$ from 100 to 600. 
We see that the central charges grow linearly in $N$. 
We plot the ratio $a/c$ vs $N$ in Figure \ref{fig:ac_SU1adj1f1fb}. 
Note also that ratio $a/c$ of the central charges of the IR SCFT in the large $N$ limit goes close to 1 but not exactly. We find this value to be strictly smaller than 1. (We have checked this numerically up to $N=2000$.) This is another indication that this theory is not quite holographically dual to Einstein-like supergravity in AdS. 

\begin{figure}[h!]
	\centering
	\includegraphics[width=9cm]{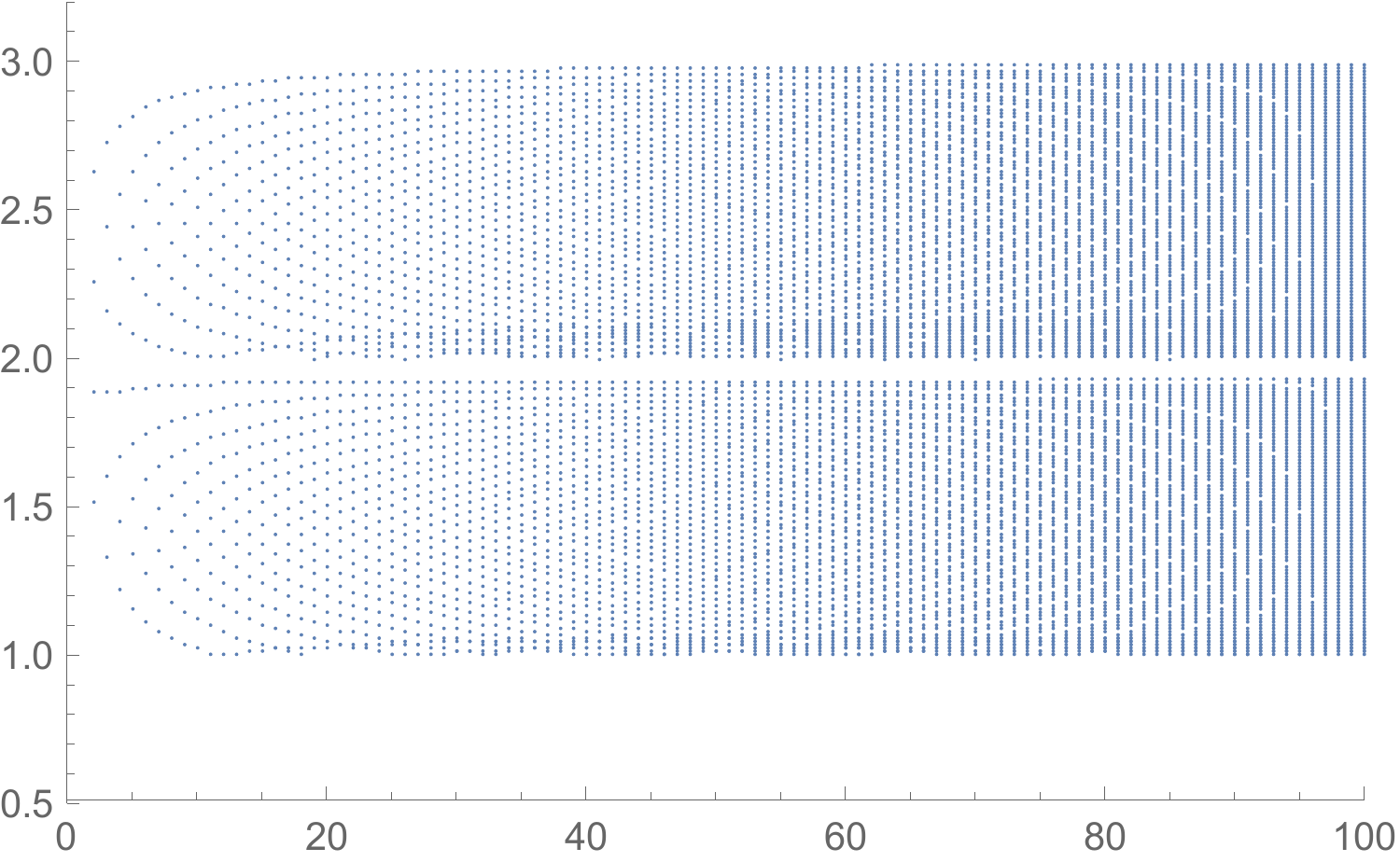}
	\caption{Dimensions of single-trace gauge-invariant operators in $SU(N)$ + 1 \textbf{Adj} + 1 ($\protect\bfund$ + $\overline{\protect\bfund})$ theory. They form a band between $1<\Delta<3$. The baryon operator is rather heavy to be seen in this plot.}
	\label{fig:spec_SU1adj1f1fb}
\end{figure}

Notice that the $R$-charge of the adjoint $\Phi$ scales as $1/N$ at large $N$, which is the main reason why we see the dense spectrum. This makes the scaling dimensions of the adjoint mesons $Q \Phi^i \widetilde{Q}$ to have a spacing of $1/N$. We plot the dimensions of the low-lying operators in Figure \ref{fig:spec_SU1adj1f1fb}.

One may notice a narrow gap ($1.92 \lesssim \Delta \lesssim 2$) in the spectrum depicted in Figure \ref{fig:spec_SU1adj1f1fb}. The lower band consists of the Coulomb branch operators $\Phi^i$ and the adjoint mesons $Q \Phi^i \widetilde{Q}$ that are not decoupled (meaning higher powers in $\Phi$), while the upper band consist of the operators corresponding to the respective flipped fields for each of the decoupled operators. Within the band, the spectrum becomes dense at large $N$. The gap appears because the light operators, given by $\Tr\Phi^i$, $Q\Phi^i \widetilde{Q}$ with $i \sim N$, do not fill the band up to $\Delta=2$. Instead, for this model, the heaviest adjoint meson operator $Q\Phi^{N-1}\widetilde{Q}$ has dimension $\Delta \simeq 1.92$. The upper part of the band consists of flip fields. The dimension of the flipped fields is given by $\Delta_{\textrm{flip}}=3-\Delta_{\CO}$, $\CO$ being the operator that decouples with its naive dimension being $0 < \Delta_{\CO} \leq 1$. Thus the dimension of the flipped fields is  bounded from below by $2$. This explains the gap between the dimensions of the adjoint mesons and the flipped fields.  

The `baryonic' operators remain heavy so that they neither decouple nor form a band. There is a single baryonic (and anti-baryonic) operator for the $N_f=1$ adjoint SQCD given as $Q (\Phi Q) (\Phi^2 Q)\cdots (\Phi^{N-1} Q)$, which lies above the `continuum band' in large $N$. They remain heavy at large $N$ with $\Delta \sim \CO(N)$. 

\begin{figure}[h]
	\centering
	\includegraphics[width=9cm]{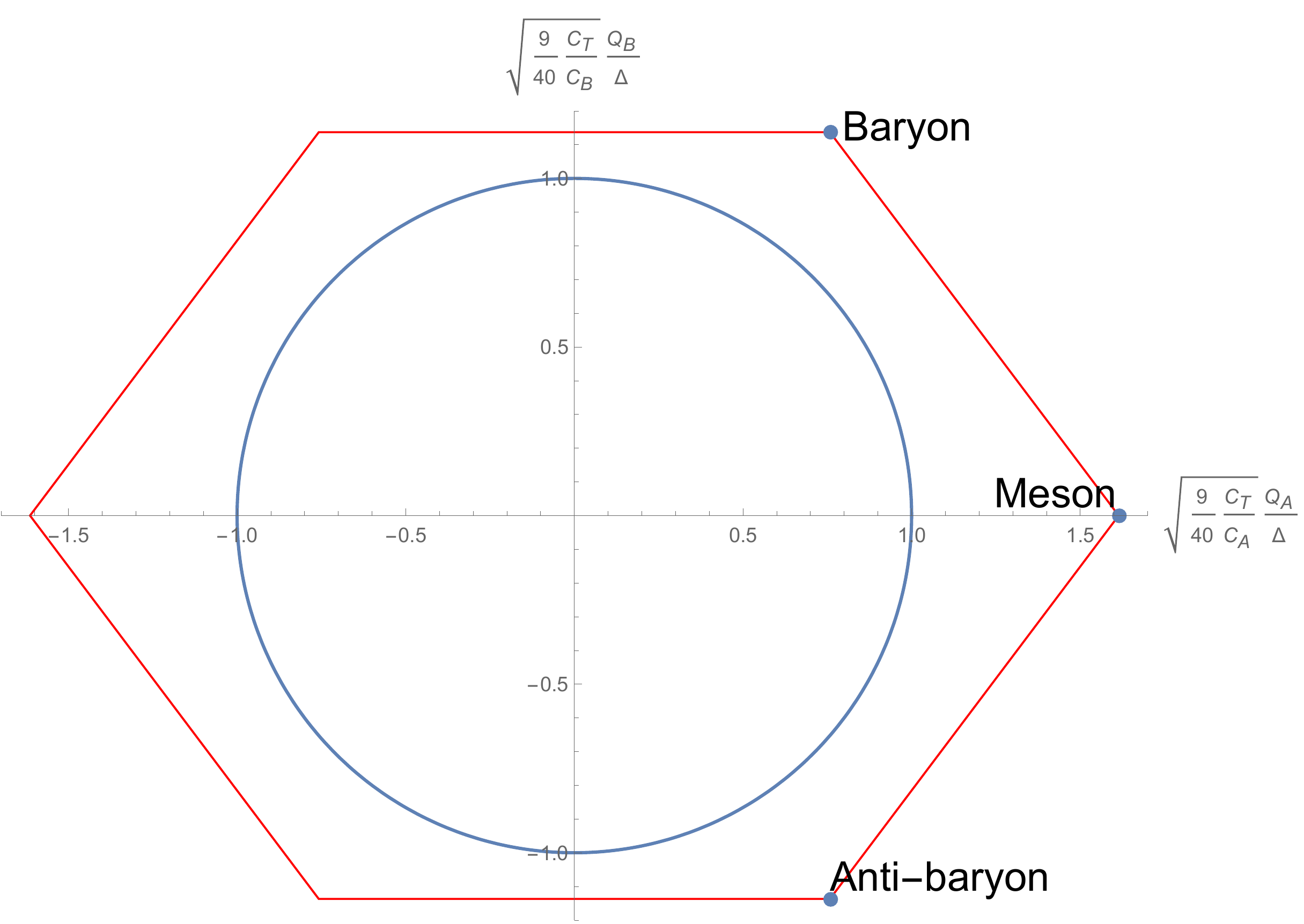}
	\caption{The figure depicts the vector space of charge-to-dimension ratios. The linear combination of lightest meson, baryon, anti-baryon and their conjugate states fill a convex hexagon. It should include the unit circle to satisfy the convex hull condition.}
	\label{fig:ch_SU1adj1f1fb}
\end{figure}
\begin{figure}[h]
	\centering
	\includegraphics[width=9cm]{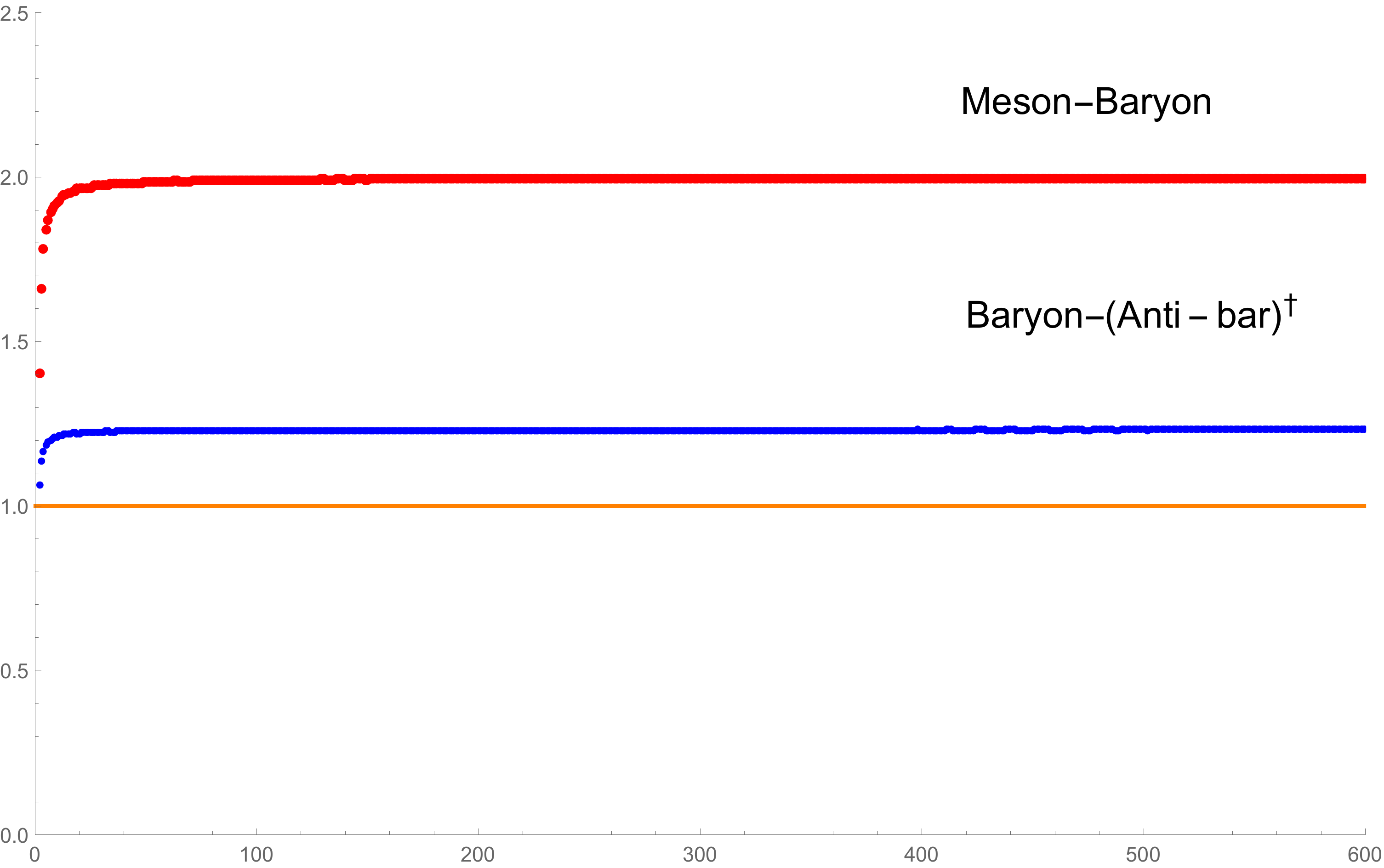}
	\caption{Checking the Weak Gravity Conjecture for $SU(N)$ with 1 adjoint and $N_f=1$. Plot of distances from the origin to the two boundary lines of convex hexagon vs $N$.}
	\label{fig:wgc_SU1adj1f1fb}
\end{figure}

Let us check the AdS version of the Weak Gravity Conjecture for this model. Consider the decay of black hole carrying an arbitrary charge with respect to $U(1)_A$ and $U(1)_B$. 
Let us consider the decay of black holes into three species of light states given by the lightest meson $Q \Phi^n\widetilde{Q}$ (for some $n$ which depends on $N$), baryon $Q (\Phi Q) (\Phi^2 Q)\cdots (\Phi^{N-1} Q)$ and the anti-baryon $\widetilde{Q} (\Phi \widetilde{Q}) (\Phi^2 \widetilde{Q})\cdots (\Phi^{N-1} \widetilde{Q})$. Any linear combination of these three states and their conjugate states with opposite charges form a hexagon in the 2d plane of $U(1)_{A,B}$ charge-to-dimension ratio space depicted in Figure \ref{fig:ch_SU1adj1f1fb}. One can easily check that $U(1)_A$ and $U(1)_B$ are mutually orthogonal.
Then checking convex-hull condition reduces to checking whether distances from origin to the two edges connecting 1) the lightest meson to the baryon, and 2) the baryon to the conjugate of anti-baryon are both larger than 1. Because of the symmetries of hexagon, distances from origin to the other lines are same to these two distances. We checked that this model satisfies the convex hull condition as is depicted in Figure \ref{fig:wgc_SU1adj1f1fb}. 

\begin{figure}[h!]
	\centering
	\includegraphics[width=9cm]{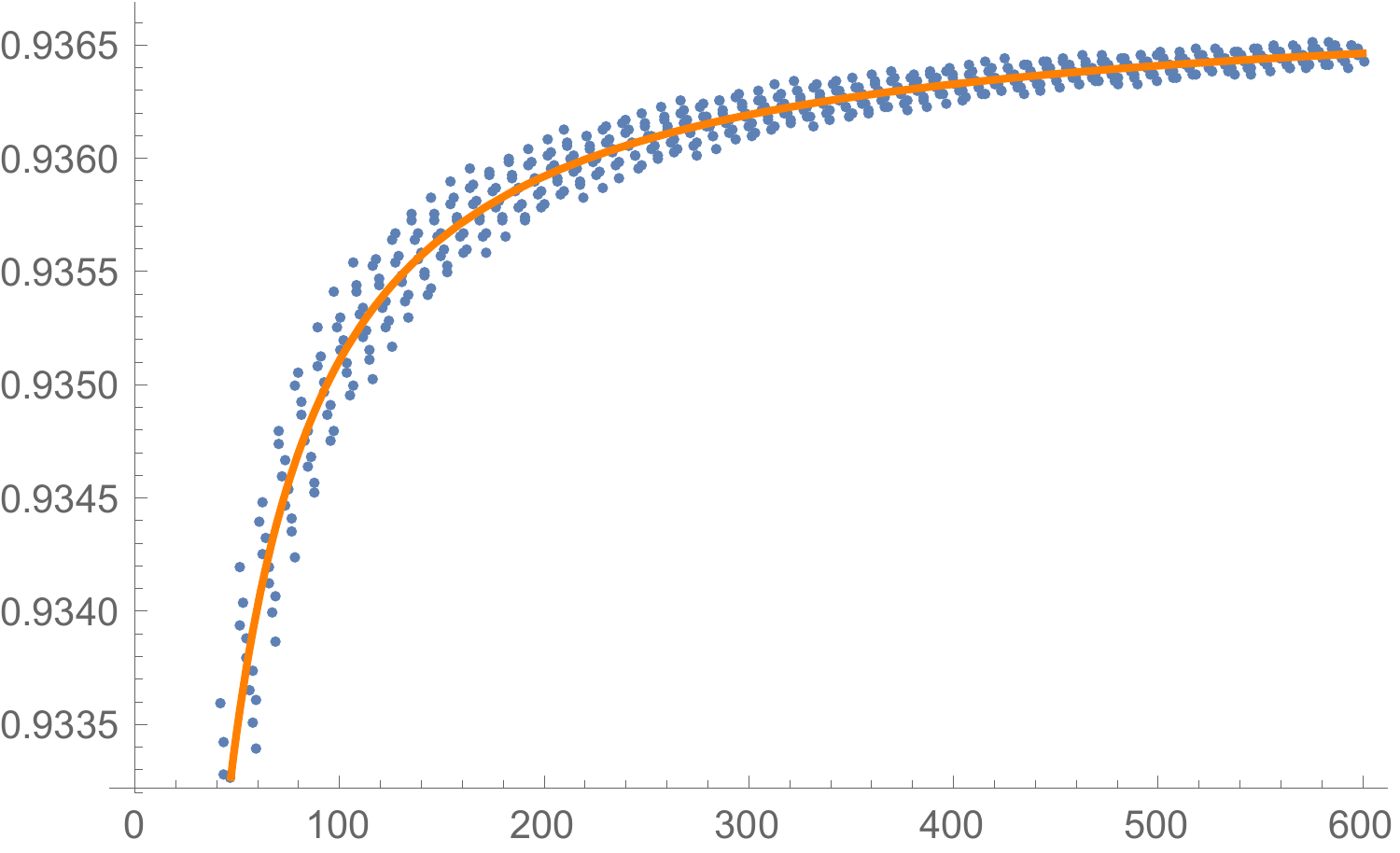}
	\caption{Plot of $a/c$ vs $N$ for the $SU(N)$ theory with 1 adjoint and $N_f=2$. The orange curve fits the plot with $a/c\sim 0.936734\, -0.162684/N$.}
	\label{fig:ac_SU1adj2f2fb}
\end{figure}

\begin{figure}[h!]
	\centering
	\includegraphics[width=9cm]{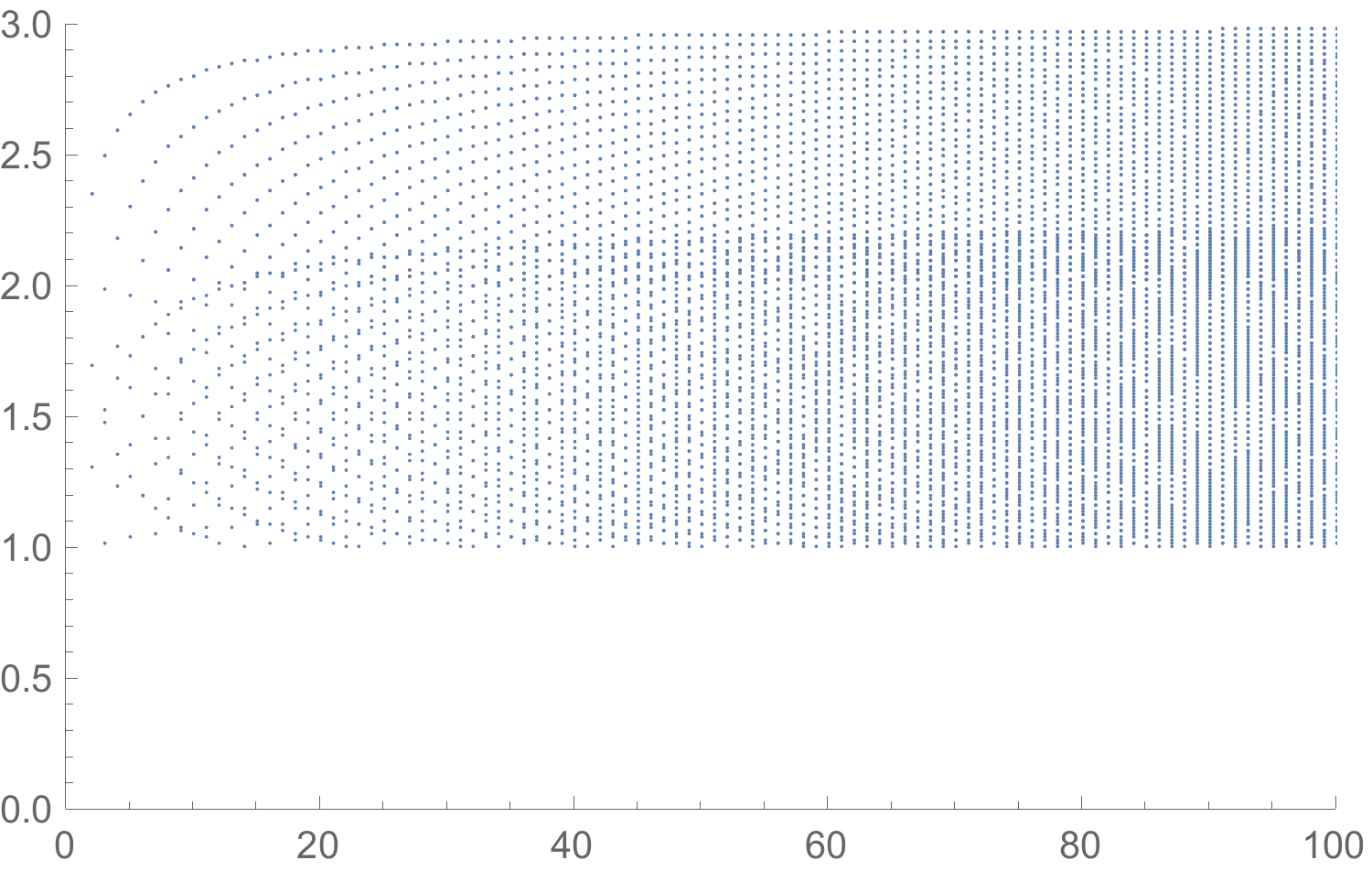}
	\caption{Dimensions of single-trace gauge-invariant operators in $SU(N)$ + 1 \textbf{Adj} + 2 ($\protect\bfund$ + $\overline{\protect\bfund})$ theory. They form a band between $1<\Delta<3$.}
	\label{fig:spec_SU1adj2f2fb}
\end{figure}

\paragraph{$N_f=2$ theory} 
Let us now consider the $N_f = 2$ theory. This case retain many of the same qualitative features as its $N_f=1$ cousin \emph{i.e.} it has a dense spectrum of light operators and displays a linear growth of central charges. In large-$N$ the central charges and the $R$-charges were numerically found to follow the trend given by:
\begin{align}
\begin{split}
&a \simeq 0.942332 \,N-1.99045 \ , \\
&c \simeq 1.00599 \,N-1.95771\ ,\\
&4\pi^4\,C_A\simeq 20.5910\,N^3+42.5137\,N^2-395.380\,N+18443.1 \ ,\\
&4\pi^4\,C_B\simeq 26.8682\,N-40.6757\ ,\\
&R_{\Phi} \simeq1.47931/N\ ,\\
&R_{Q} \simeq0.253576\, +1.16629/N \ .
\end{split}
\end{align}
We plot the ratio $a/c$ in Figure \ref{fig:ac_SU1adj2f2fb}. As was the case for the $N_f=1$ theory, we find that this time too, $a/c$ approaches a value close to 1 but stays strictly smaller than 1. Similarly, the band formed by the Coulomb branch operators, the dressed mesons and the flipped fields is shown in Figure \ref{fig:spec_SU1adj2f2fb}.

\begin{figure}[h]
	\centering
	\includegraphics[width=9cm]{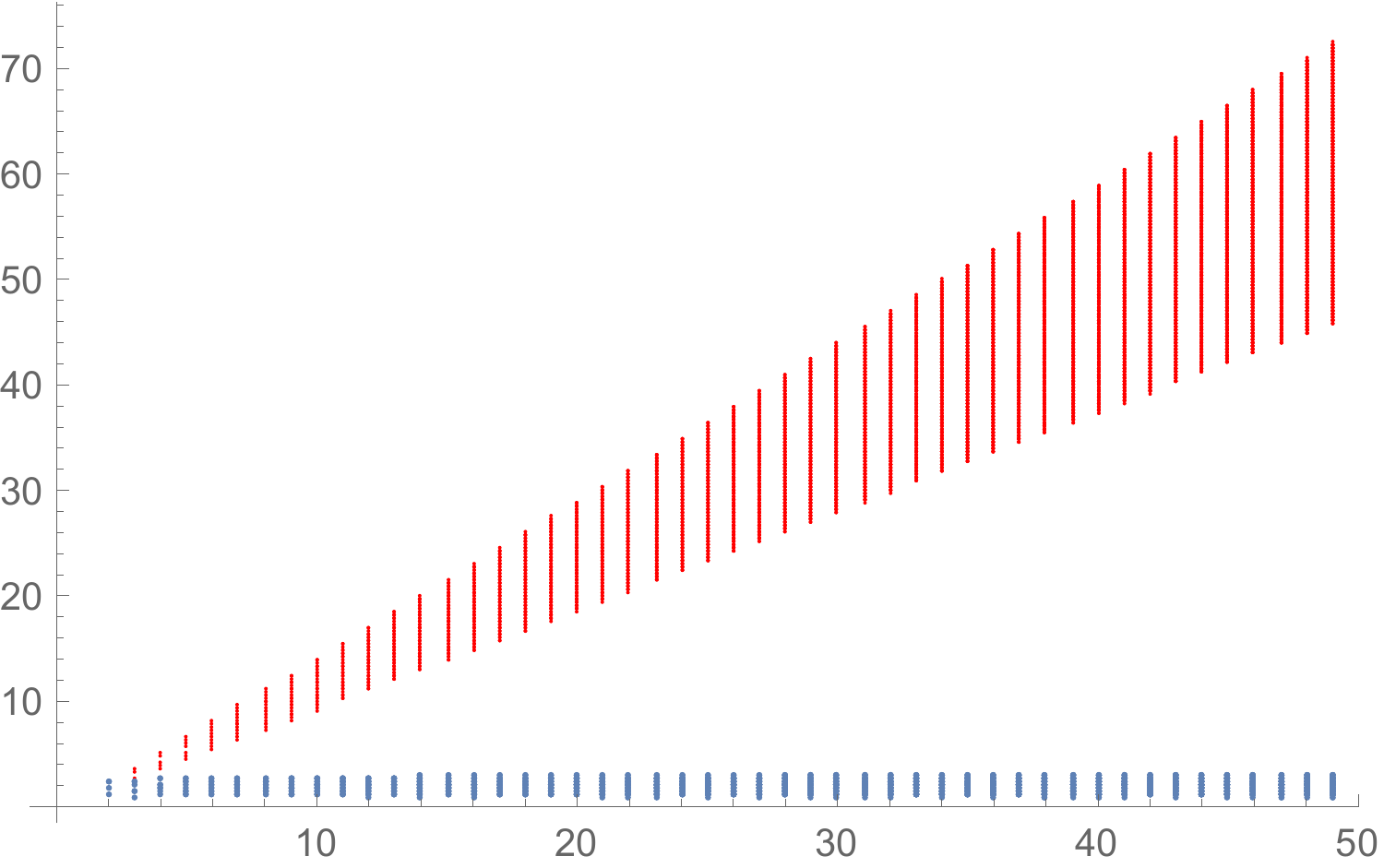}
	\caption{Dimensions of single-trace gauge-invariant operators including baryons in $SU(N)$ + 1 \textbf{Adj} + 2 ($\protect\bfund$ + $\overline{\protect\bfund})$ theory. The baryons(red) form another band above the band of Coulomb branch operators and mesons.}
	\label{fig:spec_SU1adj2f2fb_2}
\end{figure}

However, the spectrum of the $N_f=2$ theory also shows an interesting feature that was not present in the $N_f=1$ theory. Note that unlike the $N_f=1$ theory where there was just one baryon and and one anti-baryon, in the $N_f=2$ case, we have many different baryons in addition to $Q_I (\Phi Q_I) (\Phi^2 Q_I) \cdots (\Phi^{N-1} Q_I)$ with $I=1, 2$. One can form a gauge-invariant operators formed out of $N$ quarks by combining $Q_1$ and $Q_2$ to reduce the number of adjoints. For example, we have $Q_1 Q_2 (\Phi Q_1) (\Phi Q_2) \cdots (\Phi^{N/2} Q_1) (\Phi^{N/2} Q_2)$ for even $N$, which is the one with the smallest number of adjoints $N/2(N/2-1)$. Other baryonic operators can have more adjoints up to $N(N-1)/2$ which gives a width of the baryonic band to be of $\CO(N)$. 
These additional baryons form the second band above the band formed by the Coulomb branch operators, the mesonic operators and the flipped fields. We show the band formed by the baryonic operators explicitly in Figure \ref{fig:spec_SU1adj2f2fb_2}.

\begin{figure}[h]
	\centering
	\includegraphics[width=9cm]{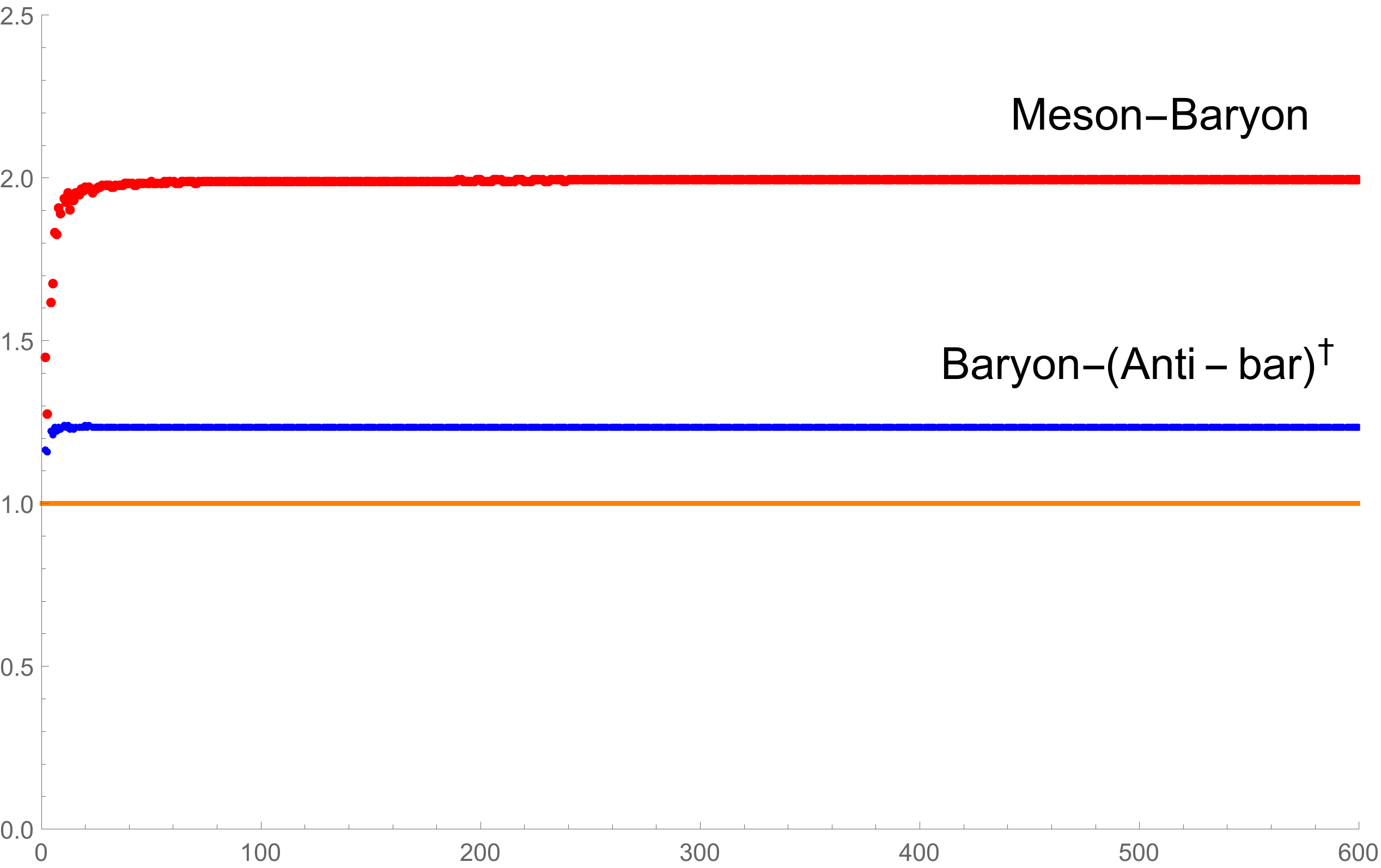}
	\caption{Checking the Weak Gravity Conjecture for $SU(N)$ with 1 adjoint and $N_f=2$. Plot of distances from the origin to the two boundary lines of convex hexagon vs $N$.}
	\label{fig:wgc_SU1adj2f2fb}
\end{figure}

Let us now check the AdS version of the Weak Gravity Conjecture for this case. As before, we would like to consider the decay of an arbitrarily charged black hole into three species of light particles corresponding to the CFT operators given by the lightest meson ($Q_I \Phi^n \widetilde{Q}_J$) for some $n$, the lightest baryon (of the form $Q^N \Phi^{N(N/2-1)}$ for $N$ even), and the lightest anti-baryon (of the form $\widetilde{Q}^N\Phi^{N(N/2-1)}$ for $N$ even). They form a hexagon on the plane of $Q_{A,B}/\Delta$ similar to the one appeared in $N_f=1$ theory. We checked that the $N_f=2$ model also satisfies the convex hull condition as is depicted in Figure \ref{fig:wgc_SU1adj2f2fb}. 

\subsection{One symmetric and $N_f$ fundamentals }
1  ($\bsym$\,+ $\overline{\bsym}$)\, + $N_f$  ( {\bfund} + {$\overline{\bfund}$} ): Let us consider the $SU(N)$ gauge theory with one 2nd rank symmetric tensor, $N_f$ fundamentals and their conjugates. There are 3 anomaly free global $U(1)$'s in addition to the $U(1)_R$ symmetry. The respective charges for the various chiral superfields are given as follows:
\begin{align}
		\begin{array}{c|c|c|c|c|c}
			& SU(N) & U(1)_S & U(1)_B & U(1)_A & R \Bstrut \\ \hline \Tstrut
			Q & \bfund & 0 & 1 & -\frac{(N+2)}{N_f} & 1-\frac{(N+2)R_S-2}{N_f} \Tstrut\\
			\widetilde{Q} & \overline{\bfund} & 0 & -1 & -\frac{(N+2)}{N_f} & 1-\frac{(N+2)R_S-2}{N_f} \Tstrut\\
			S & \bsym& 1 & 0 & 1 & R_S\Tstrut\\
			\widetilde{S} & \overline{\bsym}& -1 & 0 & 1 & R_S \Tstrut
		\end{array}
\end{align}
The gauge-invariant (single-trace) operators of this theory are given by: 
\begin{itemize}
	\item $\text{Tr}\big(S\widetilde{S}\big)^n$,\quad $n=1,\dots,N-1$
	\item $Q_I\big(\widetilde{S}S\big)^n\widetilde{Q}_J,\quad n=0,\dots,N-1$
	\item $Q_I\widetilde{S}\big(S\widetilde{S}\big)^nQ_J,\quad n=0,\dots,N-2$
	\item $\epsilon\, \mathcal{Q}^{n_1}_{I_1}\cdots\mathcal{Q}^{n_N}_{I_N}$
	\item $\epsilon\,  \epsilon\,S\dots S \big(Q_{I_1}Q_{J_1}\big)\dots\big(Q_{I_k}Q_{J_k}\big) $, \quad $k=0,\dots,N_f$
\end{itemize}
The capital letter indices, $I,J, \hdots,$ are flavor indices running from $1,\dots, N_f$. The operators listed in the 4th line above are defined in terms of dressed quarks given by
\begin{align}
\mathcal{Q}^{n}_{I}=\begin{cases}
(S\widetilde{S})^{n/2}Q_I & n=0,2,4,\dots,2N-2\\
(S\widetilde{S})^{(n-1)/2}S\widetilde{Q}_I & n=1,3,5,\dots 2N-3 \ .
\end{cases}
\end{align}
Here the operators $Q_I\widetilde{S}\big(S\widetilde{S}\big)^nQ_J$ are symmetric in their flavor indices. Also note that, in the operators $\epsilon\,  \epsilon\,S\dots S \big(Q_{I_1}Q_{J_1}\big)\dots\big(Q_{I_k}Q_{J_k}\big) $, the two color-indices of the tensor $S$ (as well as those in $\big(Q_{J_k} Q_{K_k}\big)$) are contracted with different $\epsilon$ tensors.
There will also be operators obtained by considering $ Q_I \leftrightarrow \widetilde{Q}_I$ and $S \leftrightarrow \widetilde{S}$.  
For the sake of brevity, we will not show them  here explicitly. 
We also passingly note that the operators $\epsilon\, \mathcal{Q}^{n_1}_{I_1}\cdots\mathcal{Q}^{n_N}_{I_N}$ and $\epsilon\,  \epsilon\,S\dots S \big(Q_{I_1}Q_{J_1}\big)\dots\big(Q_{I_k}Q_{J_k}\big) $ remain rather heavy and therefore never decouple.

\paragraph{$N_f=0$ case}
Let us start with the simplest example with $N_f=0$. In this case, there is only one anomaly-free global symmetry $U(1)_S$, which does not mix with the $U(1)_R$ symmetry owing to it being traceless. The $U(1)_R$ is therefore uniquely determined by the anomaly-free condition as
\begin{align}
R_S=\frac{2}{N+2}.
\end{align}
For this theory, the operator spectrum is simple. We only have the operators $\text{Tr}(S\widetilde{S})^n$ and  
$\mathrm{det}S \equiv \epsilon\,\epsilon SS\cdots S$. The latter one has dimension
\begin{align}
\Delta_{\mathrm{det} S} = \frac{3}{2}N \cdot R_S=\frac{3N}{N+2}
\end{align}
which is always greater than $1$.
On the other hand, the dimension of $\text{Tr}(S\widetilde{S})^n$ operators is given by:
\begin{align} 
\Delta_{\Tr (S\widetilde{S})^n} =  \frac{3}{2}\cdot 2n \cdot R_{S}=\frac{6n}{N+2}.
\end{align}
Among the $\text{Tr}(S\widetilde{S})^n$ operators those with dimension less than or equal to 1 decouple:
\begin{align}
\frac{6\,n}{N+2} \le 1\quad\Longrightarrow\quad  n \le \left\lfloor \frac{N+2}{6} \right\rfloor
\end{align}
The decoupled operators will be replaced by their corresponding flipped-fields.
The dimensions of the remaining $\text{Tr}(S\widetilde{S})^n$ operators 
lie in the region $1 < \Delta < 6$ at large $N$, thus forming  a dense band. 

We find that the central charges $a$ and $c$ are given as
\begin{align}
\begin{split}
a & \simeq\frac{95\,N^4+199\,N^3+39\,N^2-164\,N-88}{72\,(N+2)^3} \underset{N\gg 1}{\longrightarrow}\quad\frac{95}{72}N \ , \\
c &\simeq\frac{30\,N^4+61\,N^3+15\,N^2-36\,N-16}{24\,(N+2)^3}  \quad \underset{N\gg 1}{\longrightarrow}\quad\frac{5}{4} N\ , \\
& \qquad \frac{a}{c} \quad\underset{N\gg 1}{\longrightarrow}\quad\frac{19}{18}  \ .
\end{split}
\end{align}
The central charges grow linearly in $N$. Notice that the ratio of $a$ and $c$ \emph{does not} converge to $1$ at large $N$. 

From the spectral data one can test the WGC. The only chiral operator that is charged under $U(1)_S$ is the 
operator $\mathrm{det} S$. It has a charge-to-mass ratio given as
\begin{align}
\frac{q_S}{\Delta_{S}}=&\,\frac{N+2}{3} \ , 
\end{align}
while the flavor central charge $C_S$ is given as
\begin{align}
C_S= -\frac{9}{4\pi^4}\frac{N(N+1)}{2}\,2\,(R_S-1) =\frac{9}{4\pi^4}\,\frac{N^2(N+1)}{N+2} \ .
\end{align}
It is  easy to see that the $\mathrm{det} S$ operator indeed satisfies the WGC
\begin{align}
\frac{q_S^2}{\Delta_{S}^2} = \frac{(N+2)^2}{9}\gg\frac{40\,C_S}{9\,C_T}\sim\frac{4}{5} \ , 
\end{align}
for all $N \ge 2$. 

\begin{figure}[h!]
	\centering
	\includegraphics[width=9cm]{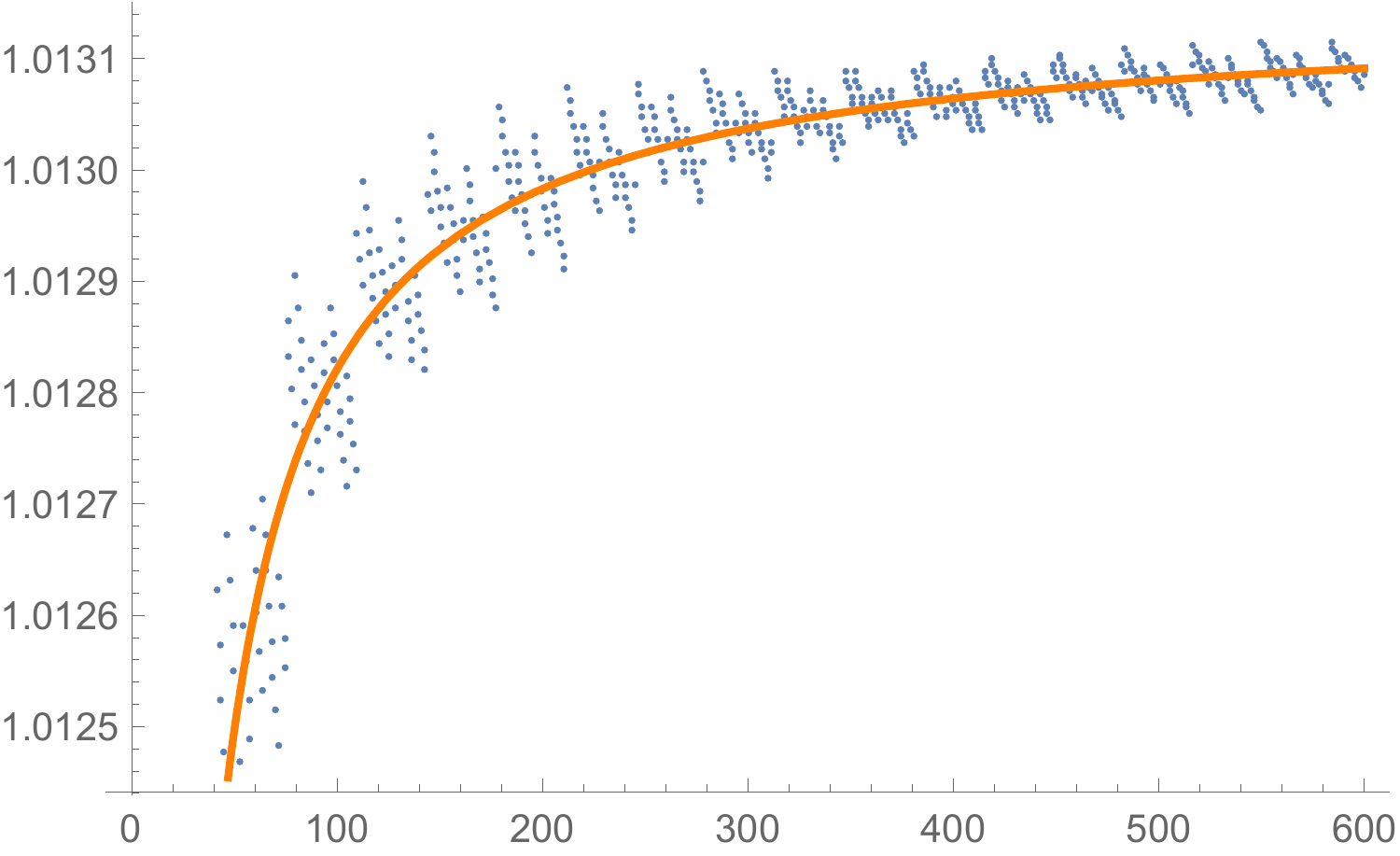}
	\caption{Plot of $a/c$ vs $N$ for the $SU(N)$ theory with one symmetric and one fundamental flavors. The orange curve fits the plot with $a/c\sim 1.01315-0.0324287/N$.}
	\label{fig:ac_SU1s1sb1f1fb}
\end{figure}

\begin{figure}[h!]
	\centering
	\includegraphics[width=9cm]{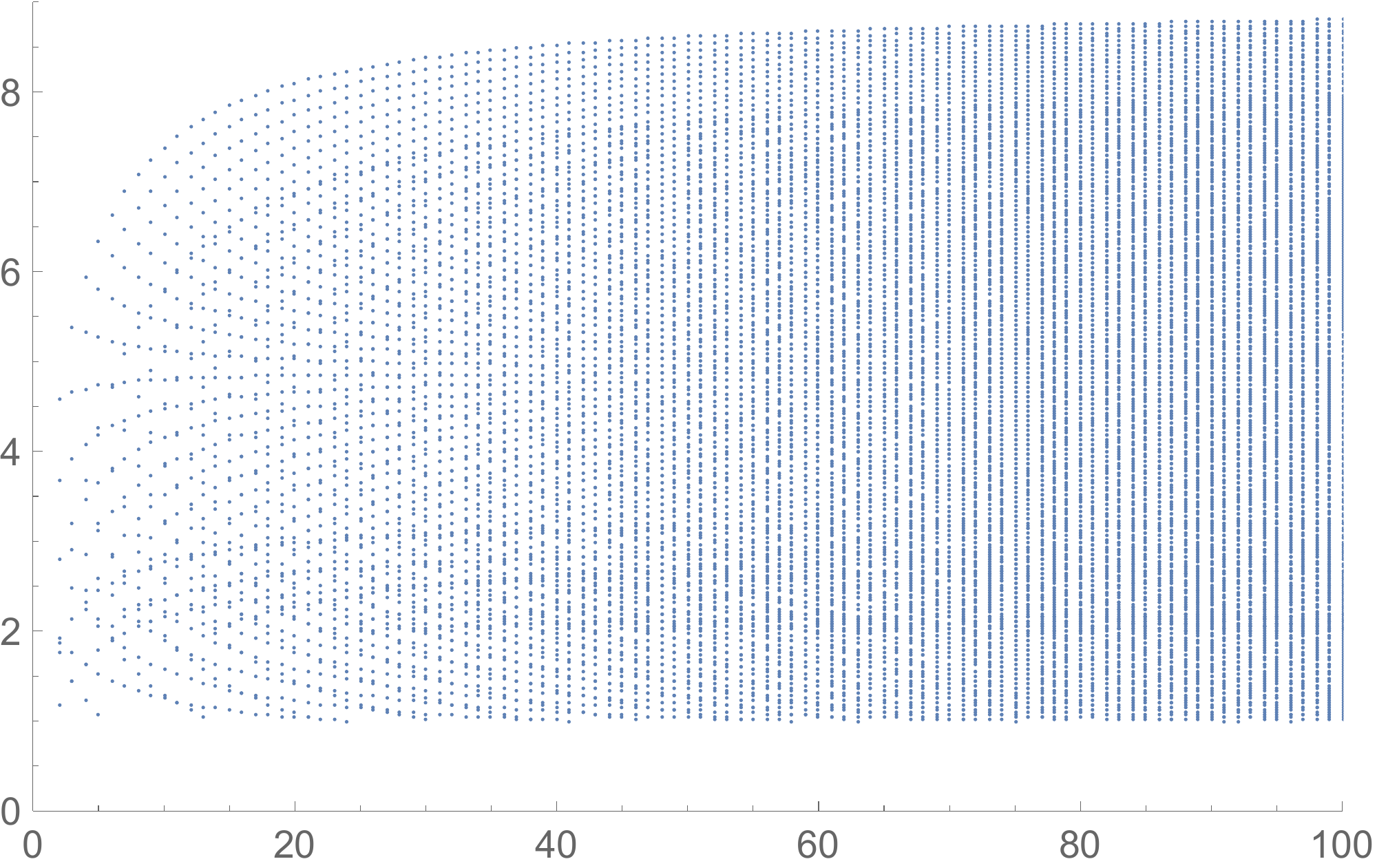}
	\caption{The spectrum of (single-trace) gauge-invariant operators for the $SU(N)$ theory with one symmetric and one fundamental flavors. It has a dense band between $1 < \Delta<9$.}
	\label{fig:spec_SU1s1sb1f1fb}
\end{figure}

\paragraph{$N_f=1$ case}
Let us consider the $N_f=1$ case. Now the IR $U(1)_R$ symmetry has to be determined by $a$-maximization. 
We find
\begin{align}
\begin{split}
&a\simeq\,1.78014 \,N-7.88976 \ , \\
&c\simeq\,1.75705 \,N-7.73346 \ , \\
&4\pi^4 C_S\simeq\,8.99994 N^2-15.9092 N+44.5183 \ , \\
&4\pi^4 C_B\simeq\,12.5806 N-14.3088 \ , \\
&4\pi^4 C_A\simeq\,12.1684 N^3+43.0439 N^2+660.400 N-26573.9 \ , \\
&R_{S}\simeq\,2.70361/N \ , \\
&R_{Q}\simeq\,0.252190\, +2.21983/N \ , 
\end{split}
\end{align}
for sufficiently large value of $N$. 
We plot the ratio of central charges $a/c$ as a function of $N$ in Figure \ref{fig:ac_SU1s1sb1f1fb}.
We also plot the spectrum of operators with dimension $1<\Delta< 9$ at the IR in Figure \ref{fig:spec_SU1s1sb1f1fb}.
Once again, they form a band of dense spectrum. The band consists of operators of the form $\text{Tr}\big(S\widetilde{S}\big)^n$, $Q_I\big(\widetilde{S}S\big)^n\widetilde{Q}_J$, $Q_I\widetilde{S}\big(S\widetilde{S}\big)^nQ_J$, $\widetilde{Q}_I S\big(\widetilde{S}S\big)^n\widetilde{Q}_J$ and the flipped fields corresponding to the decoupled operators. The operators $\mathrm{det}S \equiv \epsilon \epsilon S\dots S$, $\epsilon \epsilon S\dots S QQ$ and their conjugates also appear in the band. On the other hand, the operators $\epsilon\, \mathcal{Q}^{n_1}_{I_1}\cdots\mathcal{Q}^{n_N}_{I_N}$ have dimension $\Delta\sim \CO(N)$. They form the second band whose width is also of $\CO(N)$ similar to the one appeared at the $N_f=2$ theory in \ref{sec:SUnAdj}. 

Let us now check the WGC. The three $U(1)$ symmetries are orthogonal to each other, \emph{i.e.} $\text{Tr}\,T_i T_j=0$ for $i\neq j$. We need to find a set of super-extremal particles (operators) that allows arbitrary charged extremal black holes to decay. 
To this end, we compute the charge-to-dimension ratio of all gauge-invariant chiral operators that we listed in the beginning of this subsection. We take this comprehensive approach for this theory in contrary to finding just a set of operators satisfying the convex hull condition in the previous section. This will make sure that we are not missing any chiral operators in the theory. 

\begin{figure}[h]
	\centering
	\subcaptionbox{}{
	\includegraphics[width=4cm]{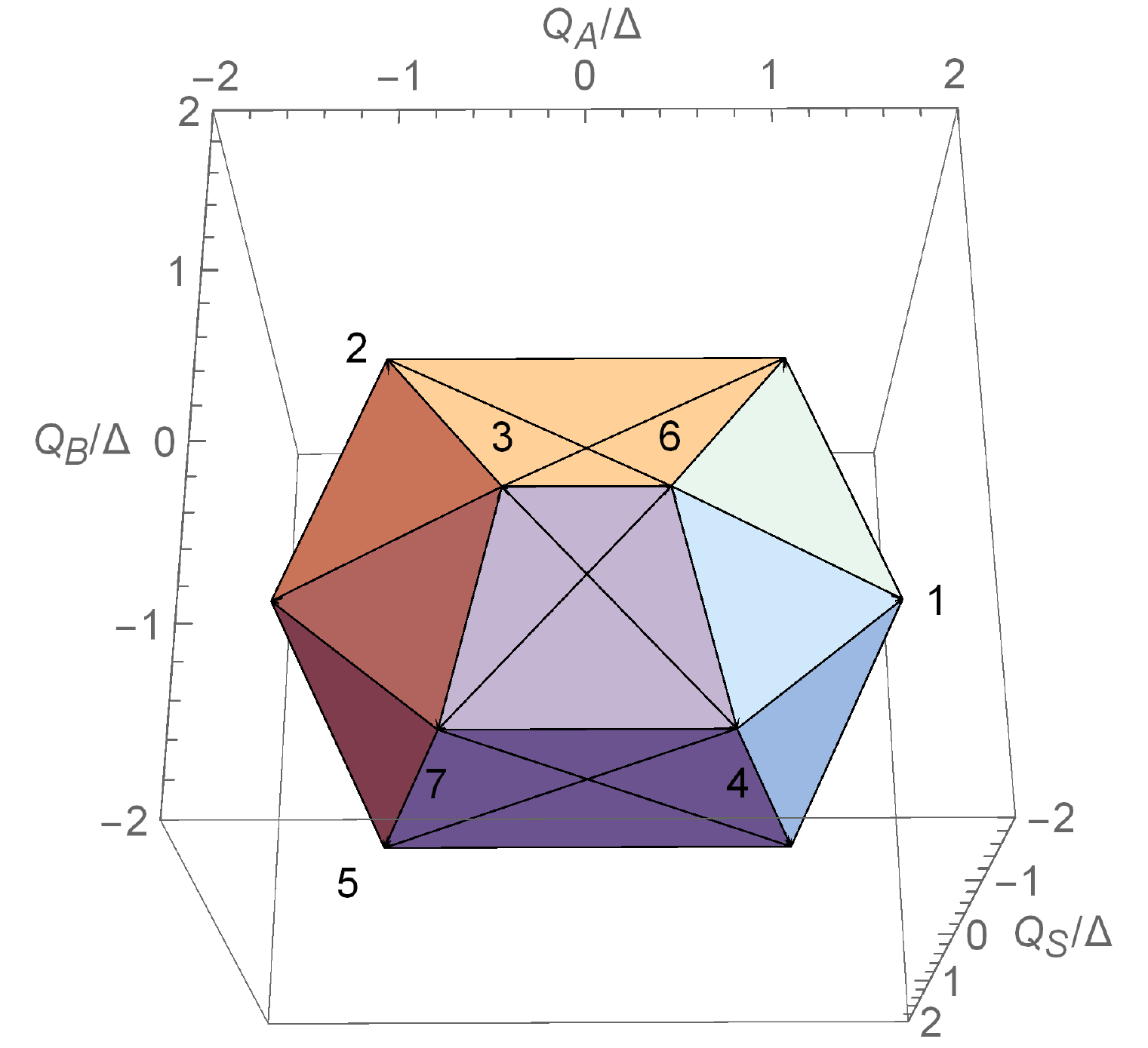}}
	\subcaptionbox{}{
	\includegraphics[width=4cm]{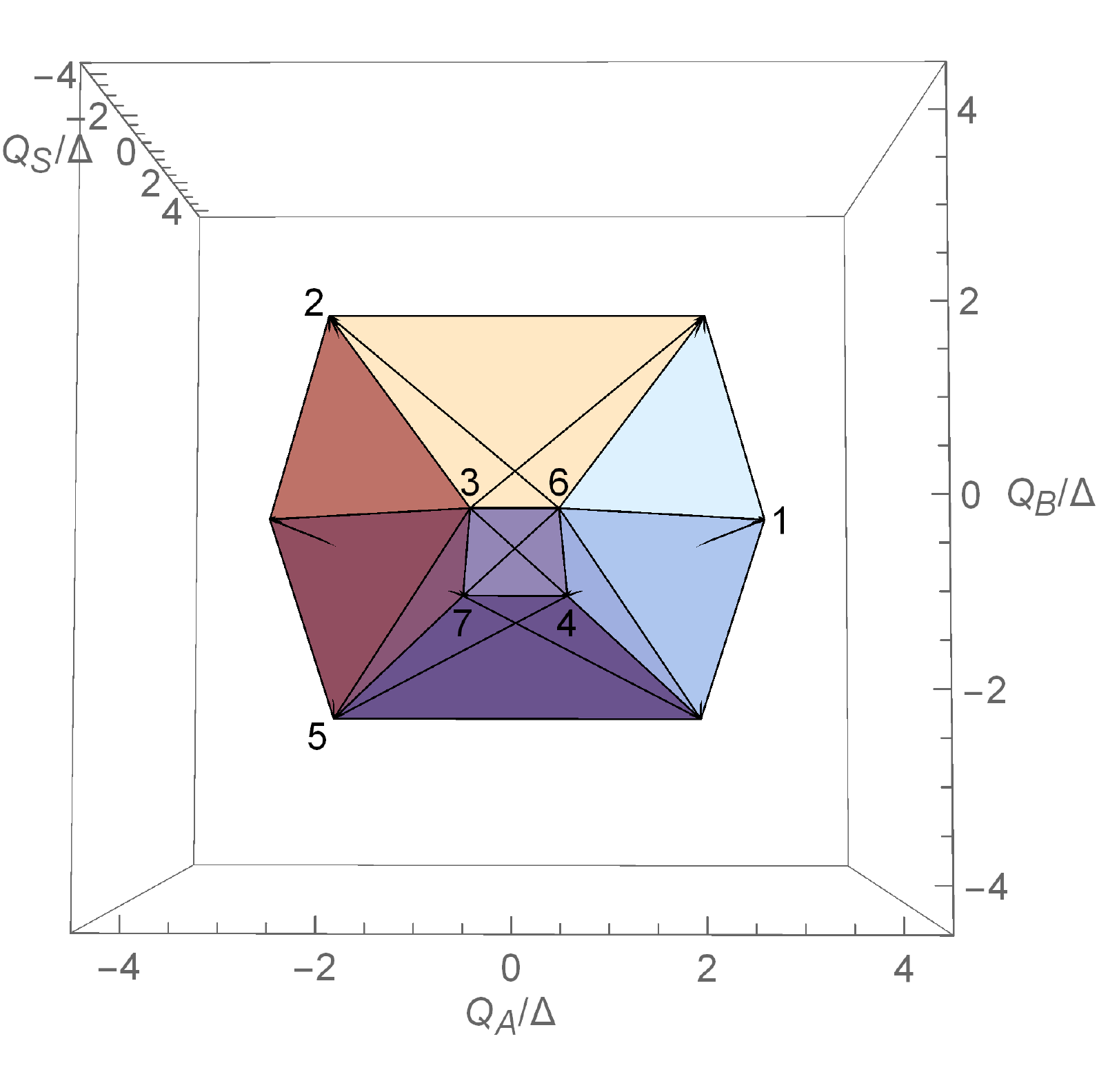}}
	\subcaptionbox{}{
	\includegraphics[width=4cm]{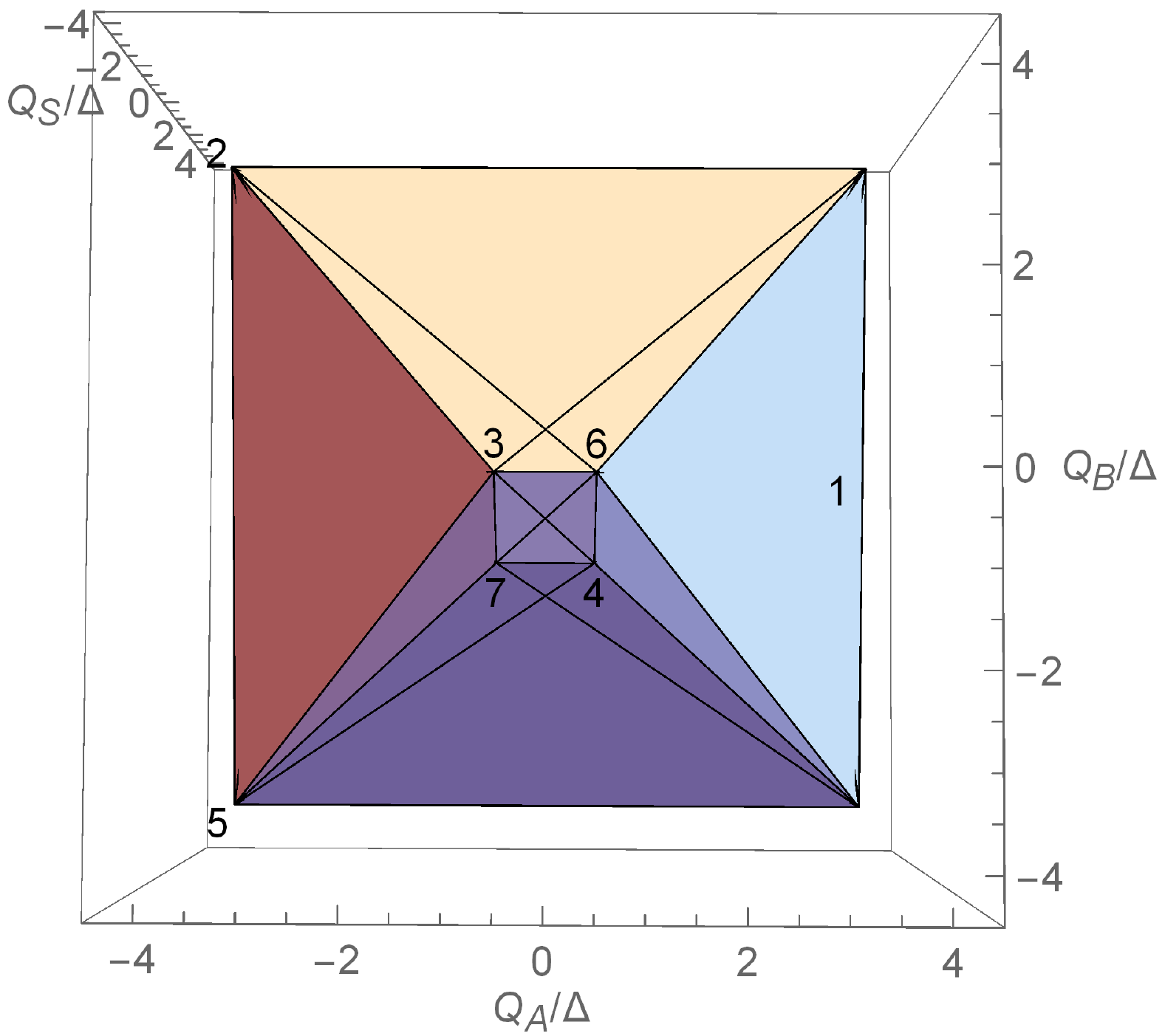}}
	\caption{Charge-to-dimension ratio space for $SU(N)$ theory with 1 symmetric and 1 fundamental flavors. The linear combination of the charge-to-dimension vector of gauge-invariant operators (and their conjugates) fill the polyhedron (a) for $N\leq 8$. For $N \geq 9$, (b) and (c) alternates as the number of decoupled operators of the form $Q_I\big(\widetilde{S}S\big)^n\widetilde{Q}_J$ changes. We normalized each axes by $\sqrt{\frac{9}{40}\frac{C_T}{C_F}}$.}
	\label{fig:ch_SU1s1sb1f1fb}
\end{figure}

The operators of the form $\text{Tr}\big(S\widetilde{S}\big)^n$ and $Q_I\big(\widetilde{S}S\big)^n\widetilde{Q}_J$ and their complex conjugates are charged only to $U(1)_A$, therefore lie on the $Q_A/\Delta$-axis in the charge-to-dimension space. These two sets of operators are distributed on the line segment, where its end points are given by the lightest operator among them, and its complex conjugate. We label this operator as ``1" in Figure \ref{fig:ch_SU1s1sb1f1fb} (a),(b) and (c). 
Next we consider the operators $Q_I\widetilde{S}\big(S\widetilde{S}\big)^nQ_J$ in the charge-to-dimension space. They are distributed on the line segment between the two endpoints given by the smallest $n$ (among the ones that are not decoupled) and the one with the largest $n$. The former is labeled by ``2" in Figure \ref{fig:ch_SU1s1sb1f1fb} (a),(b) and (c), while the latter is inside the convex polyhedron and we do not mark. 
At the same time the operators of the form $\widetilde{Q}_I S\big(\widetilde{S}S\big)^n\widetilde{Q}_J$ are distributed on a line given by the reflection along the $Q_A/\Delta$-axis, stretching from ``5" to a point inside the polyhedrons. 

The operators of the form $\epsilon\,  \epsilon\,S\dots S \big(Q_{I_1}Q_{J_1}\big)\dots\big(Q_{I_k}Q_{J_k}\big) $ are on the line between the two points corresponding to $k=0$ and $k=N_f$ that are labeled as ``3" and ``4" in Figure \ref{fig:ch_SU1s1sb1f1fb}. Similarly, the operators of the form $\epsilon\,  \epsilon\,\widetilde{S}\dots \widetilde{S} \big(\widetilde{Q}_{I_1}\widetilde{Q}_{J_1}\big)\dots\big(\widetilde{Q}_{I_k}\widetilde{Q}_{J_k}\big) $ lie on the line between vertices ``6" and ``7" in Figure \ref{fig:ch_SU1s1sb1f1fb}. 
The operators of the form $\epsilon\, \mathcal{Q}^{n_1}_{I_1}\cdots\mathcal{Q}^{n_N}_{I_N}$ lie on the line between the points corresponding to $\epsilon Q(S\widetilde{S}Q)\cdots((S\widetilde{S})^{N-1}Q)$ and $\epsilon Q(S\widetilde{Q})(S\widetilde{S})Q\cdots$, which turns out to be in the interior of the convex hulls in Figure \ref{fig:ch_SU1s1sb1f1fb} for any $N$. Then various composites constructed out of the single-trace gauge-invariant chiral operators, whose charge-to-mass ratio will fill a polyhedron whose form is shown in Figure \ref{fig:ch_SU1s1sb1f1fb} for different $N$. 

\begin{figure}[h]
	\centering
	\includegraphics[width=9cm]{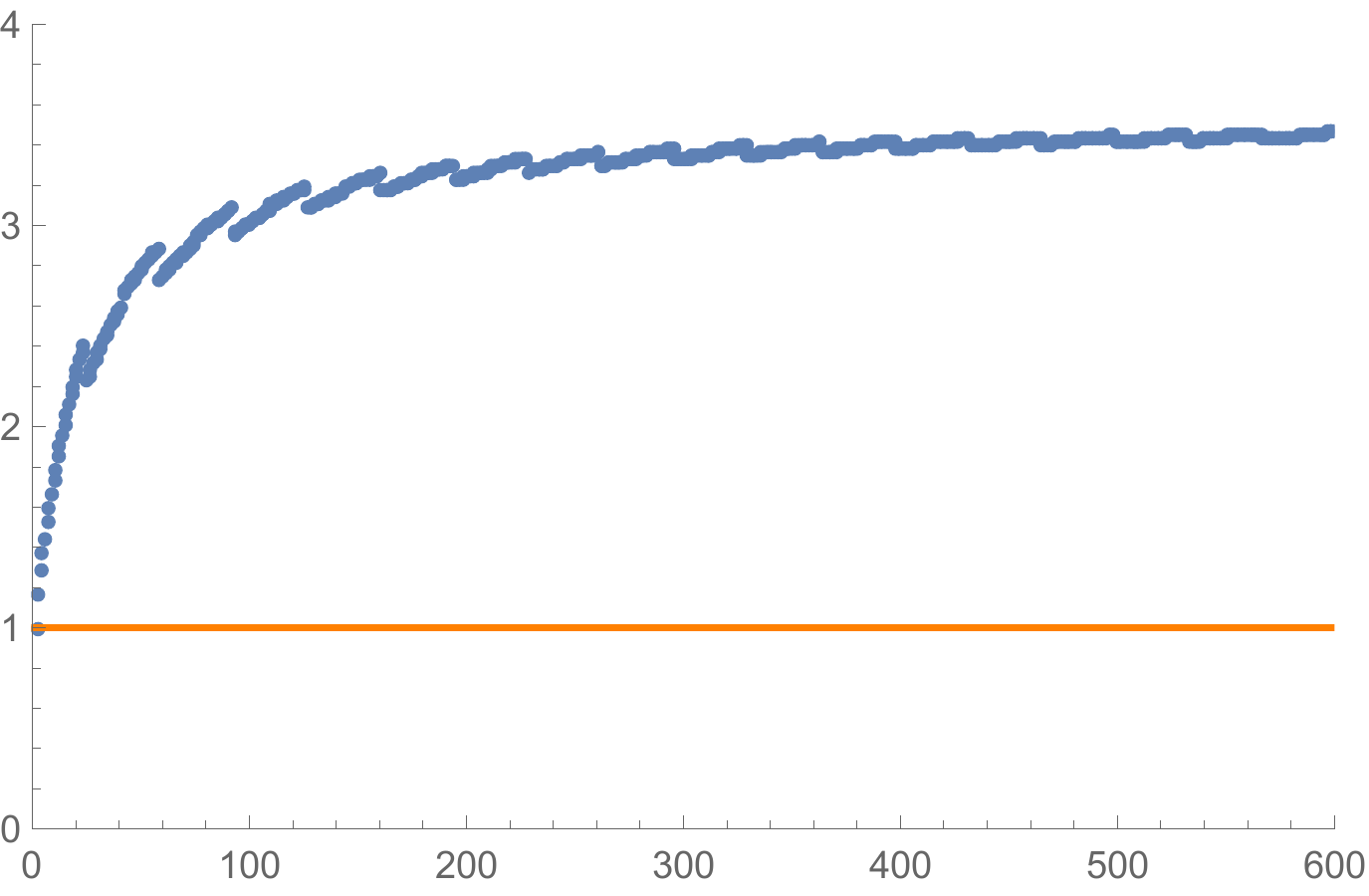}
	\caption{Testing the WGC for $SU(N)$ theory with 1 symmetric and 1 fundamental flavor. Plot of the shortest distance from the origin to the boundary surfaces of the polyhedron in charge-to-dimension ratio space vs $N$.}
	\label{fig:wgc_SU1s1sb1f1fb}
\end{figure}

Now the WGC convex hull condition is satisfied if the distance from the origin to the closest surface is greater than 1. Indeed we find this model satisfies the convex hull condition for arbitrary $N$ as is depicted in Figure \ref{fig:wgc_SU1s1sb1f1fb}.
Note that we are actually doing far more than just identifying one set of particles that allow black holes to decay. We find the entire set of gauge-invariant chiral operators form a polyhedron (with its interior included) in the charge-to-mass space. 

\subsection{One anti-symmetric and $N_f$ fundamentals}
1 \banti\, + 1 $\overline{\banti}$\, + $N_f$  ( {\bfund} + {$\overline{\bfund}$} ): 
Let us consider $SU(N)$ theory with 1 anti-symmetric, $N_f$ fundamentals and their complex conjugate representations. 
There are 3 anomaly free global $U(1)$'s and the $U(1)_R$ symmetry under which the matter multiplets are charged as follows:
\begin{align}
		\begin{array}{c|c|c|c|c|c}
			& SU(N) & U(1)_S & U(1)_B & U(1)_A & R  \\ \hline
			Q & \bfund & 0 & 1 & -\frac{(N-2)}{N_f} & 1-\frac{(N-2)R_A+2}{N_f} \Tstrut\\
			\widetilde{Q} & \overline{\bfund}& 0 & -1 & -\frac{(N-2)}{N_f} & 1-\frac{(N-2)R_A+2}{N_f} \Tstrut\\
			A & \banti & 1 & 0 & 1 & R_A \Tstrut\\
			\widetilde{A} & \overline{\banti}& -1 & 0 & 1 & R_A \Tstrut
		\end{array}
\label{eq:op_SU1a1ab4f4fb}
\end{align}
Notice that only $U(1)_A$ will mix with the $U(1)_R$ symmetry as $U(1)_S$ and $U(1)_B$ are traceless. 
The gauge-invariant operators in this theory are: 
\begin{itemize}
\item $\text{Tr}\,\big(A\widetilde{A}\big)^n,\quad n=1,\dots,\left\lfloor\frac{N-1}{2}\right\rfloor$
\item $\widetilde{Q}_I\big(A\widetilde{A}\big)^nQ_J$,\quad$n=0,\dots\left\lfloor\frac{N}{2}\right\rfloor-1$
\item $Q_I\widetilde{A}\big(A\widetilde{A}\big)^nQ_J$,\quad $n=0,\dots,\left\lfloor\frac{N-1}{2}\right\rfloor-1$
\item $\epsilon\,A^nQ_{I_1}\dots Q_{I_{N-2n}}$,\quad $n=\left\lceil\frac{N-N_f}{2}\right\rceil,\dots,\left\lfloor\frac{N}{2}\right\rfloor$
\end{itemize}
Here the subscripts $I, J$ are flavor indices running from $1,\dots,N_f$. Once again for the sake of brevity, we have not shown the chiral operators transforming in conjugate representation of the ones explicitly mentioned above. Note that the third and the fourth operators have anti-symmetric flavor indices.
When $N$ is even, the fourth line includes the Pfaffian operator $\mathrm{Pf}A \equiv \epsilon A A \cdots A$. 

Upon applying the condition in \eqref{eq:dense_conj} one might expect the theories with $N_f\geq 3$ to show a dense spectrum. However, our analysis reveals that after decoupling half of the $\text{Tr}(A\widetilde{A})^n$ operators, $a$-maximization fails to find a real R-charge but a complex number in $N_f=3$ theory. We therefore conclude that $N_f=3$ theory does not flow to an interacting SCFT in the infrared. Our analysis is consistent with the IR interacting SCFT for $N_f \ge 4$. 
We checked that $N_f=4,5$ indeed have the dense spectrum in the IR. 

\begin{figure}[h]
	\centering
	\includegraphics[width=9cm]{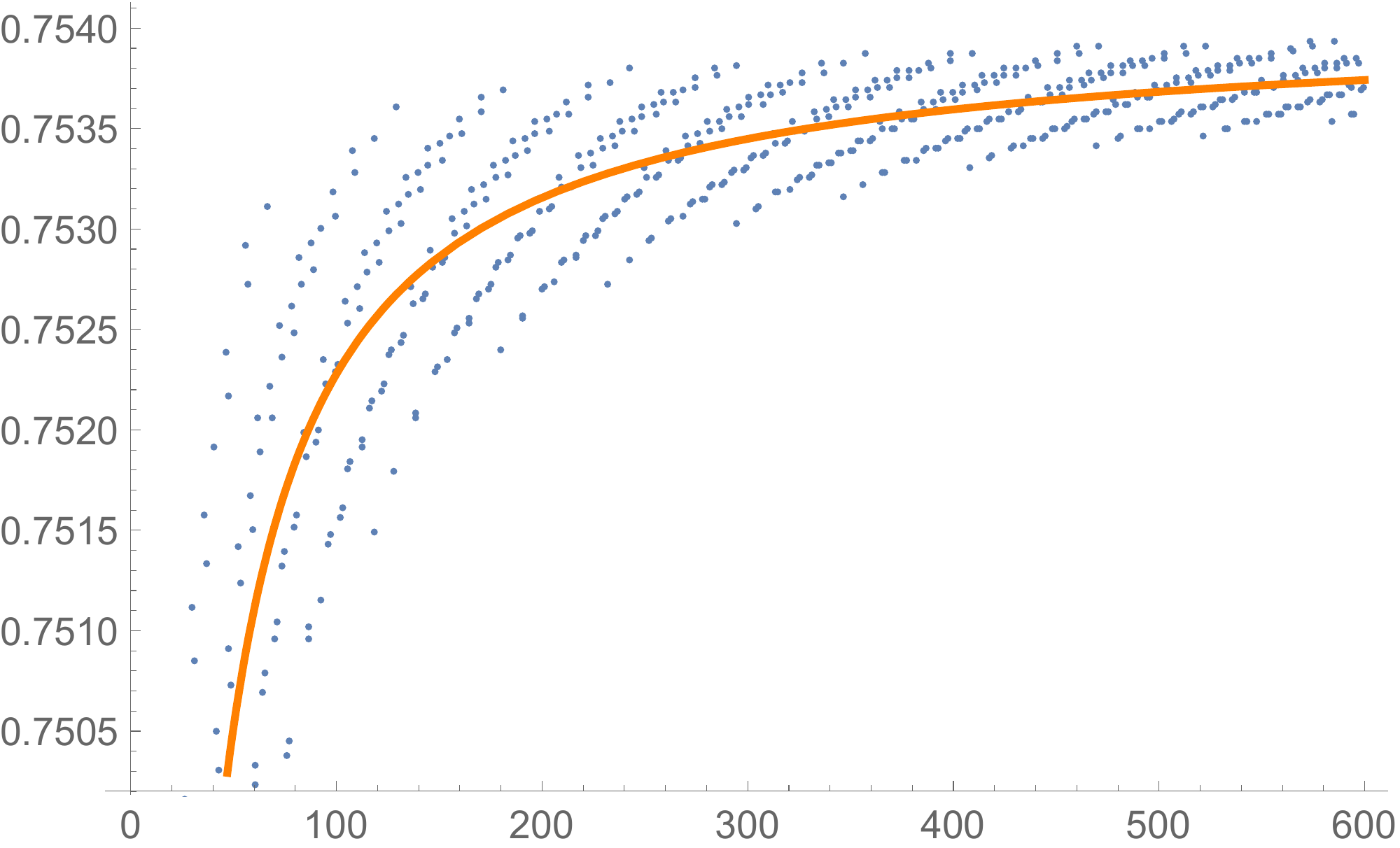}
	\caption{Plot of $a/c$ vs $N$ for the $SU(N)$ theory with 1 anti-symmetric and 4 fundamental flavors. The orange curve fits the plot with $a/c\sim0.754034 -0.175642/N$.}
	\label{fig:ac_SU1a1ab4f4fb}
\end{figure}

\begin{figure}[h]
	\centering
	\includegraphics[width=9cm]{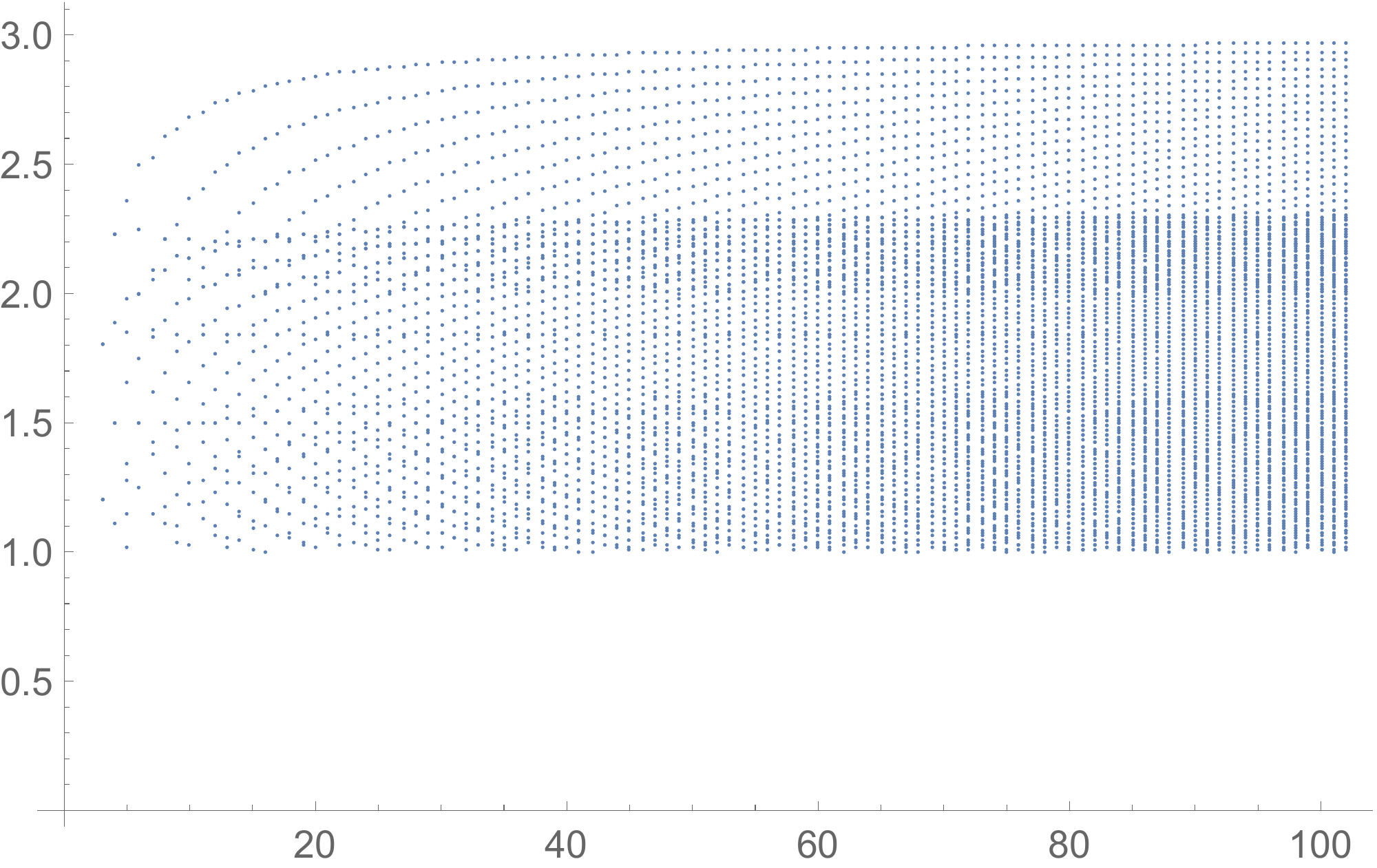}
	\caption{The low-lying spectrum of $SU(N)$ theory with 1 anti-symmetric and 4 fundamental flavors. We find a band of (single-trace) gauge-invariant operator spectrum between $1<\Delta< 3.$}
	\label{fig:spec_SU1a1ab4f4fb}
\end{figure}

\paragraph{$N_f=4$ theory}
Let us now consider the theory with \emph{i.e.} $N_f=4$.
We iterate the $a$-maximization procedure while keeping in mind to decouple the operators that violate the unitarity bounds. Subsequently, we find the central charges and the $R$-charges at large $N$ to be 
\begin{align}
\begin{split}
&a\sim0.516787 \,N+0.213983 \ ,\\
&c\sim0.685367 \,N+0.442017 \ , \\
&4\pi^4C_S\sim9.17285 N^2-24.1146 N+93.2270 \ , \\
&4\pi^4C_B\sim37.2900 N+76.7848\ ,\\
&4\pi^4C_A\sim1.46420 N^3+0.112855 N^2+36.4306 N-1662.43\ ,\\
&R_{A}\sim1.08519/N\ ,\\
&R_{Q}\sim0.228548\, +0.569326/N \ .
\end{split}
\end{align}
We plot $a/c$ vs $N$ in Figure \ref{fig:ac_SU1a1ab4f4fb} and the spectrum of gauge-invariant operators in Figure \ref{fig:spec_SU1a1ab4f4fb}. The operators $\text{Tr}(A\widetilde{A})^n$, $\widetilde{Q}_I(A\widetilde{A})^nQ_J$, $Q_I\widetilde{A}\big(A\widetilde{A}\big)^nQ_J$ as well as $\epsilon\,A^nQ_{I_1}\dots Q_{I_{N-2n}}$ and the flipped fields form a dense band of spectrum between $1<\Delta<3$. 
We also note that the Pfaffian operator $\mathrm{Pf}A \equiv \epsilon A^{N/2}$ always decouple when $N$ is even.

\begin{figure}[h]
	\centering
	\subcaptionbox{}{
	\includegraphics[width=6cm]{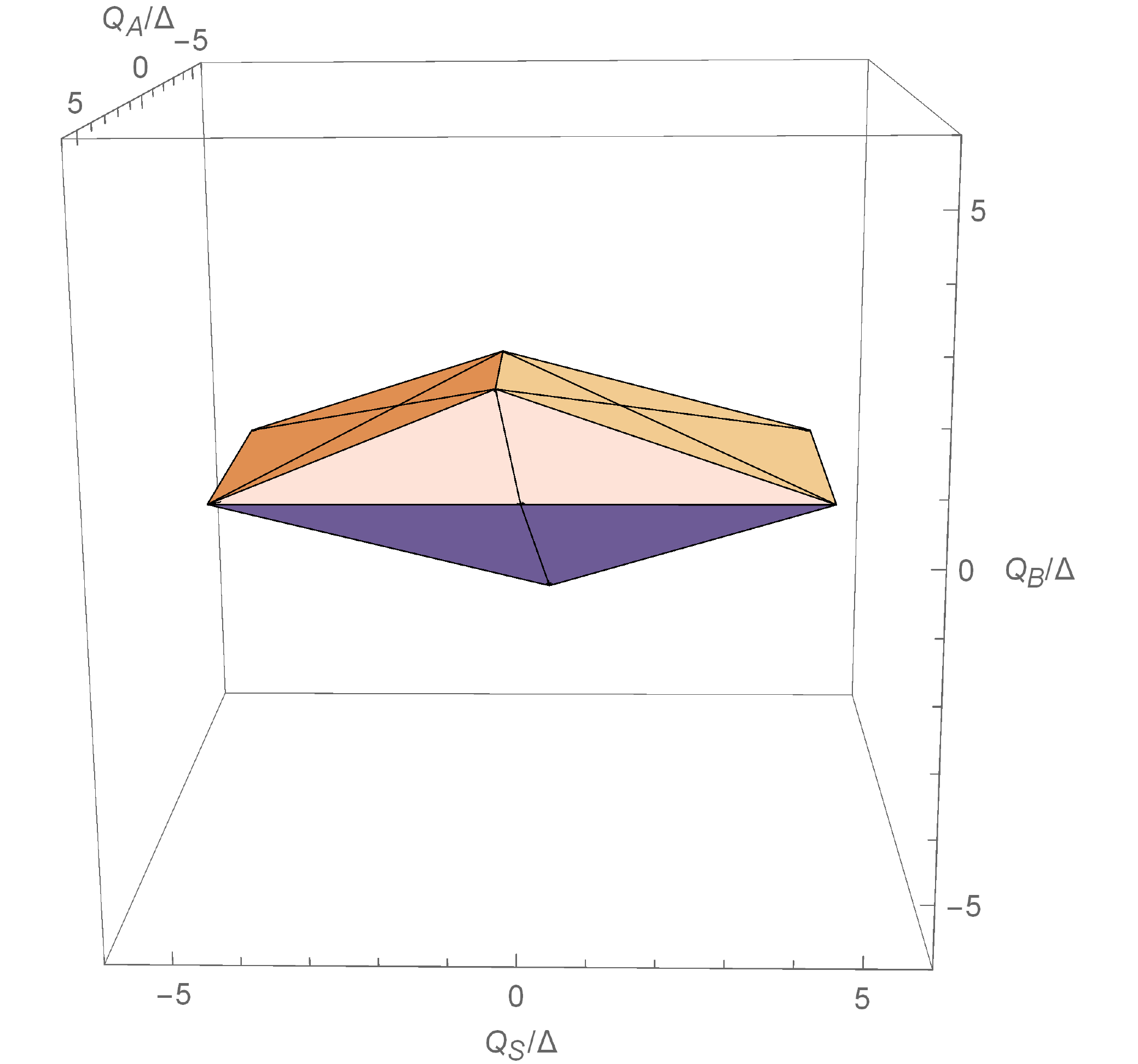}}
	\subcaptionbox{}{
	\includegraphics[width=5.5cm]{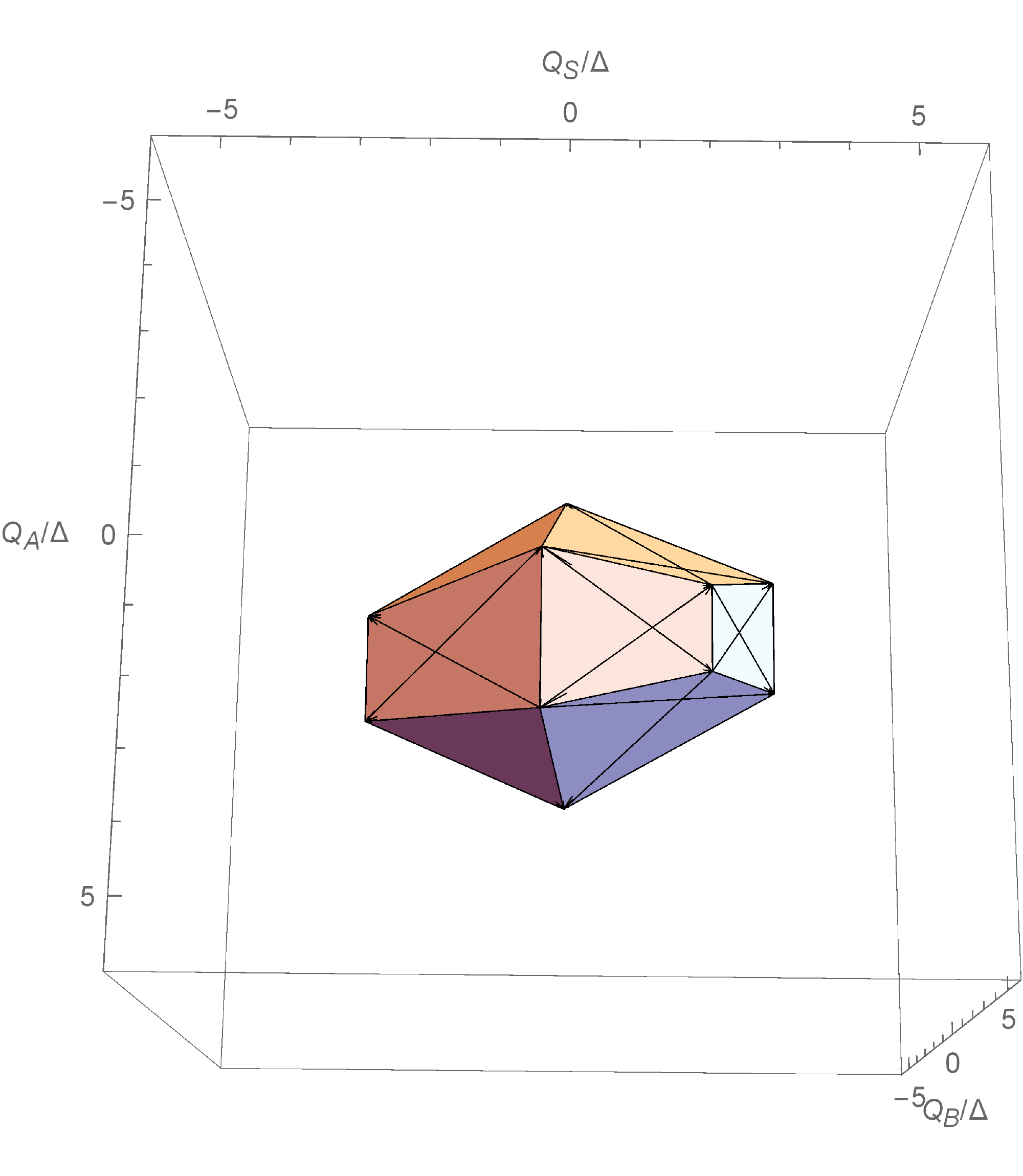}}	
	\caption{Testing the WGC for the $SU(N)$ theory with 1 anti-symmetric and 4 fundamental flavors. We construct a polyhedron in the charge-to-mass space with (a) 8 surfaces and 8 vertices for even $N\geq10$ and (b) 14 surfaces and 14 vertices for odd $N$. Each vertices corresponds to a particular charged operator.}
	\label{fig:ch_SU1a1ab4f4fb}
\end{figure}

\begin{figure}[h!]
	\centering
	\includegraphics[width=9cm]{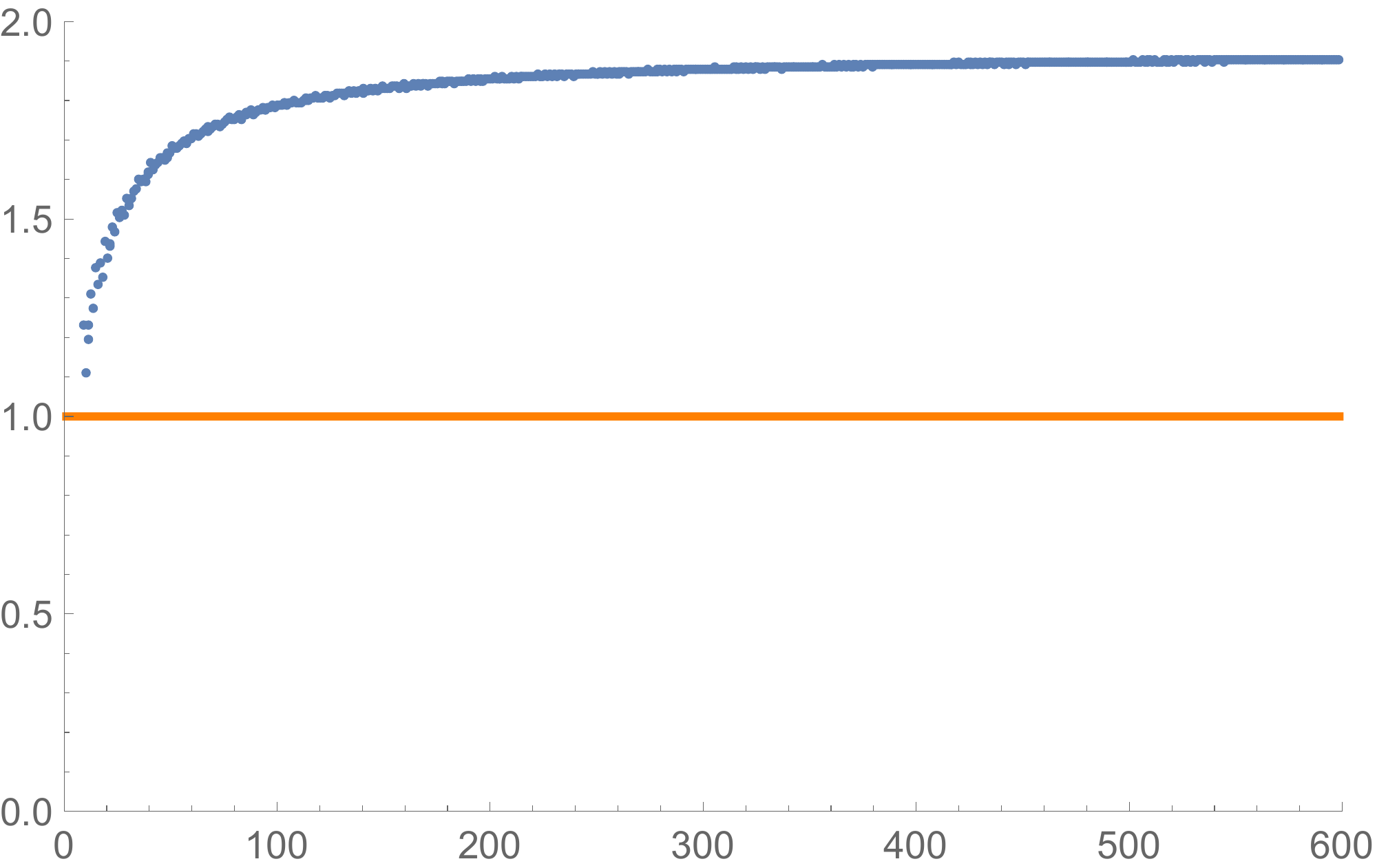}
	\caption{Testing the WGC for $SU(N)$ theory with 1 anti-symmetric and 4 fundamental flavors. Plot of the shortest distance from the origin to the boundary of convex polyhedron vs $N$ at $9\leq N\leq 600$.  }
	\label{fig:wgc_SU1a1ab4f4fb}
\end{figure}

Now, let us test whether this model satisfies the Weak Gravity Conjecture. The 3 $U(1)$ symmetries we listed above are orthogonal to each other, thus can be used as a basis of the vector space formed by the charge-to-dimension ratios. 
As before, we consider the AdS black hole decaying to the light states that are dual to gauge-invariant operators. 
At the end, any operators that are composed of gauge-invariant chiral operators and their complex conjugates fill the polyhedron in the charge-to-dimension space that are shown in Figure \ref{fig:ch_SU1a1ab4f4fb} for different $N$. 
We checked that the $SU(N)$ theory with 1 anti-symmetric and 4 fundamental flavors satisfies the weak gravity conjecture for sufficiently large $N$ by computing the shortest distance from the origin to the surface of the polyhedron.\footnote{For a small value of $N = 2,4,6,8$, we had to use convex polyhedra different from the rest and considered individually. We skip them for brevity.}
We find that the Weak Gravity Conjecture is satisfied for $N\geq6$.
The result is depicted in Figure \ref{fig:wgc_SU1a1ab4f4fb}.

\begin{figure}[h]
	\centering
	\includegraphics[width=9cm]{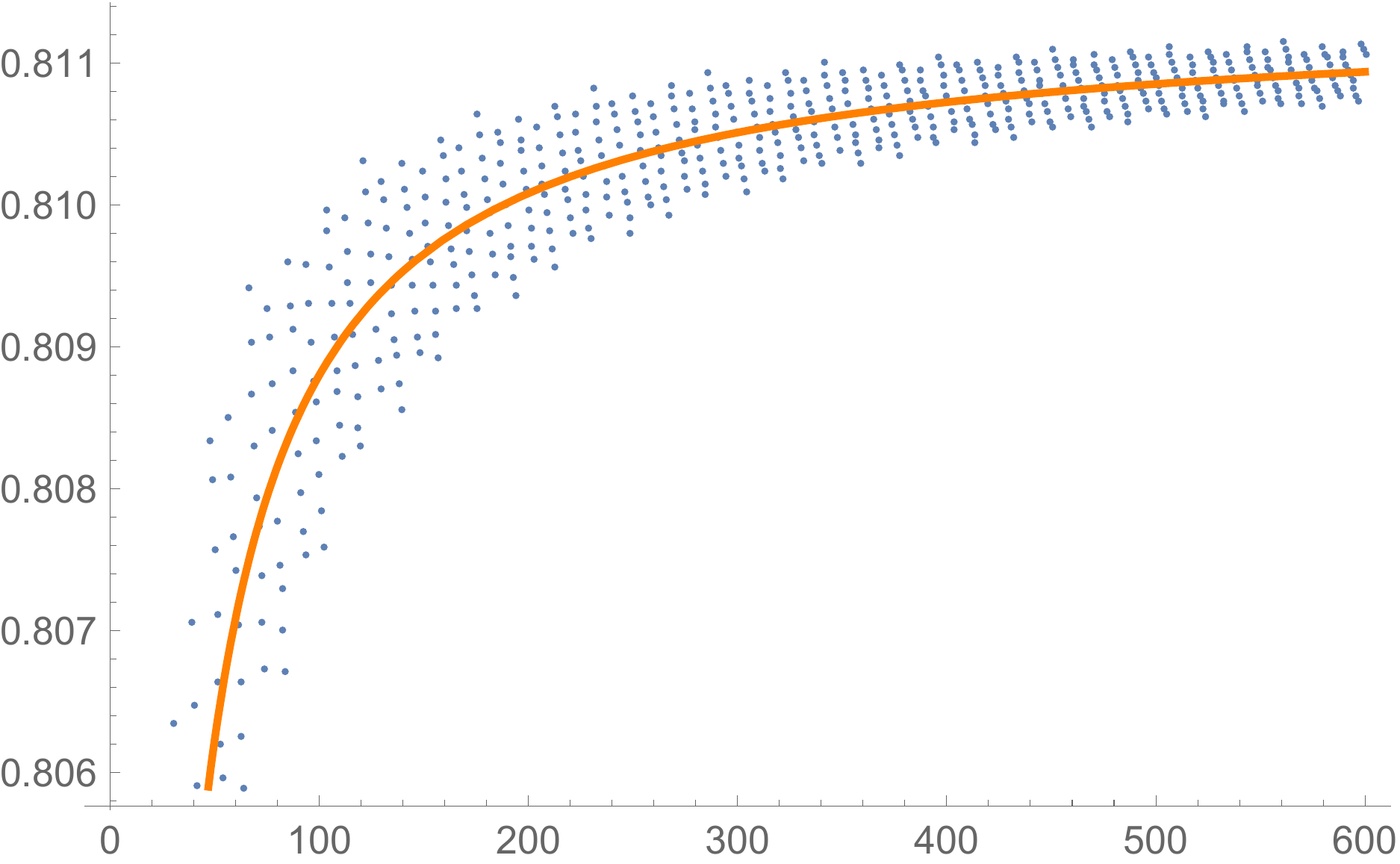}
	\caption{Plot of $a/c$ vs $N$ for the $SU(N)$ theory with 1 anti-symmetric and 5 fundamental flavors. The orange curve fits the plot with $a/c\sim0.811367\, -0.25668/N$.}
	\label{fig:ac_SU1a1ab5f5fb}
\end{figure}

\paragraph{$N_f=5$ theory}
We also study the $N_f=5$ theory. It doesn't show any qualitative difference from the case with $N_f=4$. Its central charges have the following behavior in the large-$N$ limit:
\begin{align}
\begin{split}
&a\sim0.980785 \,N-1.28501\ , \\
&c\sim1.20883 \,N-1.21142\ , \\
&4\pi^4C_S\sim8.99993 N^2-29.6979 N+84.4768\ , \\
&4\pi^4C_B\sim38.2109 N+130.812\ , \\
&4\pi^4C_A\sim1.54507 N^3+1.60691 N^2+61.4041 N-1676.66\ , \\
&R_{A}\sim1.82005/N\ , \\
&R_{Q}\sim0.233418\, +1.17044/N\ . 
\end{split}
\end{align}
In Figure \ref{fig:ac_SU1a1ab5f5fb} we show the plot of $a/c$ vs $N$ graph for the $N_f=5$ case. 
We see that the ratio of central charges asymptotes to a value smaller than 1. 

\begin{figure}[h]
	\centering
	\includegraphics[width=9cm]{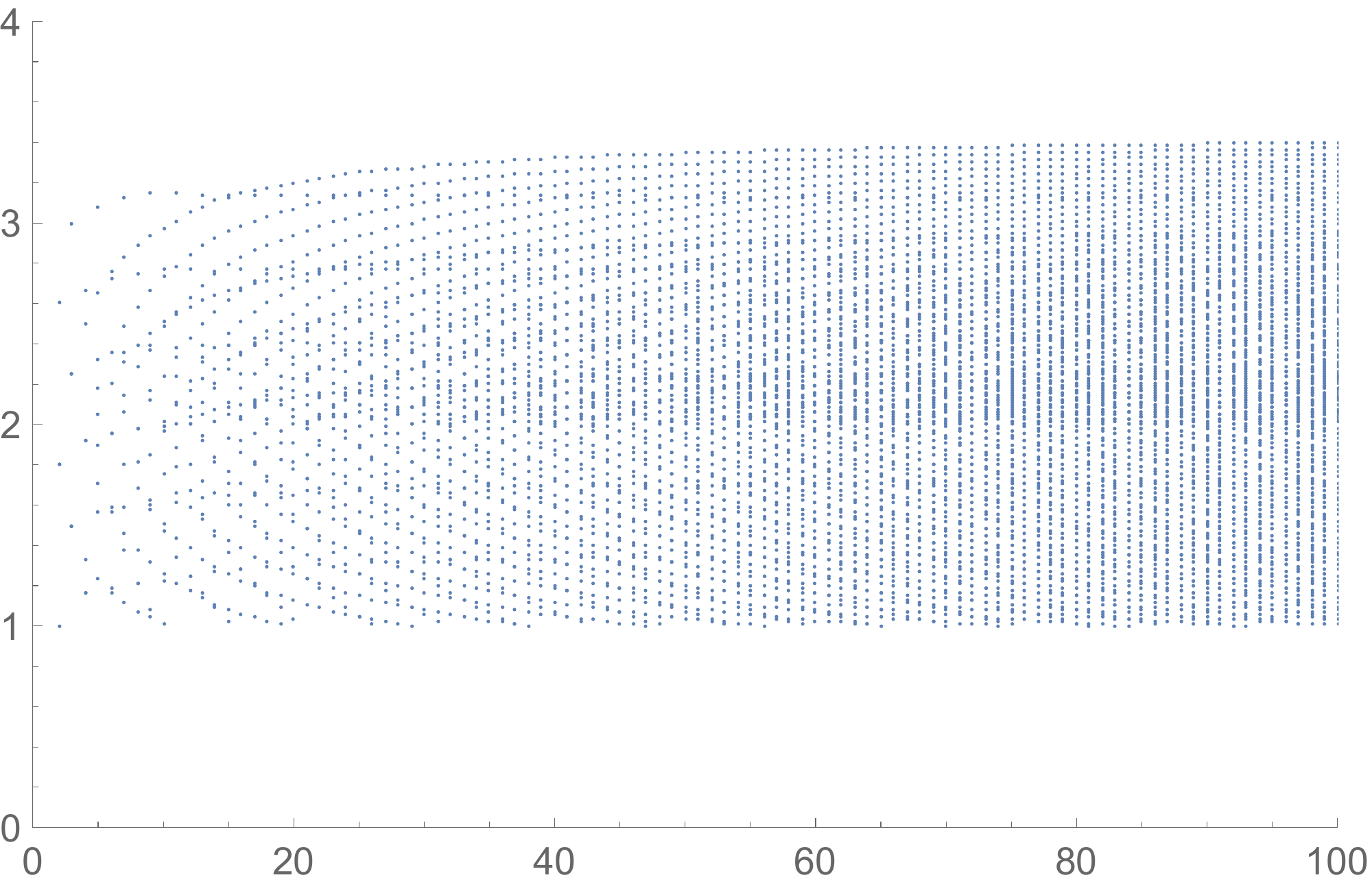}
	\caption{The low-lying spectrum of $SU(N)$ theory with 1 anti-symmetric and 5 fundamental flavors. We find a band of (single-trace) gauge-invariant operator spectrum between $1<\Delta< 4.$}
	\label{fig:spec_SU1a1ab5f5fb}
\end{figure}

For the $N_f=5$ theory, all the non-decoupled gauge-invariant operators as well as the flipped fields for the decoupled operators form a dense band with their dimensions lying between $1<\Delta<4$ as shown in Figure \ref{fig:spec_SU1a1ab5f5fb}. 

\begin{figure}[h]
	\centering
	\subcaptionbox{}{
	\includegraphics[width=6cm]{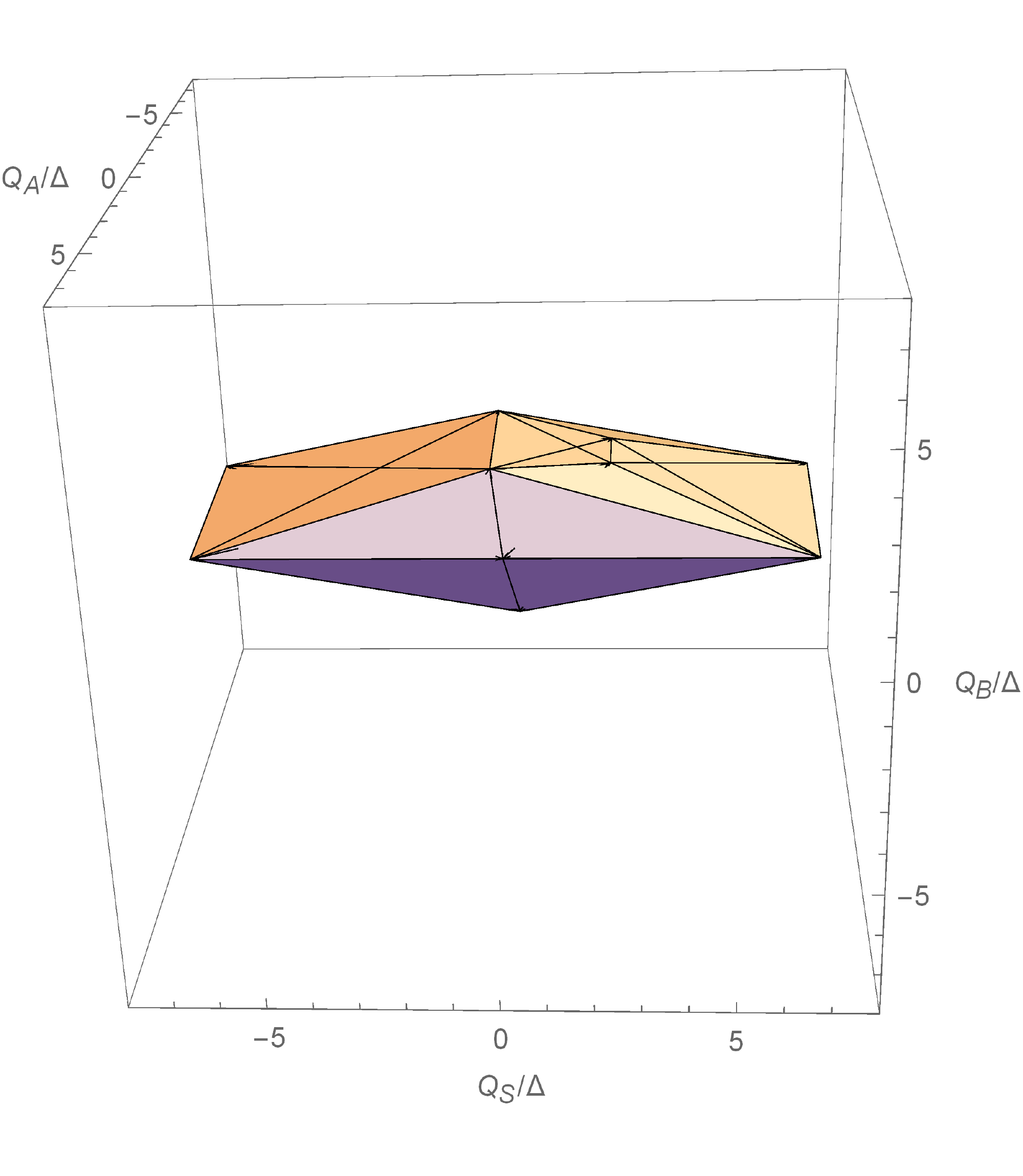}}
	\subcaptionbox{}{
	\includegraphics[width=6cm]{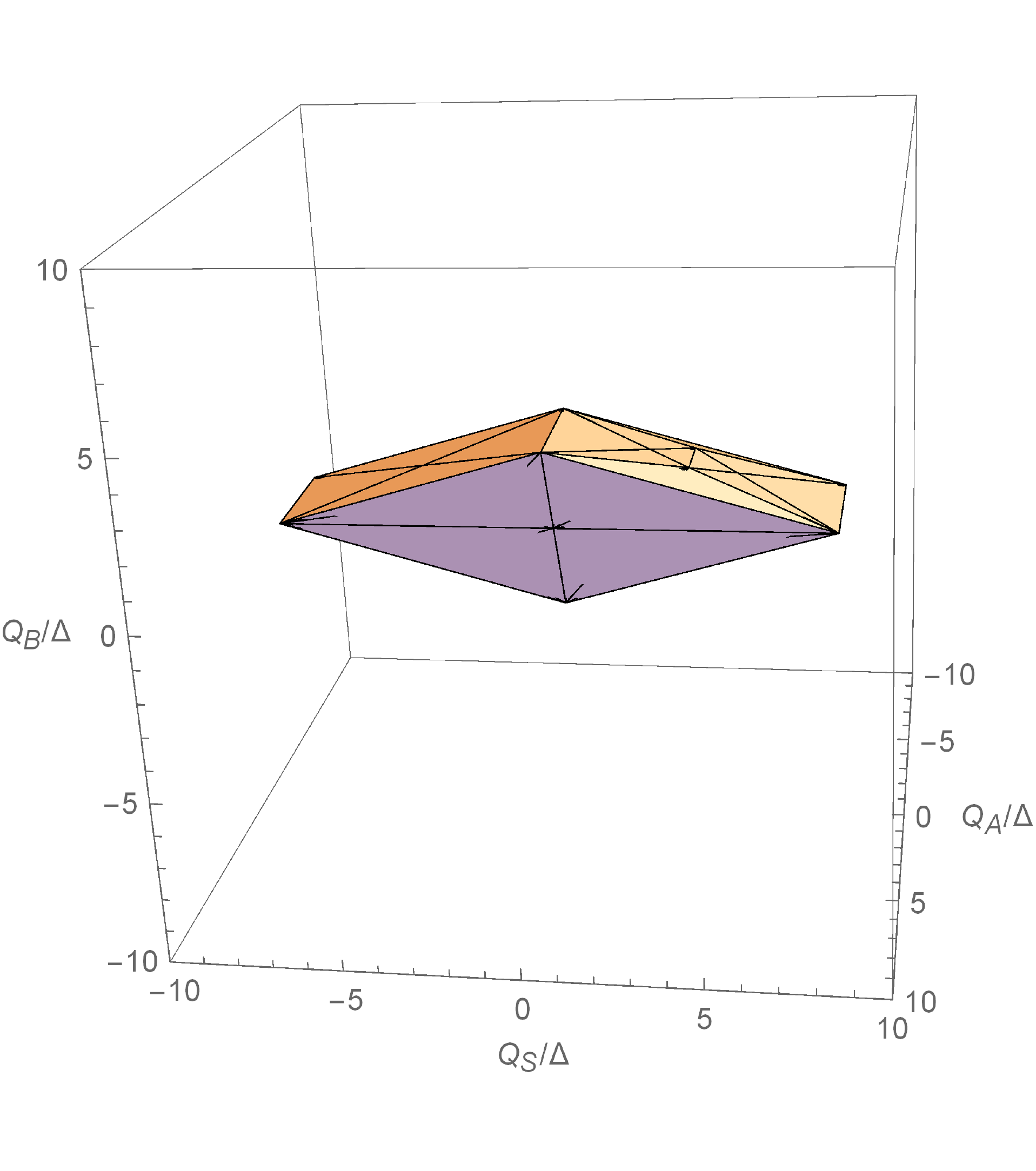}}
	\caption{Convex polyhedron formed by gauge-invariant operators for the $SU(N)$ theory with 1 anti-symmetric and 5 fundamental flavors. It has (a) 18 surfaces and 14 vertices for $5\leq N\leq 99$ and (b) 12 surfaces with 12 vertices for $N\geq 100$.}
	\label{fig:ch_SU1a1ab5f5fb}
\end{figure}

We now conisder the convex polyhedrons in the charge-to-dimension space corresponding to the arbitrary composite operators made out of gauge-invariant chiral operators and their complex conjugates. For $5\leq N \leq 99$, the convex polyhedron has a shape similar to that shown in Figure \ref{fig:ch_SU1a1ab5f5fb} (a) but has different shape for $N\geq 100$ as depcited in Figure \ref{fig:ch_SU1a1ab5f5fb} (b).
We checked the convex hull condition by computing the shortest distance from the origin to the boundary of convex polyhedron for $5\leq N\leq600$. We had to individually consider the cases for $N\leq 4$ and found that they don't satisfy the convex hull condition even after including all the single-trace chiral ring operators.
\begin{figure}[h]
	\centering
	\includegraphics[width=9cm]{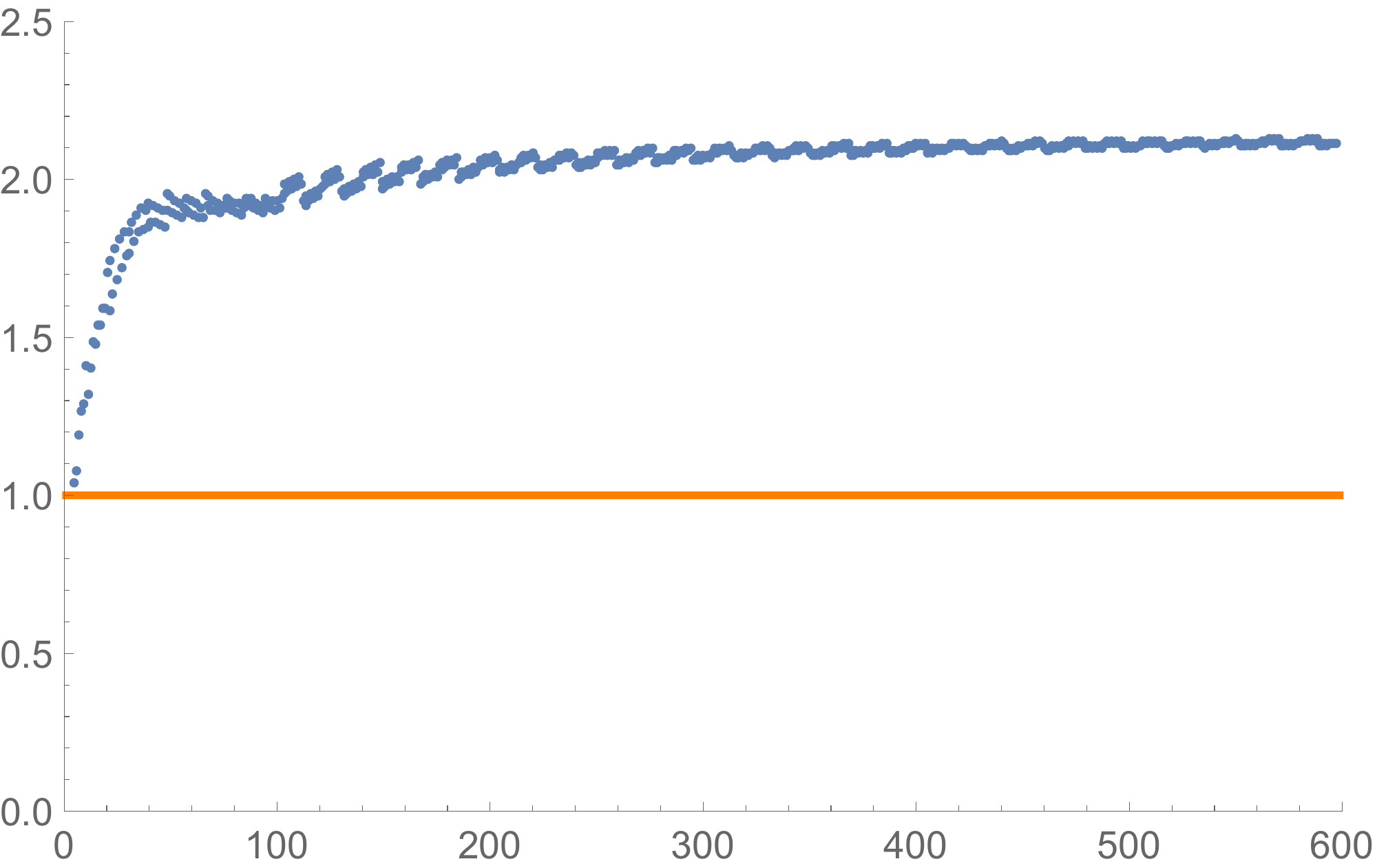}
	\caption{Testing the WGC for $SU(N)$ theory with 1 anti-symmetric and 5 fundamental flavors. The shortest distance from the origin to the boundary of convex polyhedron as a function of $N$ for $5\leq N \leq 600$. }
	\label{fig:wgc_SU1a1ab5f5fb}
\end{figure}

\subsection{1 $S$ + 1 $\bar{A}$ + 8 $\bar{\square}$ + $N_f$ ($\square$ + $\bar{\square}$)} \label{sec:TrFviolating}
1 \bsym\, + 1 $\overline{\banti}$\, + 8 $\overline{\bfund}$\, + $N_f$  ( {\bfund} + {$\overline{\bfund}$} ): 
Let us consider a chiral $SU(N)$ gauge theory with 1 symmetric, 1 anti-symmetric bar and 8 anti-fundamentals. We  add $N_f$ fundamental and anti-fundamentals on top of it. 
This theory has 3 anomaly-free flavor $U(1)$ symmetries. The charges for the matter fields are given as follows:
\begin{align}
		\begin{array}{c|c|c|c|c|c}
			& SU(N) & U(1)_S & U(1)_A & U(1)_B & U(1)_R \\ \hline
			Q & \bfund & -\frac{N+2}{2(N_f+4)} & -\frac{N-2}{2(N_f+4)} & N_f+8 & R_Q \Tstrut\\
			\widetilde{Q} & \overline{\bfund} & -\frac{N+2}{2(N_f+4)} & -\frac{N-2}{2(N_f+4)} & -N_f & 1-\frac{N_f(R_Q-1)+(N+2)(R_S-1)+(N-2)(R_A-1)}{N_f+8}\Tstrut\\
			S & \bsym & 1 & 0 & 0 & R_S\Tstrut\\
			\widetilde{A} & \overline{\banti} & 0 & 1 & 0 & R_A \Tstrut
		\end{array}
\end{align}
We now list the schematic form of the gauge-invariant operators in this theory:
\begin{itemize}
\item $\mathrm{Tr}\big(S\widetilde{A}\big)^{2n}$,\quad $n=1,\dots,\left\lfloor\frac{N-1}{2}\right\rfloor$
\item $\widetilde{Q}_{\tilde{I}}\big(S\widetilde{A}\big)^{n}Q_J$,\quad $n=0,\dots,N-2$
\item $Q_I\widetilde{A}\big(S\widetilde{A}\big)^{2n+1}Q_J$,\quad $n=0,\dots,\left\lfloor\frac{N-1}{2}\right\rfloor-1$
\item $Q_I\widetilde{A}\big(S\widetilde{A}\big)^{2n}Q_J$,\quad $n=0,\dots,\left\lfloor\frac{N}{2}\right\rfloor-1$
\item $\widetilde{Q}_{\tilde{I}}\big(S\widetilde{A}\big)^{2n}S\widetilde{Q}_{\tilde{J}}$,\quad $n=0,\dots,\left\lfloor\frac{N}{2}\right\rfloor-1$
\item $\widetilde{Q}_{\tilde{I}}\big(S\widetilde{A}\big)^{2n+1}S\widetilde{Q}_{\tilde{J}}$,\quad $n=0,\dots,\left\lfloor\frac{N-1}{2}\right\rfloor-1$
\item $\epsilon\,\widetilde{A}^n\widetilde{Q}_{\tilde{I}_1}\dots \widetilde{Q}_{\tilde{I}_{N-2n}}$,\quad $n=\left\lceil\frac{N-N_f-8}{2}\right\rceil,\dots,\left\lfloor\frac{N}{2}\right\rfloor$. 
\item $\epsilon\,\epsilon\, S^{N-n}\big(Q_{I_{1}}Q_{J_1}\big)\dots\big(Q_{I_{n}}Q_{J_{n}}\big)$,\quad$n=0,\dots, N_f$
\item $\epsilon\,(S\widetilde{A}S)^{(N-n)/2}\CQ^{k_1}_{I_1}\dots \CQ^{k_n}_{I_n}$ . 
\end{itemize}
Here the subscripts $I,J$ are the flavor indices running from $1,\dots,N_f$ for fundamental chiral fields $Q$, while $\tilde{I},\tilde{J}$ are the flavor indices for anti-fundamental chiral fields $\widetilde{Q}$ running from $1,\dots,N_f+8$. The operators $\epsilon\,(S\widetilde{A}S)^{(N-n)/2}\CQ^{k_1}_{I_1}\dots \CQ^{k_n}_{I_n}$ in the last line are defined in terms of the dressed quarks $\CQ^n_I$ given by
\begin{align}
\mathcal{Q}^{n}_{I}=\begin{cases}
(S\widetilde{A})^{n/2}Q_I & n=0,2,4,\dots\\
(S\widetilde{A})^{(n-1)/2}S\widetilde{Q}_I & n=1,3,5,\dots\ .
\end{cases}
\end{align}
We note that not all of $\epsilon\,(S\widetilde{A}S)^{(N-n)/2}\CQ^{k_1}_{I_1}\dots \CQ^{k_n}_{I_n}$ are independent single-trace operators.\footnote{We studied the set of $\epsilon\,(S\widetilde{A}S)^{(N-n)/2}\CQ^{k_1}_{I_1}\dots \CQ^{k_n}_{I_n}$ by computing the Hilbert series for $N=3,4,5$ and found relations among them.} 
 It turns out these operators are heavy enough and never decouple, so do not affect our analysis via $a$-maximization.

\begin{figure}[h!]
	\centering
	\includegraphics[width=9cm]{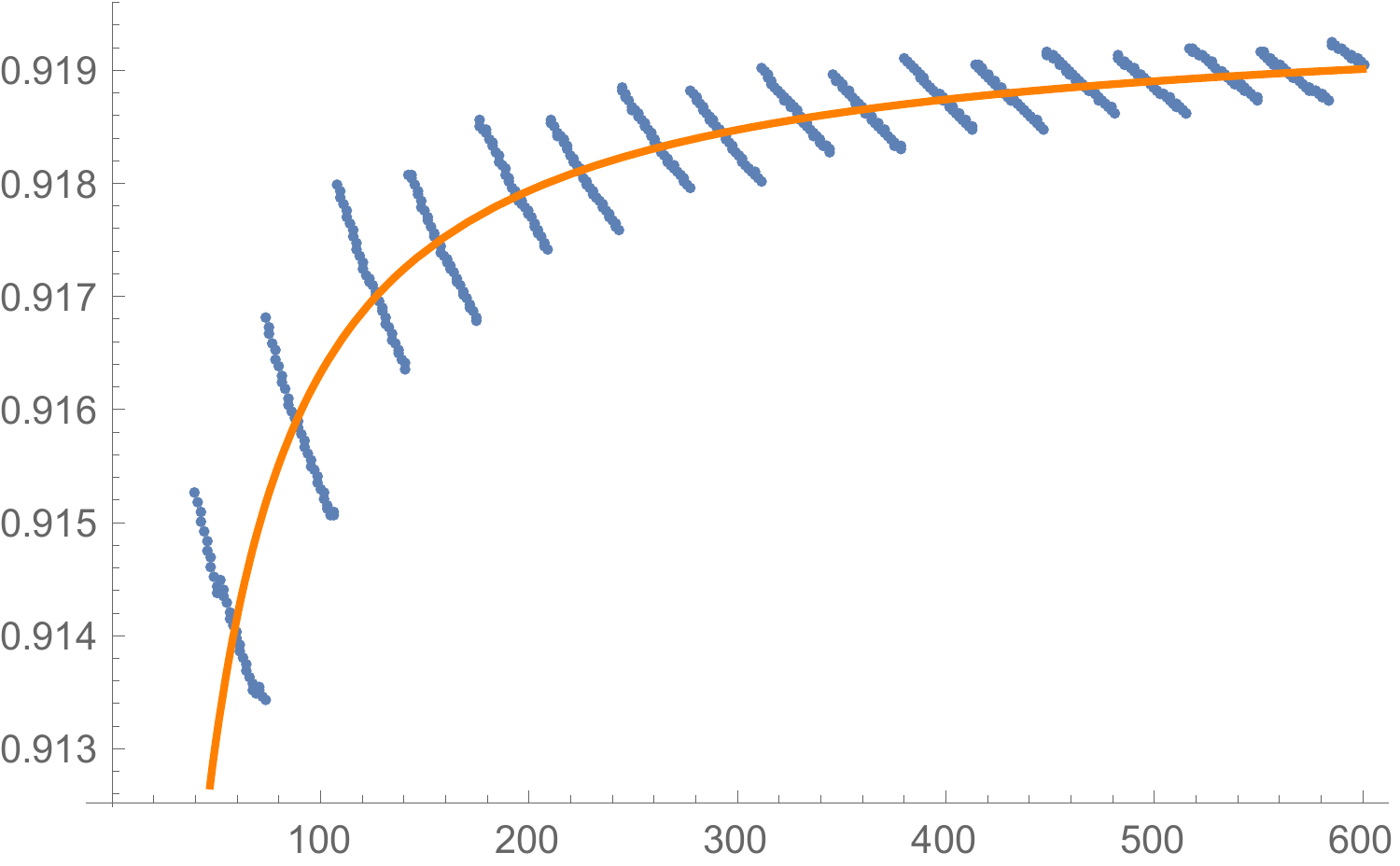}
	\caption{Plot of $a/c$ vs $N$ for the $SU(N)$ theory with 1 symmetric, 1 anti-symmetric, and 8 anti-fundamentals. The orange curve fits the plot with $a/c\sim 0.919548\,-0.322605/N$.}
	\label{fig:ac_SU1s1ab8fb}
\end{figure}

\paragraph{$N_f= 0$ case}
Let us start with the simplest case with $N_f=0$. Notice that we still have 8 anti-fundamental matters. Repeating the $a$-maximization procedure, we obtain
\begin{align}
\begin{split}
&a\simeq1.84737 N-6.74182 \ , \\
&c\simeq2.00913 N-6.68587 \ , \\
&R_S\simeq2.51330/N \ , \\
&R_{\widetilde{A}}\simeq3.42858/N \ , \\
&R_{\widetilde{Q}}\simeq0.244789\, +2.37709/N \ .
\end{split}
\end{align}
As in the previous cases, we see that the central charges grow linearly in $N$. We plot the ratio $a/c$ vs $N$ in Figure \ref{fig:ac_SU1s1ab8fb}. 
The ratio of the central charges asymptotes to a value smaller than 1 in the large $N$ limit. 

\begin{figure}[h]
	\centering
	\includegraphics[width=9cm]{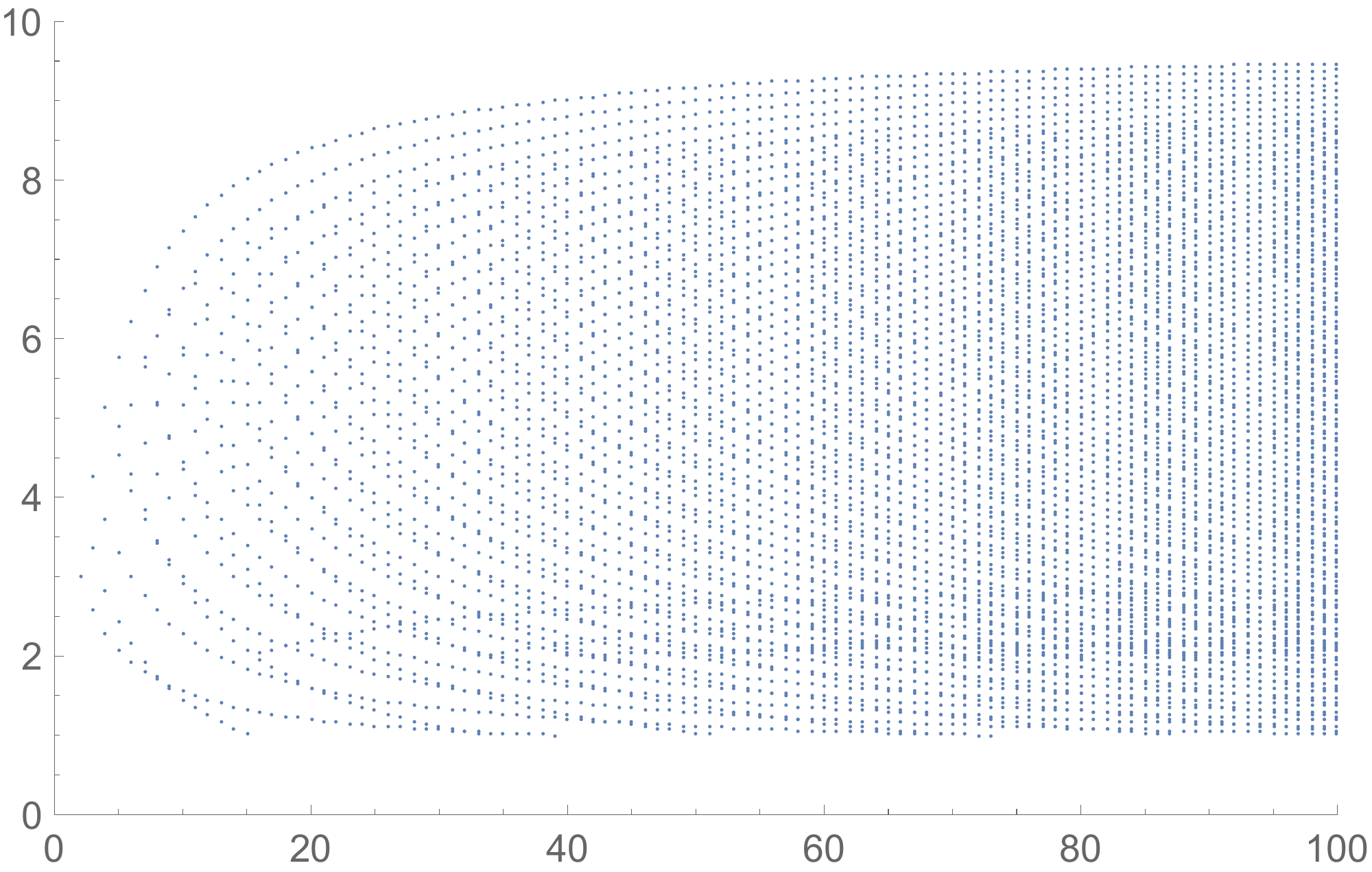}
	\caption{The spectrum of gauge-invariant operators for the $SU(N)$ theory with 1 symmetric, 1 anti-symmetric, and 8 anti-fundamental matters. They form a dense band between $1 < \Delta<10$.}
	\label{fig:spec_SU1s1ab8fb}
\end{figure}

In the large-$N$ limit, the $R$-charges for the symmetric and anti-symmetric matters go to zero. On the other-hand, the $R$-charges for the anti-fundamentals continue to be of $\CO(1)$. 
Therefore, the operators made only out of the symmetric and anti-symmetric tensors or the ones with only a few number of anti-fundamental chiral fields, can go below the unitarity bound and will have to be removed using the flip fields. We plot the dimensions of the unitary gauge-invariant operators (including the flip fields) in Figure \ref{fig:spec_SU1s1ab8fb}.
We see that there is a dense band between $1 < \Delta<10$. 
Every single-trace gauge-invariant operators (that are not decoupled) except the ones of the form $\epsilon\,(S\widetilde{A}S)^{(N-n)/2}\CQ^{k_1}_{I_1}\dots \CQ^{k_n}_{I_n}$ as well as the flip fields corresponding to the decoupled operators join the band. There is no operator of the form $\epsilon\,(S\widetilde{A}S)^{(N-n)/2}\CQ^{k_1}_{I_1}\dots \CQ^{k_n}_{I_n}$ in the $N_f=0$ theory.

Now, let us check the WGC convex hull condition. The $U(1)_S$ and $U(1)_A$ are inconvenient charge basis to check because they are not mutually orthogonal. We will therefore switch to a basis of $U(1)$ flavor symmetries given by $U(1)_1=U(1)_S-U(1)_A$ and $U(1)_2 = U(1)_S + U(1)_A+\lambda U(1)_1$ with $\lambda = - {(N+1)}/{(N+2)}$. Note that the chiral superfields $S,\ \widetilde{A}\ \text{and} \ \widetilde{Q}$ have charges $1,-1 \ \text{and} \ -1/2$ respectively such that the operators $\Tr (S \widetilde{A})^{2n}$ and $\widetilde{Q}(S\widetilde{A})^nS\widetilde{Q}$ are not charged with respect to $U(1)_1$.
The superfields $S, \widetilde{A}, \widetilde{Q}$ have charges ${1}/{(N+2)},  {(2N+3)}/{(N+2)}, \ \text{and} \ {(- N^2 + 2)}/{4(N+2)}$ respectively with respect to $U(1)_2$. The flavor central charges of $U(1)_{1,2}$ are 
\begin{align}
\begin{split}
&4\pi^4C_1\simeq8.99995 N^2-13.5611 N+33.7170 \ , \\
&4\pi^4C_2\simeq2.58951 N^3+16.9040 N^2-1284.91 N+57464.1 \ . 
\end{split}
\end{align}

\begin{figure}[h!]
	\centering
	\includegraphics[width=9cm]{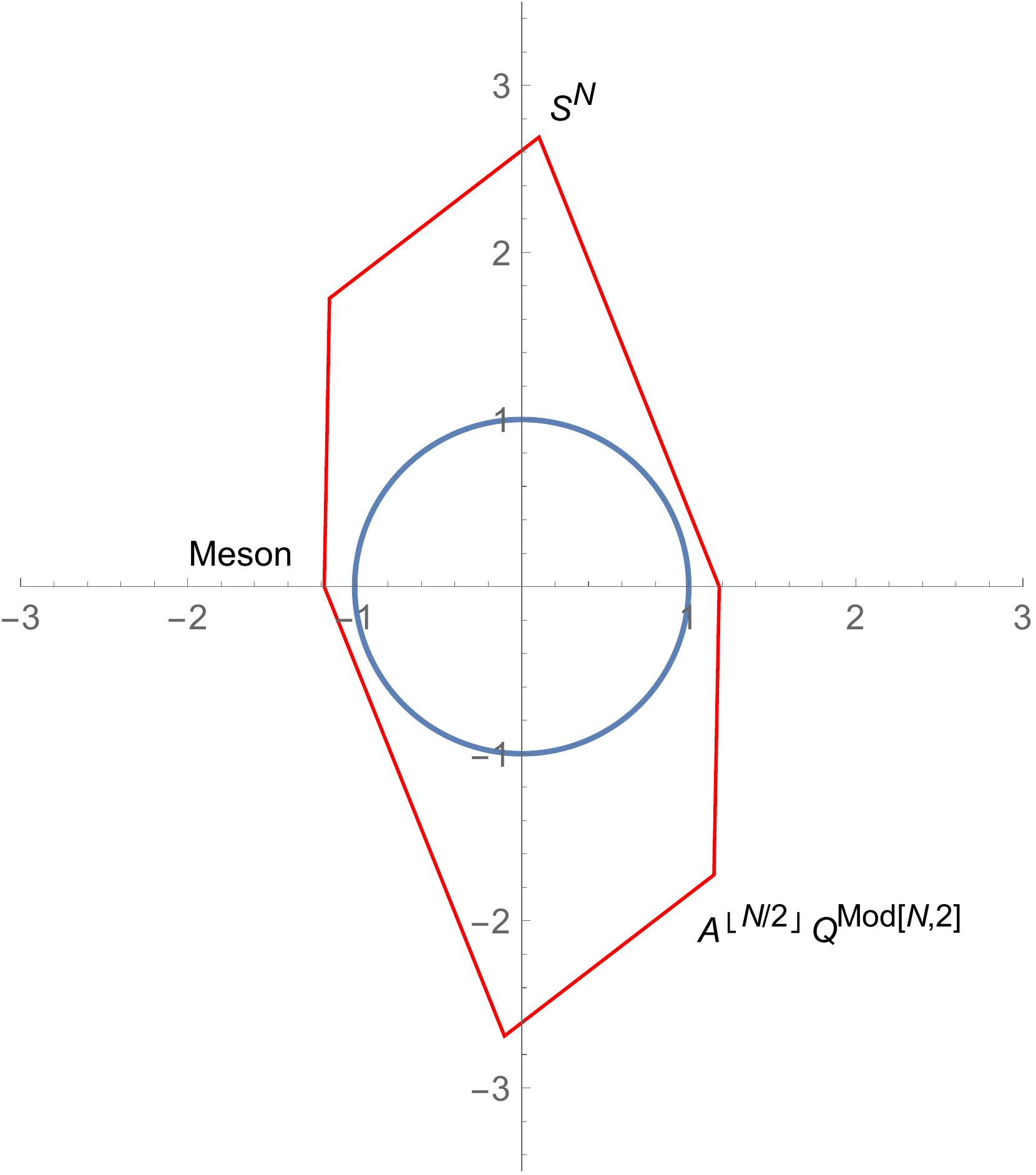}
	\caption{Testing the WGC for $SU(N)$ theory with 1 symmetric, 1 anti-symmetric and 8 anti-fundamentals. Convex hexagon formed by linear combination of lightest $\widetilde{Q}(S\widetilde{A})^nS\widetilde{Q}$, $S^N$, and $A^{\lfloor N/2\rfloor}Q^{\text{Mod}[N,2]}$ and their aniti-chiral conjugate pairs on the charge-to-dimension space normalized by $\sqrt{\frac{9}{40}\frac{C_T}{C_{1,2}}}$. It must include a unit circle to satisfy the convex hull condition.}
	\label{fig:ch_SU1s1ab8fb}
\end{figure}

\begin{figure}[h!]
	\centering
	\includegraphics[width=9cm]{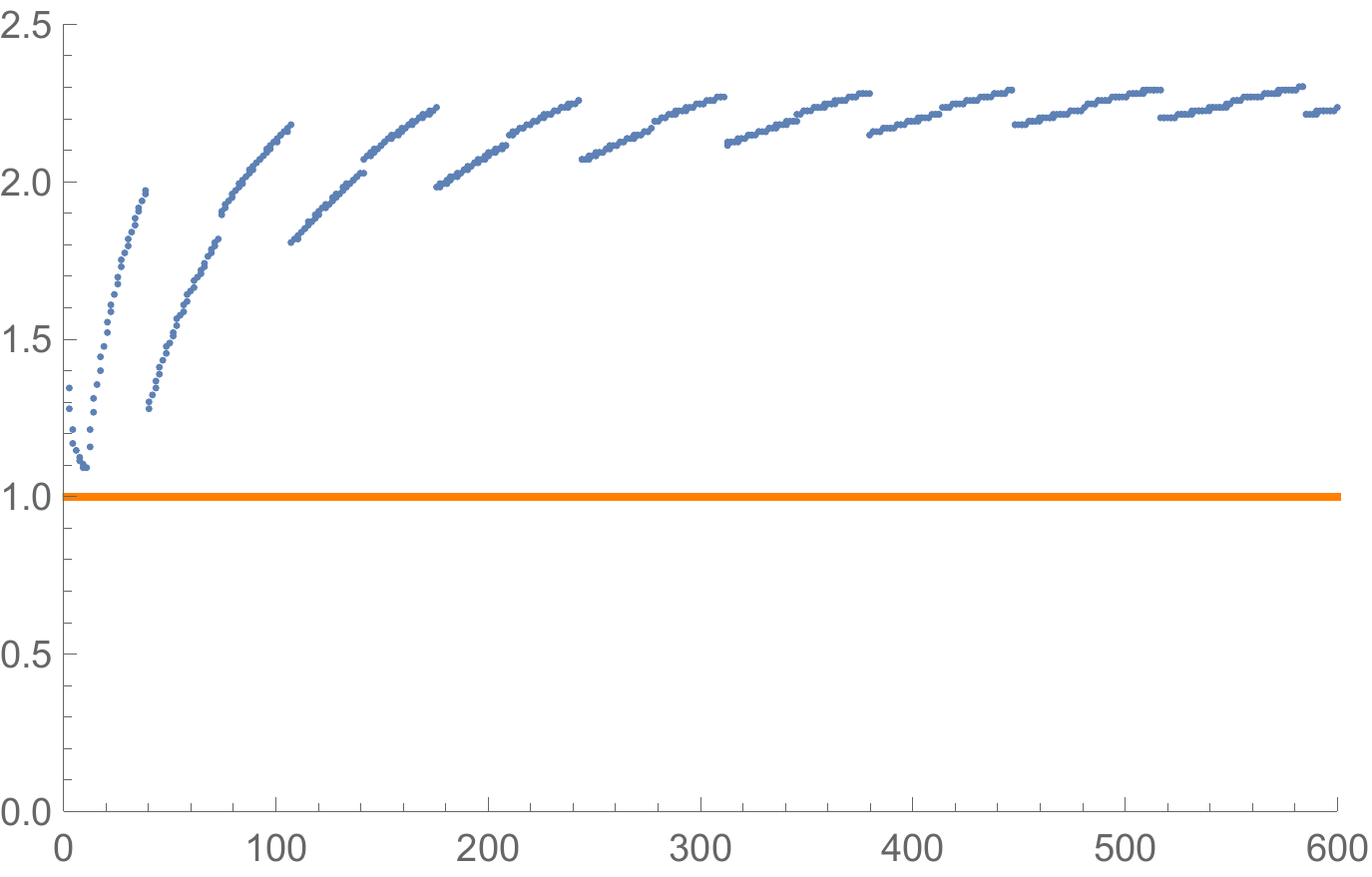}
	\caption{Testing WGC for the $SU(N)$ theory with 1 symmetric, 1 anti-symmetric, and 8 anti-fundamental matters. Plot of the shortest distance from the origin to the boundary of the convex hexagon vs $N$.}
	\label{fig:wgc_SU1s1ab8fb}
\end{figure}

Then we check the convex hull condition for these $U(1)_{1,2}$ symmetries. We consider the lightest operators among the form of $\widetilde{Q}(S\widetilde{A})^nS\widetilde{Q}$, $S^N$, and $A^{\lfloor N/2\rfloor}Q^{\text{Mod}[N,2]}$ and their anti-chiral conjugate pairs.\footnote{For small $N$, the lightest operator of the form $\Tr (S\widetilde{A})^n$ has larger charge-to-dimension ratio on $U(1)_1$. In such cases we replace the operator $\widetilde{Q}(S\widetilde{A})^nS\widetilde{Q}$ with $\Tr (S\widetilde{A})^n$.} As depicted in Figure \ref{fig:ch_SU1s1ab8fb}, they form a convex hexagon in the plane spanned by charge-to-dimension ratios.
The convex hull condition is satisfied when the hexagon encloses the unit circle. We checked that this is indeed the case by computing the shortest distance from the origin to the boundary of the hexagon as shown in Figure \ref{fig:wgc_SU1s1ab8fb}.

\begin{figure}[h!]
	\centering
	\includegraphics[width=9cm]{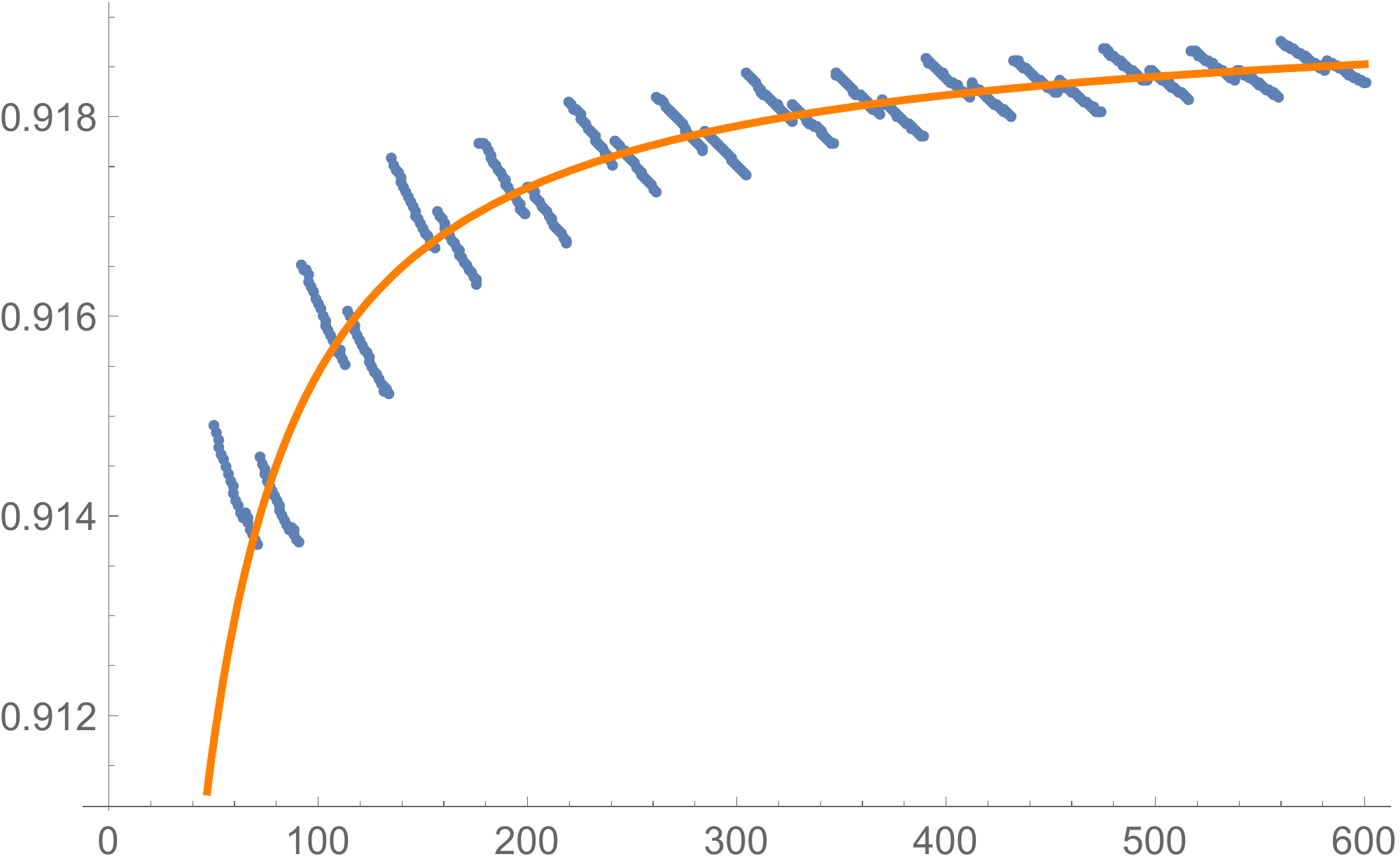}
	\caption{Plot of $a/c$ vs $N$ for the $SU(N)$ theory with 1 symmetric, 1 anti-symmetric, 1 fundamental and 9 anti-fundamentals. The orange curve fits the plot with $a/c\sim 0.919143\, -0.370927/N$.}
	\label{fig:ac_SU1s1ab1f9fb}
\end{figure}

\begin{figure}[h]
	\centering
	\includegraphics[width=9cm]{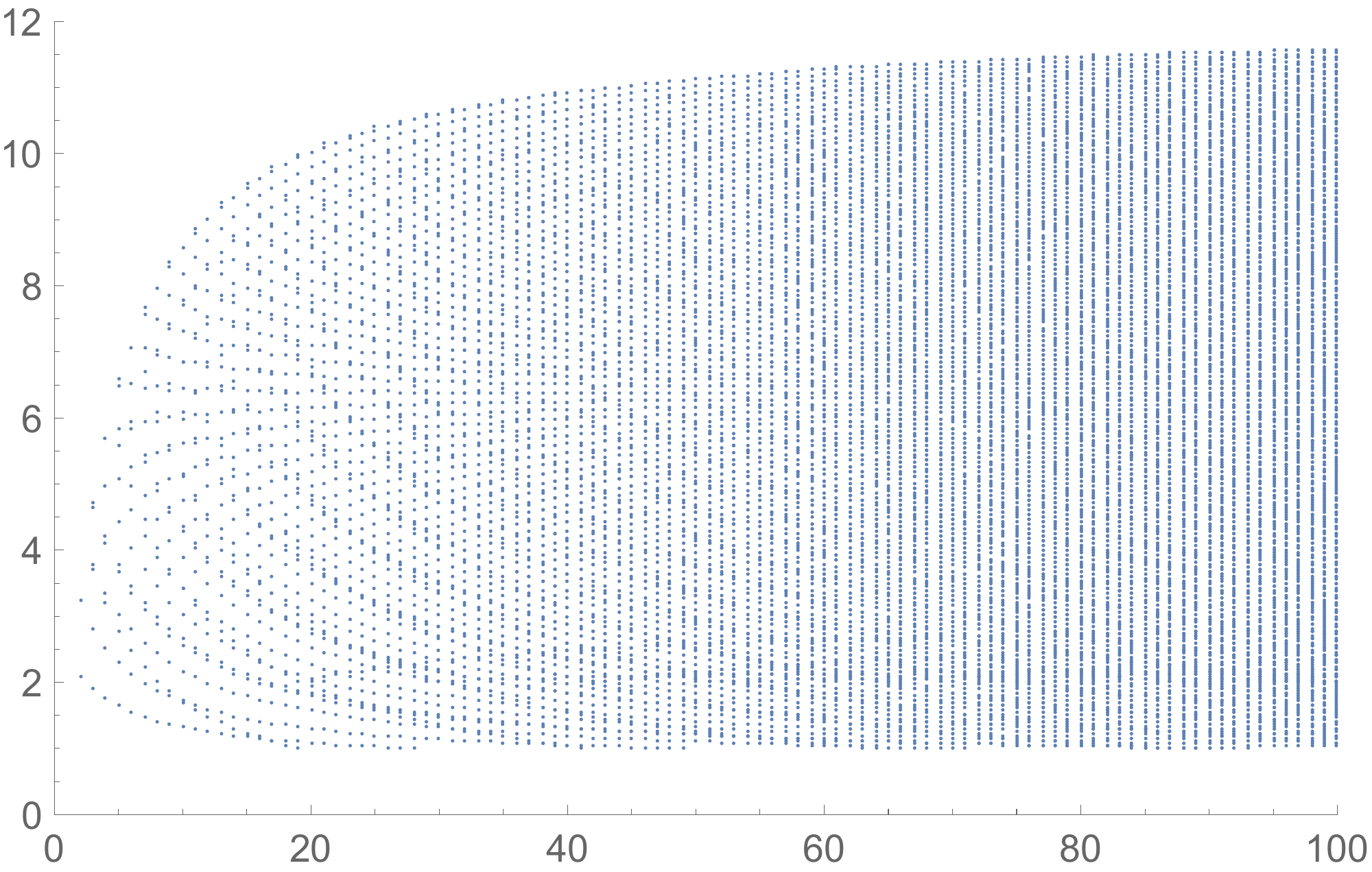}
	\caption{The spectrum of single-trace gauge-invariant operators for the $SU(N)$ theory with 1 symmetric, 1 anti-symmetric, 1 fundamental and 9 anti-fundamental matters. They form a dense band between $1 < \Delta<12$.}
	\label{fig:spec_SU1s1ab1f9fb}
\end{figure}

\paragraph{$N_f=1$ case}
We now move on to study the $N_f=1$ theory in a similar way. It turns out that some of the operators that decouple carry non-trivial charges with respect to flavor symmetries, thus causing the $U(1)_B$ symmetry to be no longer traceless in the IR. After carefully accounting for these decoupled operators, we switch to an orthogonal basis of the flavor symmetries given by 
\begin{align}
\begin{split}
&U(1)_1=U(1)_S-U(1)_A \ , \\
&U(1)_2=U(1)_S+U(1)_A+\lambda_{21}U(1)_1+\lambda_{23}U(1)_3 \ , \\
&U(1)_3=U(1)_B+\lambda_{31}U(1)_1 \ , 
\end{split}
\end{align}
where the coefficients are 
\begin{align}
\begin{split}
&\lambda_{31}=-\frac{\text{Tr}\,Q_1Q_B}{\text{Tr}\,Q_1^2} \ , \\
&\lambda_{21}=-\frac{\text{Tr}\,Q_1(Q_S+Q_A)}{\text{Tr}\,Q_1^2} \ , \\
&\lambda_{23}=-\frac{\text{Tr}\,Q_3(Q_S+Q_A)}{\text{Tr}\,Q_3^2}=-\frac{\text{Tr}\,(Q_S+Q_A)Q_B+\lambda_{31}\text{Tr}\,Q_1(Q_S+Q_A)}{\text{Tr}\,Q_B^2-\lambda_{31}\text{Tr}\,Q_1Q_B} \ .
\end{split}
\end{align}
The central charges, flavor central charges and the $R$-charge are now given by
\begin{align}
\begin{split}
&a\simeq2.30737 N-10.3265 \ , \\
&c\simeq2.51064 N-10.3477 \ , \\
&4\pi^4C_1\simeq8.99985 N^2-23.8115 N+86.4607 \ , \\
&4\pi^4C_2\simeq2.08849 N^3+2.56454 N^2+1698.21 N-67199.5 \ , \\
&4\pi^4C_3\simeq531.794 N-415.541\ , \\
&R_S\simeq3.24381/N \ , \\
&R_{\widetilde{A}}\simeq4.15231/N \ , \\
&R_{Q}\simeq0.244599\, +2.68770/N \ , \\
&R_{\widetilde{Q}}\simeq0.244697\, +2.90899/N \ .
\end{split}
\end{align}
The $a/c$ asymptotes to 0.9191 at large $N$, which is smaller than 1 as ploted in Figure \ref{fig:ac_SU1s1ab1f9fb}.
The low-lying gauge-invariant operators form a dense band between $1<\Delta<12$. Every operators we listed in the beginning of this subsection as well as the flip fields corresponding to the decoupled operators join the band, except some of the operators of the form $\epsilon\,(S\widetilde{A}S)^{(N-n)/2}\CQ^{k_1}_{I_1}\dots \CQ^{k_n}_{I_n}$ with large $n$. 
We plot the spectrum without $\epsilon\,(S\widetilde{A}S)^{(N-n)/2}\CQ^{k_1}_{I_1}\dots \CQ^{k_n}_{I_n}$ in Figure \ref{fig:spec_SU1s1ab1f9fb}.

\begin{figure}[h!]
	\centering
	\subcaptionbox{}{
	\includegraphics[width=4cm]{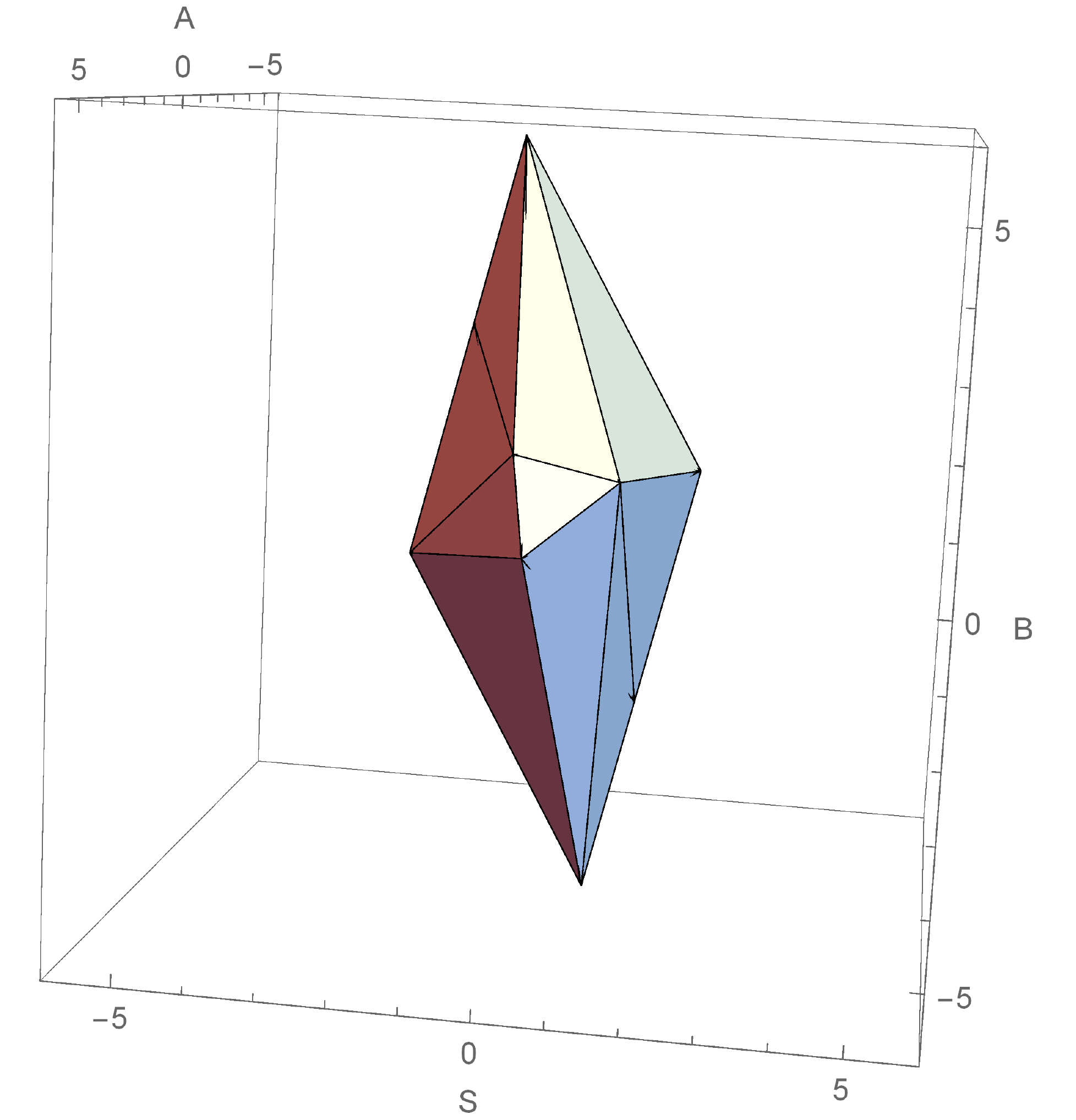}}
	\subcaptionbox{}{
	\includegraphics[width=4cm]{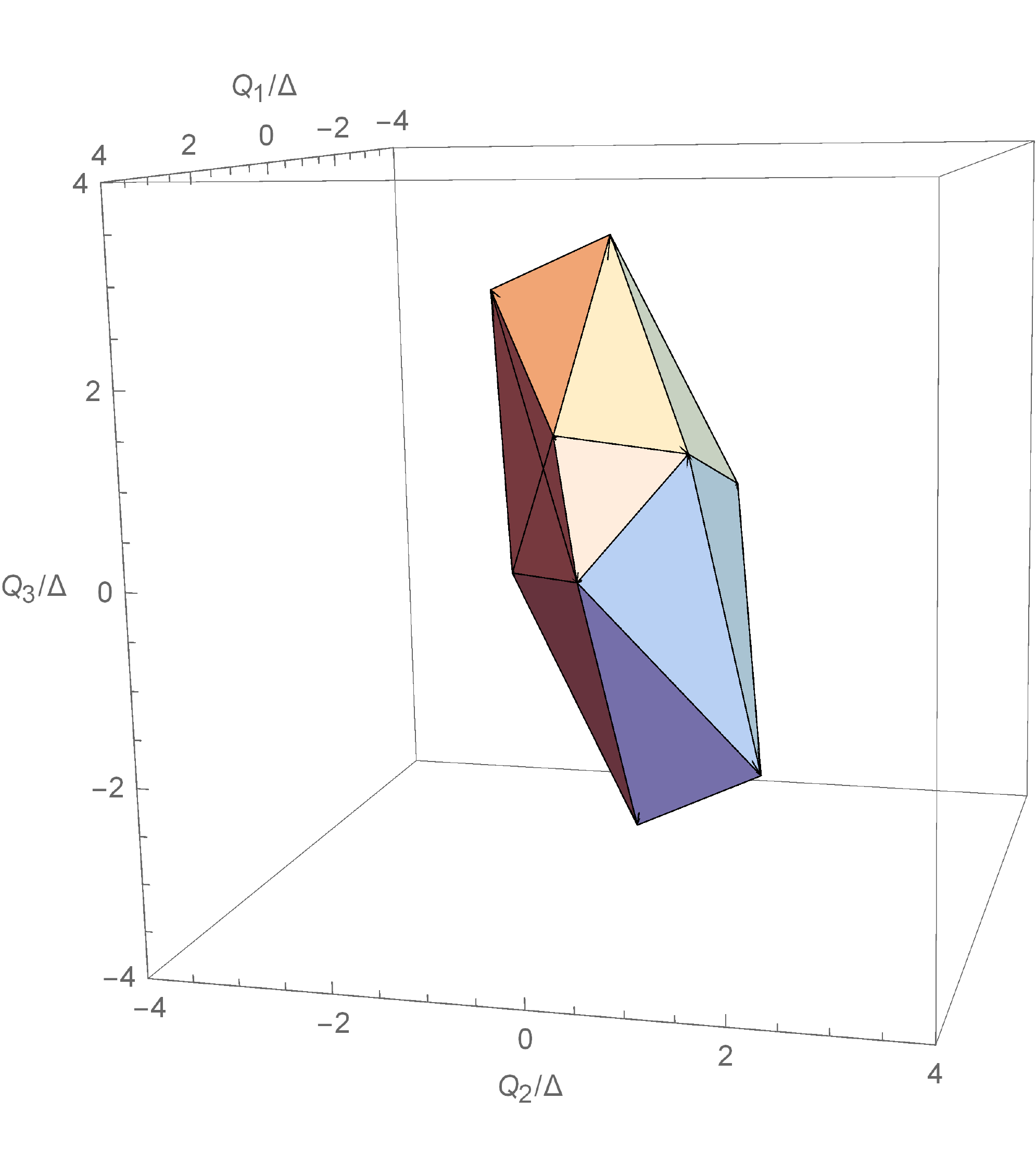}}
	\subcaptionbox{}{
	\includegraphics[width=4cm]{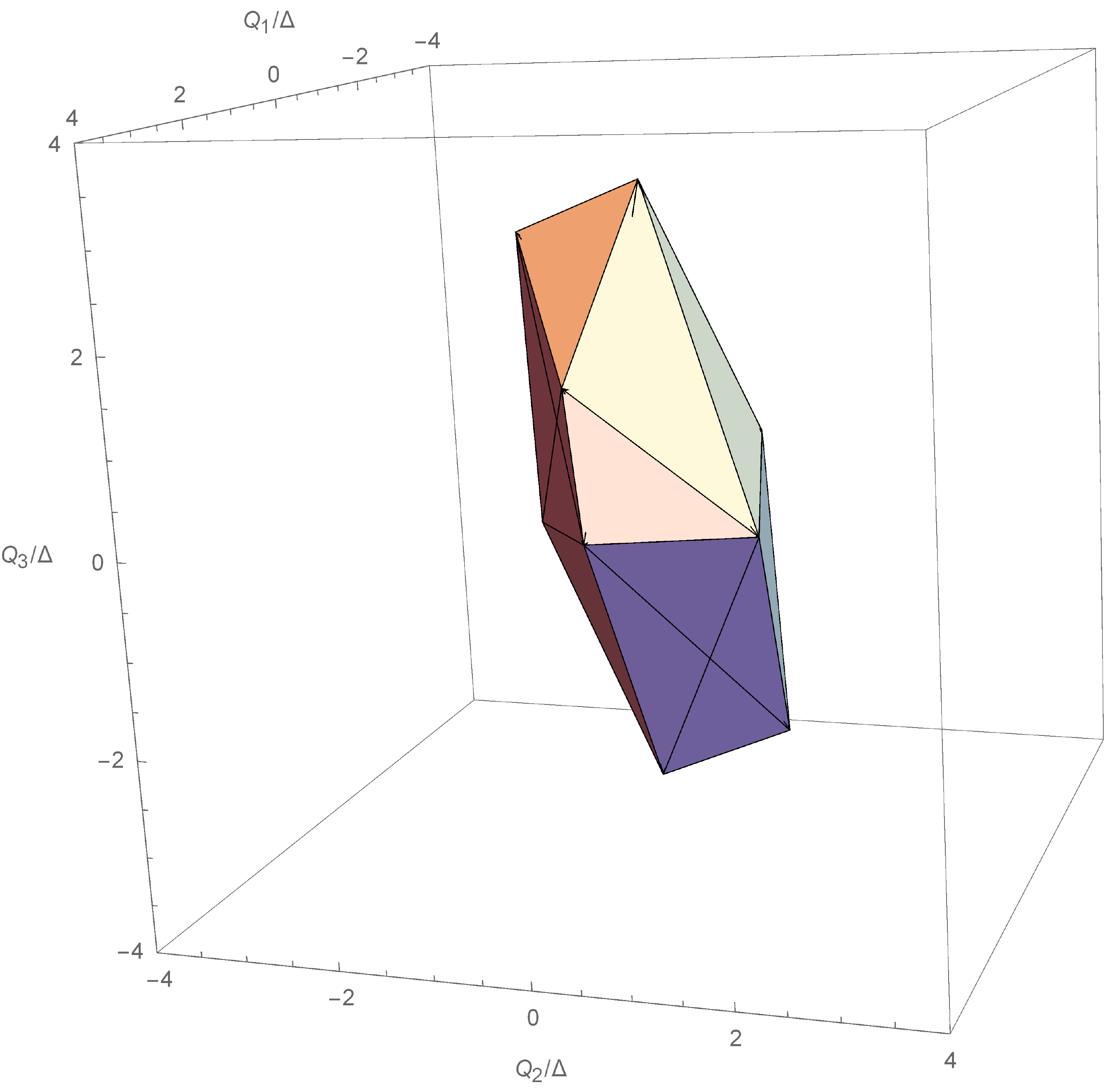}}	
	\caption{Checking the WGC for the chiral $SU(N)$ theory with $N_f=1$. Charge-to-mass ratio of light states. They fill a polyhedron with (a) 16 surfaces and 10 vertices for $N\geq 29$, (b) 18 surfaces and 12 vertices for odd $N\leq 29$ and (c) 16 surfaces and 12 vertices for even $N\leq 28$.}
	\label{fig:ch_SU1s1ab1f9fb}
\end{figure}

With the spectral data we can check the WGC. Once again we consider the operators consist with gauge-invariant chiral operators on the charge-to-dimension space. For sufficiently large $N\geq29$, they fill a convex polyhedron as shown in Figure \ref{fig:ch_SU1s1ab1f9fb}(a). For $N<29$, the `topology' of the polyhedrons are different depending upon whether $N$ is odd or even. These have been shown in Figure \ref{fig:ch_SU1s1ab1f9fb}(b) and Figure \ref{fig:ch_SU1s1ab1f9fb}(c) respectively.

\begin{figure}[h!]
	\centering
	\includegraphics[width=9cm]{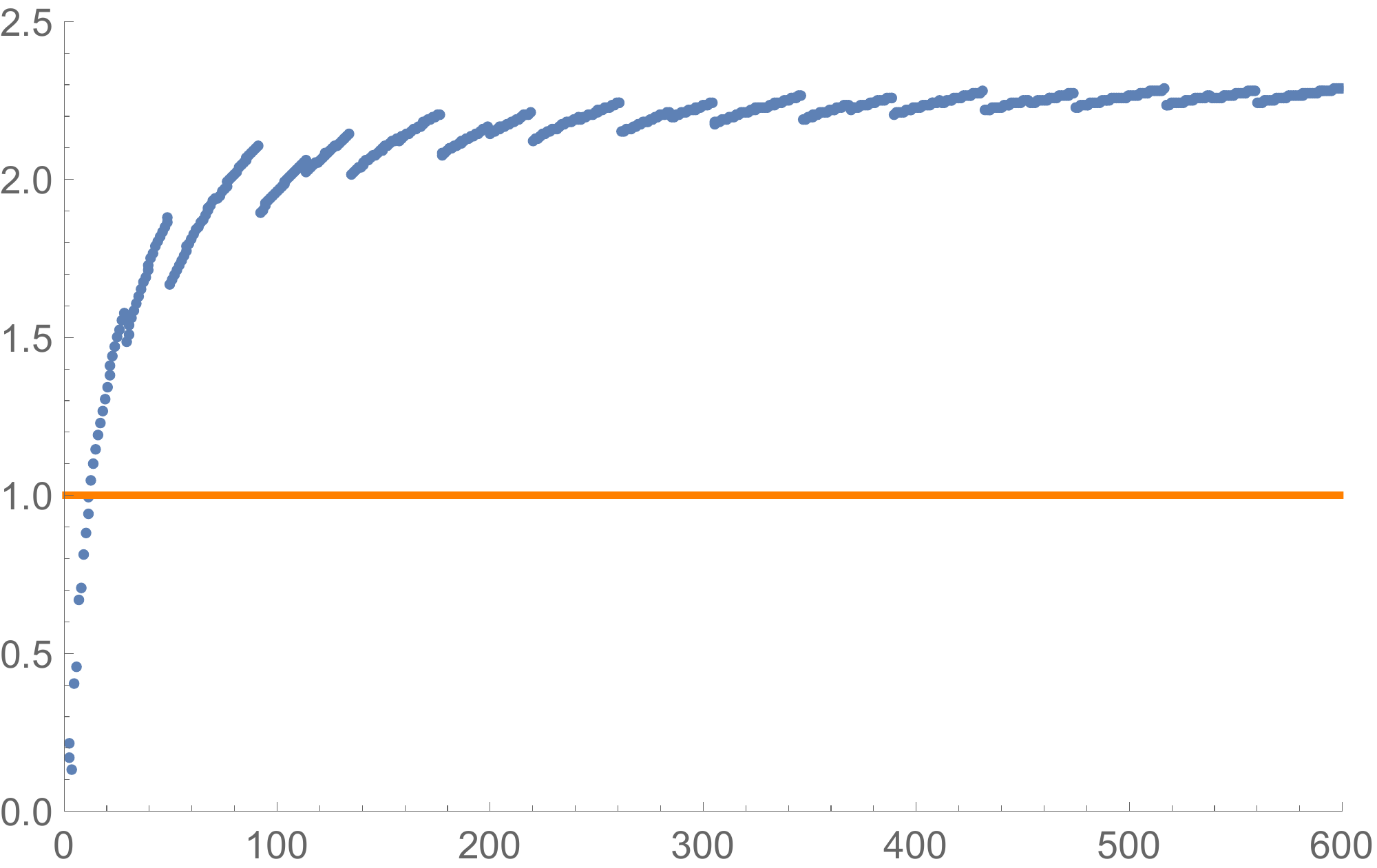}
	\caption{Checking the WGC for the chiral $SU(N)$ theory with $N_f=1$. Plot of the shortest distance from origin to the boundary surfaces of polyhedron in charge-to-dimension space. The chiral ring operators do not satisfy the WGC at $N\leq 12$.}
	\label{fig:wgc_SU1s1ab1f9fb}
\end{figure}

One may check the convex hull condition by computing the shortest distance from the origin to the boundary surfaces of polyhedron. We checked this for $N=2,\dots,600$, and found that the convex hull condition is satisfied for $N>12$. The results are depicted in Figure \ref{fig:wgc_SU1s1ab1f9fb}.

\section{SO(N) theories} \label{sec:SO}

Let us perform a similar analysis on $SO(N)$ gauge theories. The two main difference from $SU(N)$ gauge theories are 1) $SO(N)$ gauge theories are free from gauge anomalies\footnote{We must take care of the gauge anomaly if we consider the spinor representations in $Spin(N)$. However, we don't consider them in this paper because their degrees of freedom are much larger than the rank-2 anti-symmetric tensor at large $N$.}, and 2) the rank-2 anti-symmetric representation is also the adjoint representation. 
The large $N$ limit restricts the matter representations to be either symmetric ($\ssym$), anti-symmetric ($\santi$) or fundamental ($\sfund$). In order for the gauge theory to be asymptotically free, the following condition must be satisfied:
\begin{align}
\begin{split}
(N+2)\times N_{\ssym}+(N-2)\times N_{\santi}+N_{\sfund}\,\leq 3(N-2)
\end{split}
\label{eq:BDcond}
\end{align}
Here $N_{\mathbf{R}}$ denotes the number of chiral multiplets in representation $\mathbf{R}$. 
All possible $SO(N)$ theories that can have large $N$ limit with a fixed flavor symmetry are listed in Table \ref{tab:SOlist}.
\begin{table}[htbp]
	\centering
	\begin{tabular}{|c|cc|c|}
		\hline 
		Theory & $\beta_{\textrm{matter}}$  & dense spectrum & conformal window \\\hline \hline
		1 \bsym\, + $N_f$ \bfund& $\sim \,N$  & Y & $0\le N_f \le 2N-8^*$\Tstrut\\
		1 \banti\, + $N_f$ \bfund & $\sim \,N$  & Y & $1\le N_f \le 2N-4^*${\rule[-2ex]{0pt}{-3.0ex}}\Tstrut\\
		 \hline
		2 \bsym\, + $N_f$ \bfund & $\sim 2\,N$  & N & $0\le N_f \le N-10^* $\Tstrut\\
		1 \bsym\, + 1 \banti\, +$N_f$ \bfund& $\sim 2\,N$ & N & $0\le N_f \le N-6^*$\Tstrut\\
		2 \banti\, + $N_f$ \bfund& $\sim 2\,N$  & N & $0\le N_f<N-2 $\Tstrut\\
		3 \banti & $\sim 3\,N$ & N & $*$\Tstrut \\ \hline
	\end{tabular}
	\caption{List of all possible $SO(N)$ theories with large $N$ limit with a fixed flavor symmetry. $\beta_{\textrm{matter}}$ denotes the contribution to the 1-loop beta function from the matter multiplets for $N_f \ll N$. The last column denotes the condition for the theory to flow to a superconformal fixed point. For the cases with *, the theory does not flow when $N_f$ saturates the upper bound, but possess non-trivial conformal manifold. The first two classes of theories exhibit dense spectrum for $N_f \ll N$.}
	\label{tab:SOlist}
\end{table}
The first two of these exhibit a dense spectrum of operators. We study these theories in detail in the following sub-sections. 

\subsection{1 symmetric and $N_f$ fundamentals}
1 \bsym\, + $N_f$  \bfund \,: This theory has 1 anomaly-free global symmetry that we call $U(1)_B$ under which the symmetric field $S$ and fundamental $Q$ have charges given by $1$ and $-(N+2)/N_f$.
The schematic form of the gauge-invariant operators are
\begin{itemize}
	\item $\text{Tr} S^n$,\quad $n=1,\dots,N$
	\item $Q_I S^n  Q_J,\quad n=0,\dots,N-1$
\end{itemize}
Here the indices $I, J$ runs from 1 to $N_f$. 
For this model, we obtain a non-trivial fixed point for $N_f \ge 0$. 

\paragraph{$N_f=0$ case} 
Let us start with the simplest case. There is no fundamental chiral multiplet and the $R$-charge is already determined by anomaly-free condition 
to be $R_S = \frac{4}{N+2}$. There is no anomaly-free (continuous) flavor symmetry. The classical $U(1)$ flavor symmetry acting on $S$ is anomalous and therefore breaks down to $\mathbf{Z}_{2N+4}$ .
The only gauge-invariant (single-trace) operator 
is of the form $ {\rm Tr} S^n$ with $n=1, 2, \ldots, N$, with it's dimension being
\begin{align}
 \Delta_{S^n} = \frac{3}{2} n R_S=\frac{6 n}{N+2} \ .
\end{align}
Some of these operators can decouple along the RG flow since they violate the unitarity bound when
\begin{align}
n< \left\lfloor\frac{N+2}{6} \right\rfloor \ . 
\end{align}
When this happens, we introduce a flip field, which would have dimension $3 - \frac{6n}{N+2}$ to remove the decoupled operator. We see that at large $N$, the Coulomb branch operators and the flip fields fill a band of conformal dimensions given by $1 < \Delta < 6$. 

At large $N$, the central charges $a$ and $c$ are given as
\begin{align}
\begin{split}
a&\simeq\frac{3059 N^6+2688 N^5-5508 N^4-4000 N^3+2496 N^2-1408}{2304 (N+1)^3 (N+2)^2} \ , \\
c&\simeq\frac{2895 N^6+1672 N^5-7672 N^4-5424 N^3+3536 N^2+1792 N-768}{2304 (N+1)^3 (N+2)^2} \ . 
\end{split}
\end{align}
We thus see that the ratio of the central charges asymptotes to 
\begin{align}
\frac{a}{c}\quad\underset{N\gg 1}{\longrightarrow}\quad\frac{3059}{2895} \ .
\end{align}
Notice that $a/c$ is greater than 1 at large $N$. 
Unlike the case of $SU(N)$ gauge theories with 1 symmetric and 1 anti-symmetric, this model does not have a continuous flavor symmetries so we have nothing to say about the Weak Gravity Conjecture. 

\begin{figure}[h!]
	\centering
	\includegraphics[width=9cm]{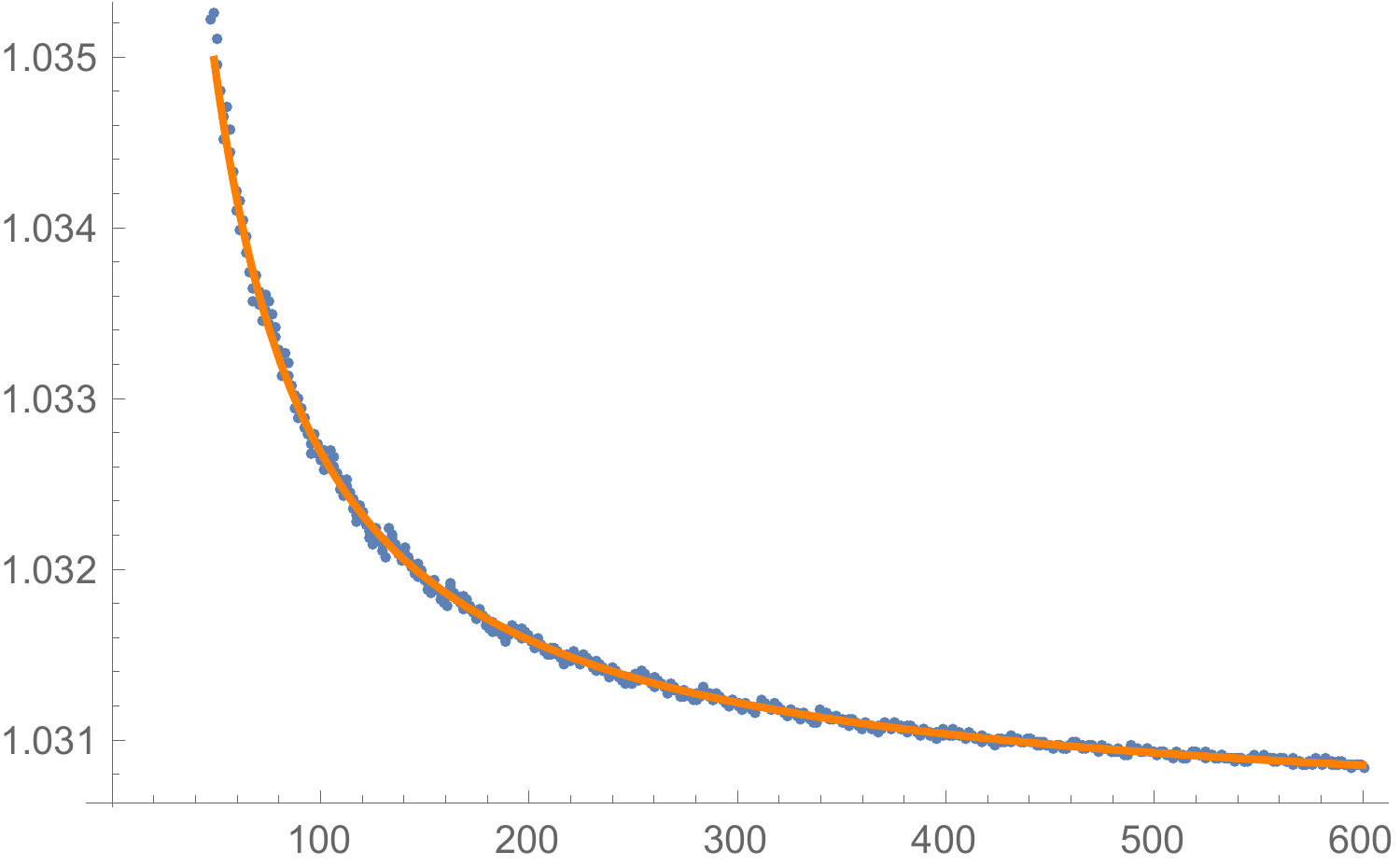}
	\caption{Plot of $a/c$ vs $N$ for the $SO(N)$ theory with 1 symmetric and 1 fundamental. The orange curve fits the plot with $a/c \sim 1.03049 +0.179993/N$.}
	\label{fig:ac_SO1s1f}
\end{figure}

\paragraph{$N_f=1$ case}
Next we move on to the case with 1 fundamental chiral multiplet. We find the asymptotic behavior of the $R$-charges and central charges $a$, $c$ as 
\begin{align}
\begin{split}
&a\simeq 1.54827 N-10.2117\ ,\\
&c\simeq 1.50239 N-10.1376\ , \\
&4\pi^4C_B\simeq6.22247 N^3+12.5168 N^2+52.0892 N-4305.74\ ,\\
&R_{S}\simeq 4.67249/N \ ,\\
&R_{Q}\simeq 0.253588\, +3.51834/N \ ,
\end{split}
\end{align}
at large $N$. 
We plot the ratio of central charges $a/c$ vs $N$ in Figure \ref{fig:ac_SO1s1f}.
Once again, we see that $a/c$ is greater than 1 and does not go to zero at large enough $N$. 

\begin{figure}[h!]
	\centering
	\includegraphics[width=9cm]{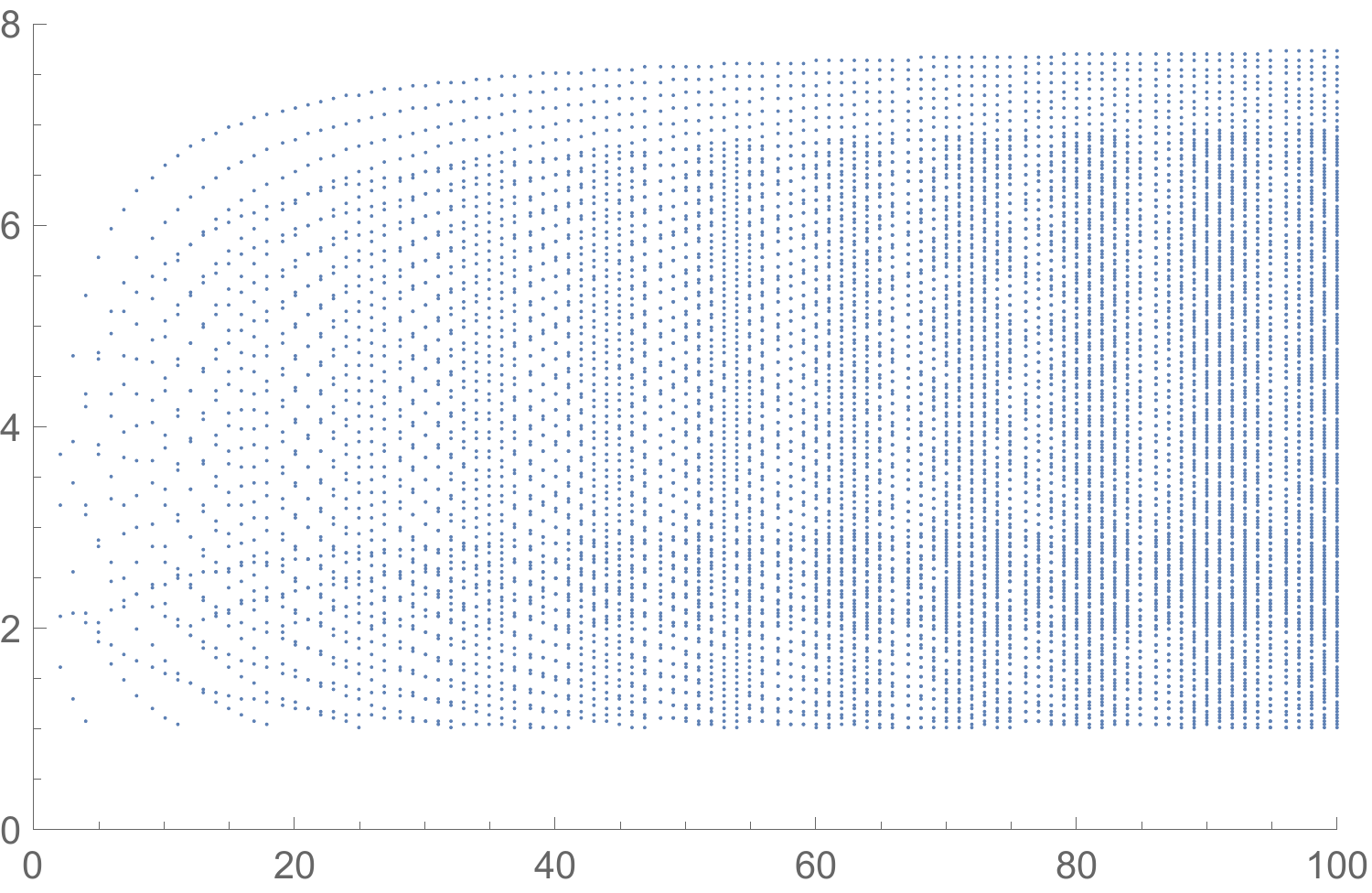}
	\caption{The spectrum of the (single-trace) gauge-invariant operators for the $SO(N)$ theory with 1 symmetric and 1 fundamental matter. They form a dense band of conformal dimension $ 1 < \Delta < 8$.}
	\label{fig:spec_SO1s1f}
\end{figure}

We see that the $R$-charge of the rank-2 tensor $S$ goes to 0 at large $N$, which makes it possible for the mesons $Q S^n Q$ to form a dense band of conformal dimensions though some of them may decouple for low $n$. Operators of form $\epsilon\epsilon S^{n-k}Q^{2k}$ are not single-trace-operators since they can be written in terms of ${\rm Tr} S^n$ and $Q_I S^n  Q_J$ by expressing $\epsilon$ in terms of  Kronecker-$\delta$'s.
We plot the conformal dimensions of single-trace gauge-invariant operators in Figure \ref{fig:spec_SO1s1f}.
We see that there is a dense band of conformal dimension $1 < \Delta < 8$. The band comprises of the mesons ($QS^n Q$), the operators ${\rm Tr} S^n$ the flip fields for the decoupled operators. 

\begin{figure}[h!]
	\centering
	\includegraphics[width=9cm]{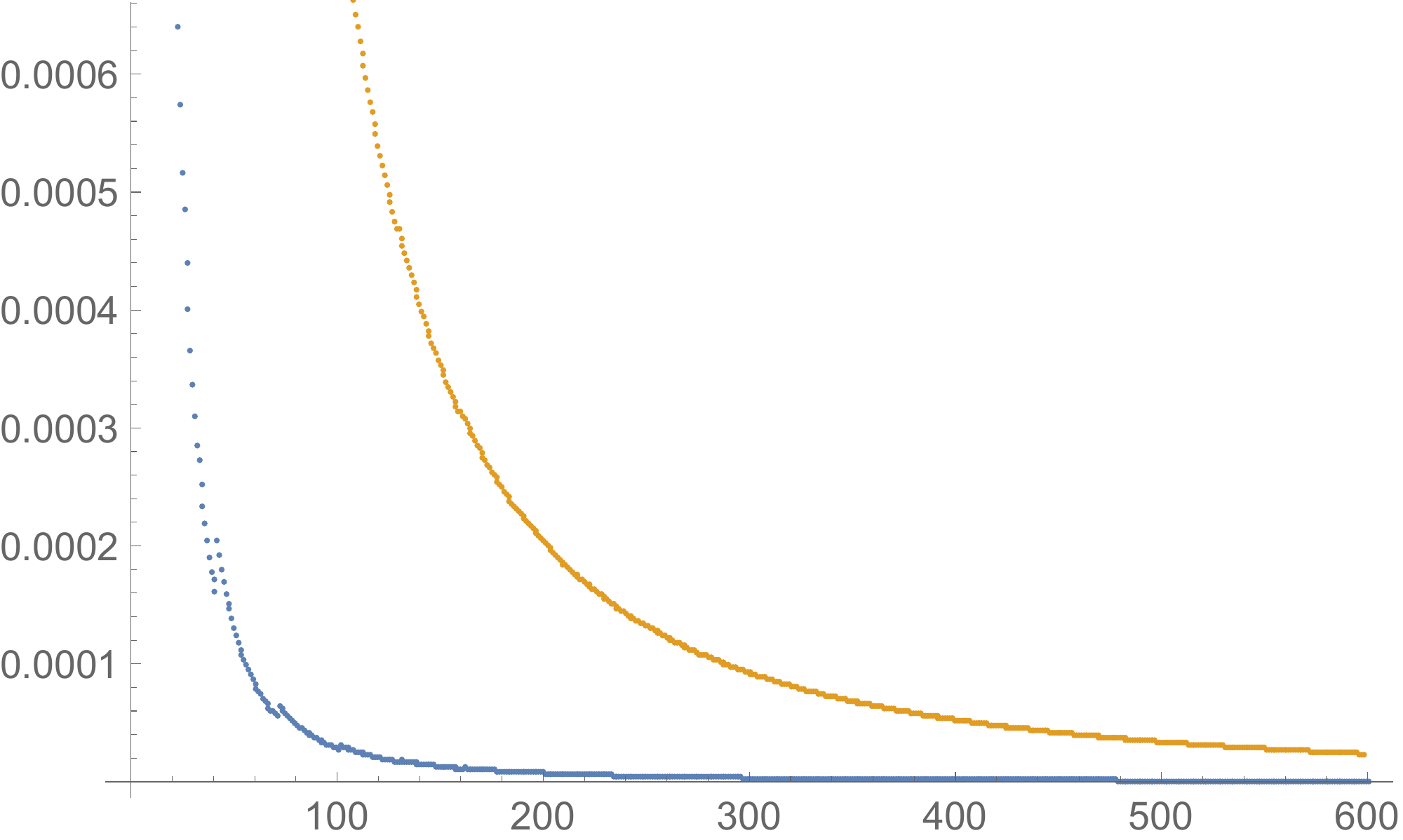}
	\caption{Testing the WGC for $SO(N)$ theory with 1 symmetric and 1 fundamental. Plot of $\frac{\Delta_{\text{mes}}^2}{q^2}$ (blue curve) for the lightest meson and $\frac{9}{40}\frac{C_B}{C_T}$ (orange curve) vs $N$.}
	\label{fig:wgc_SO1s1f}
\end{figure}

Now we can check WGC from the spectral data. 
We find that for all values of $N$, the lightest operator of form $Q S^n Q$ 
has minimal dimension-to-charge ratio, which is given by the blue curve in Figure \ref{fig:wgc_SO1s1f}. 
We see that this model indeed satisfies the WGC.

\subsection{1 anti-symmetric and $N_f$ fundamentals}
1 \banti\, + $N_f$  {\bfund}: This model has 1 anomaly-free global $U(1)_B$ under which the anti-symmetric field $A$ and fundamental $Q$ have charges $1$ and $-(N-2)/N_f$ respectively. 
The schematic form of the gauge-invariant operators is given as follows:
\begin{itemize}
	\item Coulomb branch operators: $\Tr A^{2n}$, $\mathrm{Pf}A$, \quad $n=1,\dots,\left\lfloor\frac{N-1}{2}\right\rfloor$
	\item Symmetric mesons: $Q_I A^{2n} Q_J,\quad n=0,\dots,\left\lfloor\frac{N}{2}\right\rfloor$
	\item Anti-symmetric mesons: $Q_I\,A^{2n+1}\,Q_J,\quad n=0,\dots,\left\lfloor\frac{N-1}{2}\right\rfloor$
	\item $\epsilon\,A^nQ_{I_1}\dots Q_{I_{N-2n}},\quad n=\left\lceil\frac{N-N_f}{2}\right\rceil,\dots\left\lfloor\frac{N}{2}\right\rfloor-1$
\end{itemize}
where indices $I, J$ run from $1, 2, \ldots N_f$.

\begin{figure}[h]
	\centering
	\includegraphics[width=9cm]{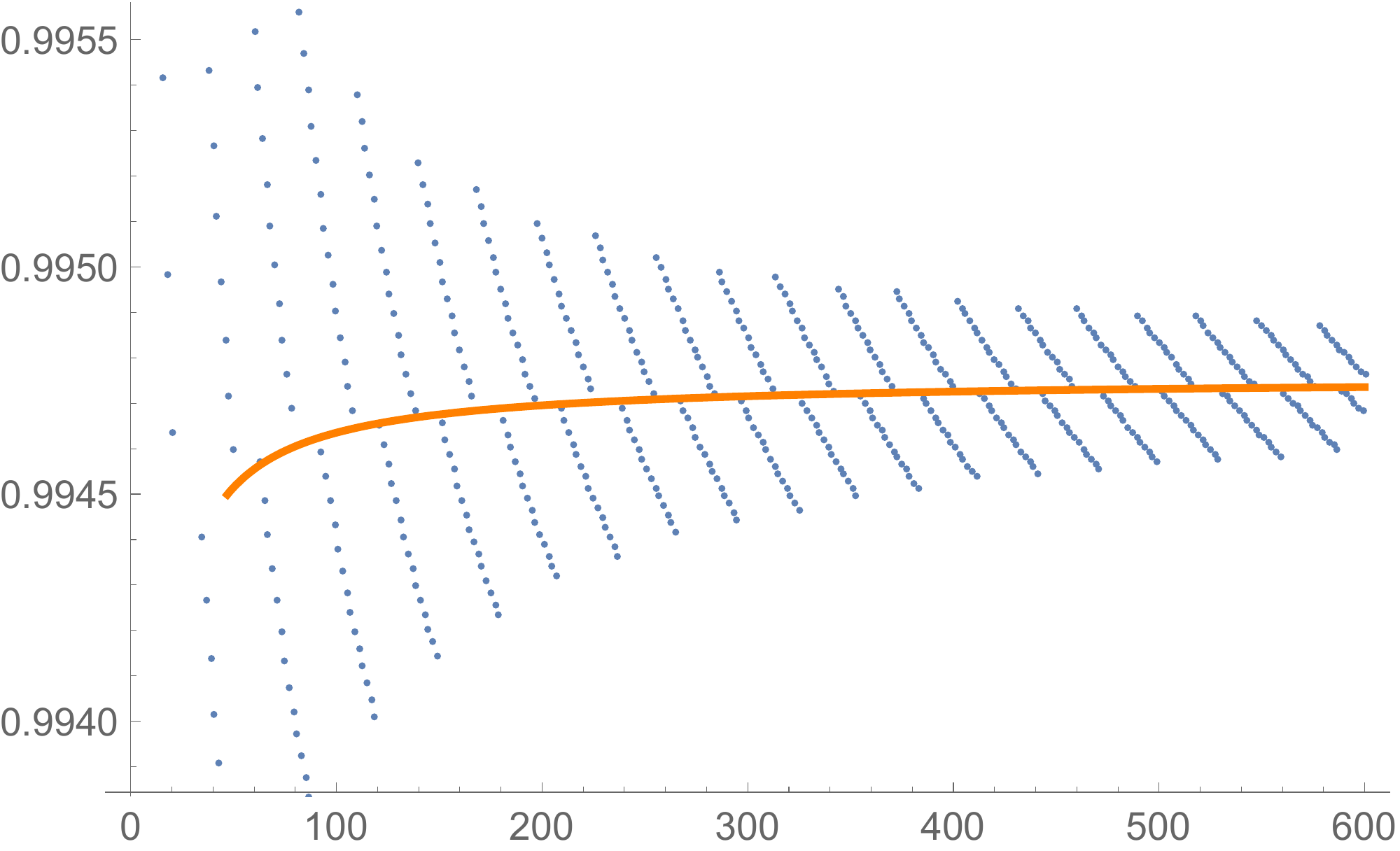}
	\caption{Plot of $a/c$ vs $N$ for the $SO(N)$ theory with 1 anti-symmetric and 1 fundamental (vector). The orange curve fits the plot with $a/c \sim 0.994756 -0.0120384/N  $.}
	\label{fig:ac_SO1a1f}
\end{figure}

The simplest model that shows a dense spectrum has $N_f=1$. Let us study this model in detail. Upon performing $a$-maximization repeatedly while removing the decoupled operators, we find the asymptotic behavior of the $R$-charges and the central charges to be 
\begin{align}
\begin{split}
&a \simeq\,0.250425 N-0.0881209 \ , \\
&c\simeq\,0.251746 N-0.0857222\ , \\
&4\pi^4C_B\simeq\,4.95300 N^3+31.3224 N^2+2.86094 N+897.928\\
&R_{A}\simeq\,0.723009/N \ , \\
&R_{Q} \simeq\begin{cases} 
0.284351+0.241116/N & \text{if}\,\, N=\text{even}\\
0.284347+0.0988993/N & \text{if}\,\, N=\text{odd}.
\end{cases}
\end{split}
\end{align}
The central charges grow linearly in $N$ and the $R$-charge of the anti-symmetric tensor goes to zero at large $N$. 
We plot the ratio of central charges $a/c$ vs $N$ in Figure \ref{fig:ac_SO1a1f}.
We find that $a/c$ remains less than 1 and does not asymptote to 1 at large $N$. 

\begin{figure}[h]
	\centering
	\includegraphics[width=8.5cm]{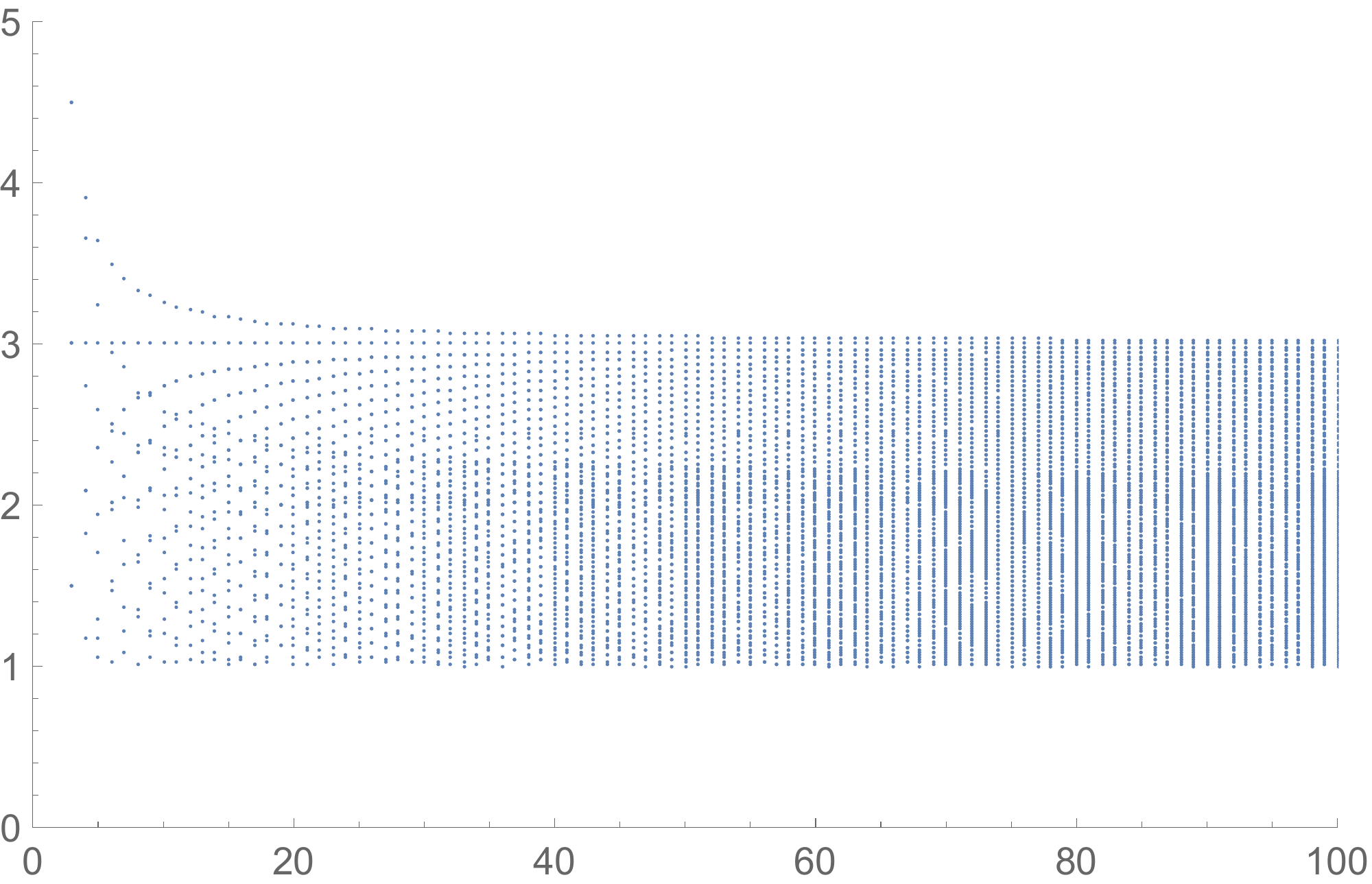}
	\caption{The spectrum of (single-trace) gauge-invariant operators for $SO(N)$ theory with 1 anti-symmetric and 1 fundamental. They form a dense band of conformal dimension $1 < \Delta < 4$ at large $N$.}
	\label{fig:spec_SO1a1f}
\end{figure}

The spectrum of low-lying gauge-invariant operator at the IR fixed point is depicted in Figure \ref{fig:spec_SO1a1f}.
They are given by the Coulomb branch operators $\text{Tr}\,A^{2n}$, the mesons $Q A^n Q$ (with $n$ even), and the operator $A^{\lfloor N/2\rfloor}Q^{\text{Mod}[N,2]}$ and the flip fields for the decoupled operators. 

\begin{figure}[h]
	\centering
	\includegraphics[width=9cm]{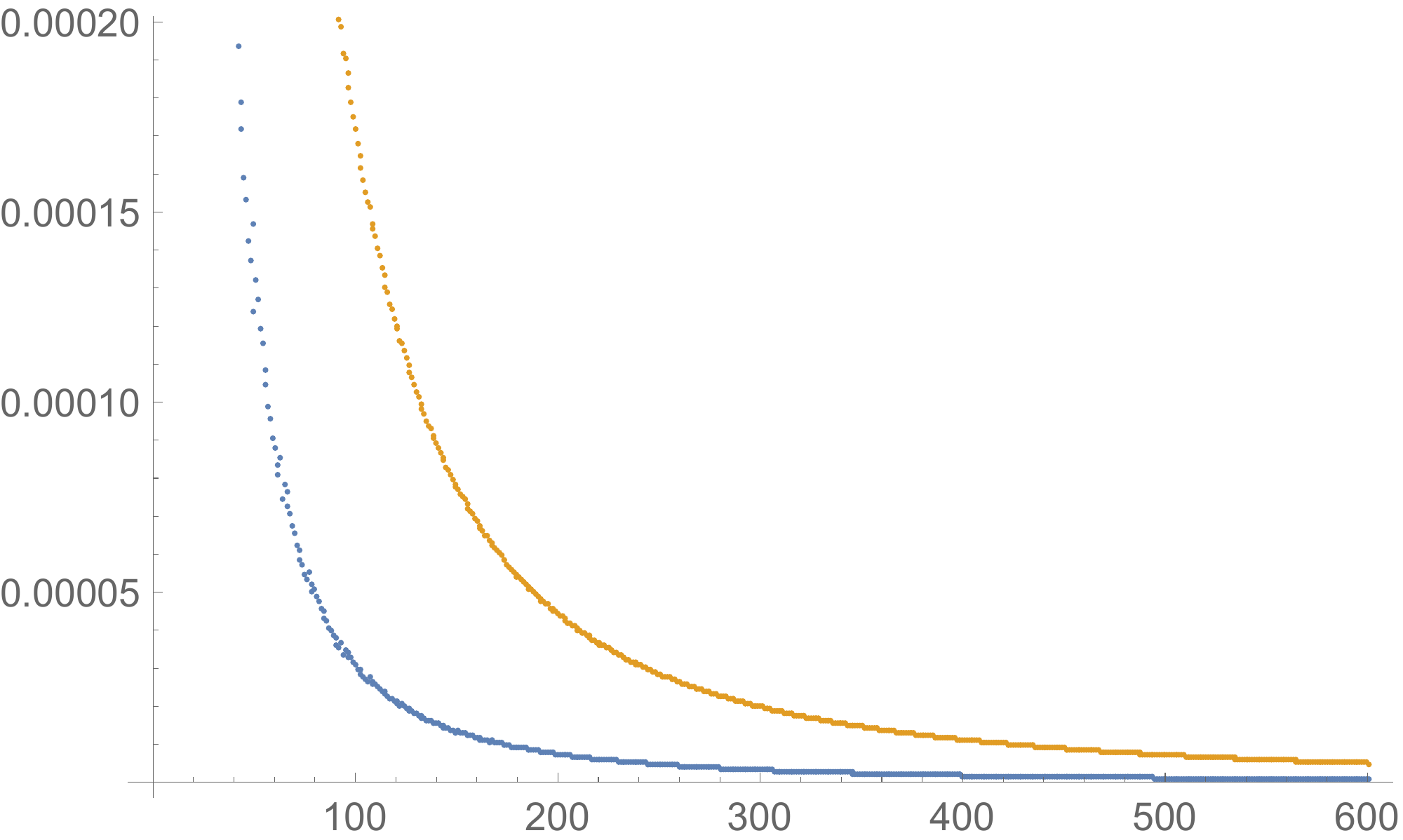}
	\caption{Testing WGC for the $SO(N)$ theory with 1 anti-symmetric and 1 fundamental. Plot of $\frac{\Delta_{\text{mes}}^2}{q^2}$ (blue curve) and $\frac{9}{40}\frac{C_B}{C_T}$ (orange curve) vs $N$.}
	\label{fig:wgc_SO1a1f}
\end{figure}

Now we can check the WGC from the spectral data. Amongst the lightest meson, the lightest Coulomb branch operator, and $\epsilon A^{\lfloor N/2\rfloor}Q^{\text{Mod}[N,2]}$, the meson turns out to have the smallest dimension-to-charge ratio, which is given by the blue curve in Figure \ref{fig:wgc_SO1a1f}. We can clearly see that this theory satisfy the WGC.

We now carry out a similar analysis of the analyze the $N_f=2$. The central charges, flavor central charge and R-charges are obtained as
\begin{align}
\begin{split}
&a\simeq0.471292 N-0.483137\ ,\\
&c\simeq0.503128 N-0.517622\ ,\\
&4\pi^4C_B\simeq0.893101 N^3+15.2127 N^2-225.497 N+8217.09 \ , \\
&R_{A}\simeq 1.50156/N\ , \\
&R_{Q}\simeq0.253511\, +0.757998/N\ .
\end{split}
\end{align}
Similar to the $N_f=1$ theory, the central charges exhibit a linear growth in $N$, and the R-charge of the anti-symmetric matter vanishes in the large $N$ limit. Also the $a/c$ does not approach 1 in the large $N$ limit.
\begin{figure}[h]
	\centering
	\includegraphics[width=9cm]{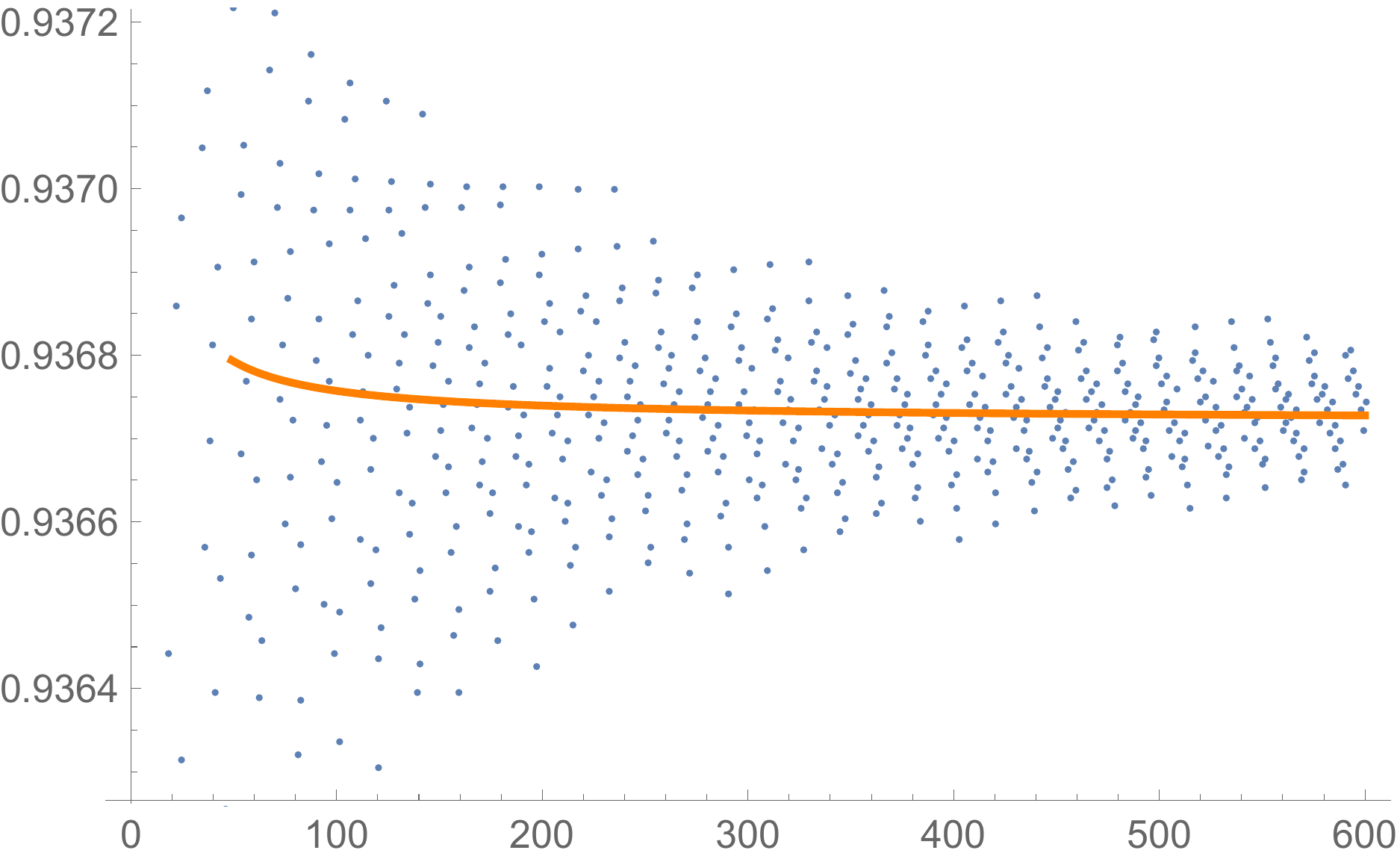}
	\caption{Plot of $a/c$ vs $N$ for the $SO(N)$ theory with 1 anti-symmetric and 2 fundamental (vector). The orange curve fits the plot with $a/c\sim 0.936722\, +0.00353576/N  $.}
	\label{fig:ac_SO1a2f}
\end{figure}

\begin{figure}[h]
	\centering
	\includegraphics[width=8.5cm]{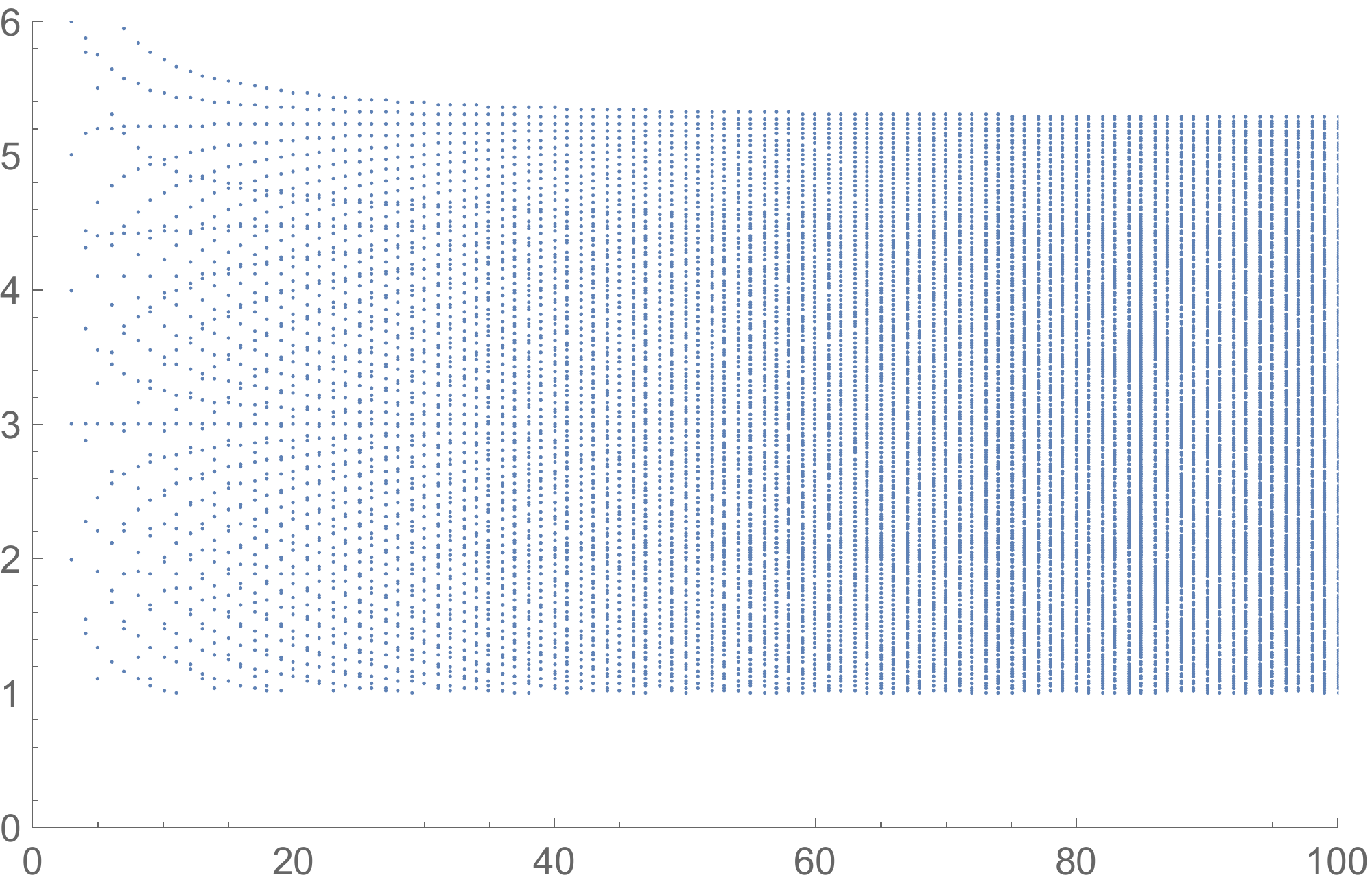}
	\caption{The spectrum of (single-trace) gauge-invariant operators for $SO(N)$ theory with 1 anti-symmetric and 2 fundamental. They form a dense band of conformal dimension $1 < \Delta < 6$ at large $N$.}
	\label{fig:spec_SO1a2f}
\end{figure}

Once again the spectrum of low-lying gauge-invariant operators forms a dense band with conformal dimensions lying in the range $1<\Delta<6$ at IR fixed point, as shown in Figure \ref{fig:spec_SO1a2f}. It comprises of the Coulomb branch operators, the mesons, the operators $\epsilon\,A^nQ_{I_1}\dots Q_{I_{N-2n}}$ and the flip fields for the decoupled operators. 

\begin{figure}[h]
	\centering
	\includegraphics[width=9cm]{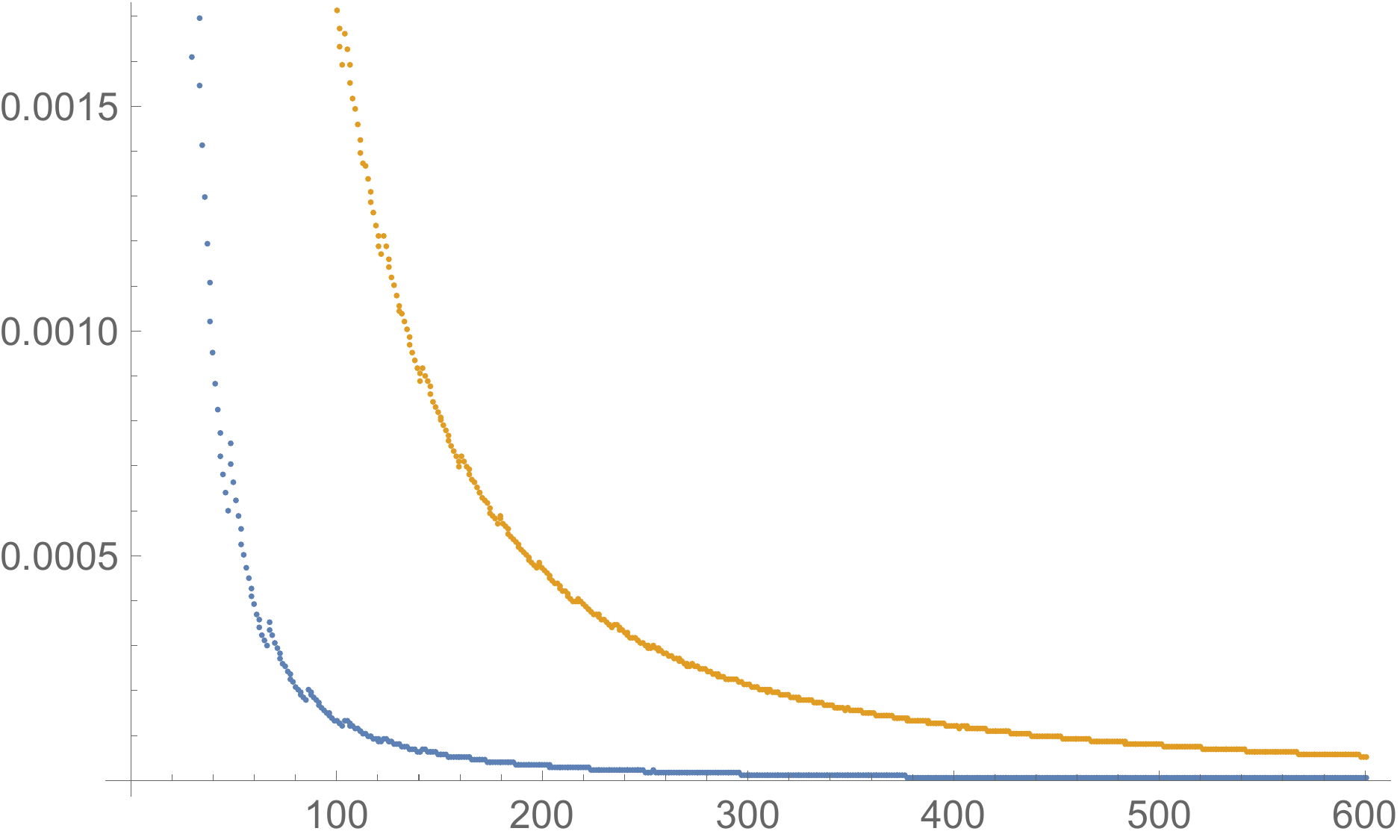}
	\caption{Testing WGC for the $SO(N)$ theory with 1 anti-symmetric and 2 fundamental. Plot of $\frac{\Delta_{\text{mes}}^2}{q^2}$ (blue curve) and $\frac{9}{40}\frac{C_B}{C_T}$ (orange curve) vs $N$.}
	\label{fig:wgc_SO1a2f}
\end{figure}

Finally we check the Weak Gravity Conjecture. It turns out that the lightest meson $Q_IA^nQ_J$ has smallest dimension-to-charge ratio among the gauge-invariant operators. We plot ${\Delta_{\text{mes}}^2}/{q^2}$ and $9{C_B}/{40 C_T}$ in blue and orange curve respectively in Figure \ref{fig:wgc_SO1a2f}. The figure clearly shows that the WGC is satisfied.

\section{Sp(N) theories} \label{sec:Sp}
We now analyze $Sp(N)$ gauge theories. The $Sp(N)$ gauge theory does not have any triangle anomaly but can suffer from a Witten anomaly \cite{Witten:1982fp}. Here we consider the theories with an even number of fundamental matters to ensure the absence of the Witten anomaly. 
As before, we restrict ourselves to the matters in symmetric (=adjoint $\ssym$), anti-symmetric ($\santi$) and fundamental ($\sfund$) representations. 
The condition of asymptotic freedom requires the number of matter multiplets, $N_\mathbf{R}$, to satisfy
\begin{align}
\begin{split}
(N+1)\times N_{\ssym}+(N-1)\times N_{\santi}+\frac{1}{2}N_{\sfund} \leq 3(N+1) \ . 
\end{split}
\label{eq:BDcond}
\end{align}
We also need to impose $N_{\sfund}$ to be even. 
We list all possible theories with large $N$ limit in Table \ref{tab:Splist}. 
\begin{table}
\centering
	\begin{tabular}{|c|cc|c|}
		\hline
		Theory & $\beta_{\textrm{matter}}$ & dense spectrum & conformal window \\\hline \hline
		1 \bsym \, + 2$N_f \, \bfund$ & $\sim N$  & Y & $1\le N_f \le 2N+2^*$\Tstrut \\
		1 \banti\, + 2$N_f \,\bfund $ & $\sim N$  & Y & $4\le N_f < 2N+4$\Tstrut  {\rule[-2ex]{0pt}{-3.0ex}}  \\
		\hline
		2 \bsym\, + 2$N_f \, \bfund $ & $\sim 2N$  & N & $0\le N_f <N+1$\Tstrut \\
		1 \bsym\, + 1 \banti\, + 2$N_f \,\bfund$& $\sim 2N$ & N & $0\le N_f\le N+3^* $\Tstrut\\
		2 \banti\, + 2\,$N_f \, \bfund$& $\sim 2N$  & N & $0\le N_f \le N+5^*$\Tstrut\\
		2 \bsym\, + 1 \banti\, + 2$N_f \,\bfund $& $\sim 2N$ & N & $0 \leq N_f <2$\Tstrut\\
		1 \bsym\, + 2 \banti\, + 2$N_f \,\bfund $ & $\sim 2N$ & N & $0 \leq N_f \le 4^*$\Tstrut\\
		3 \banti\, + 2$N_f$ \bfund & $\sim 3N$ & N & $0\le N_f \le 6^*$\Tstrut\\
		3 \bsym\, & $\sim 3N$ & N & * \Tstrut \\ \hline
	\end{tabular}
	\caption{List of all possible $Sp(N)$ theories with large $N$ limit and fixed global symmetry. $\beta_{\textrm{matter}}$ denotes the contribution to the 1-loop beta function from the chiral multiplets when $N_f \ll N$. It has to be less than $3N+3$ to be asymptotically free. The last column denotes the condition for the theory to have a superconformal fixed point. The cases with * do not flow but have non-trivial conformal manifolds (when $N_f$ saturates the upper bound). 
	The first two theories have dense spectrum for $N_f\ll N$. }
	\label{tab:Splist}
\end{table}
We will focus on the two theories that exhibit dense spectrum at large $N$. 

\subsection{1 symmetric and $2N_f$ fundamentals}
1 \bsym\, + 2$N_f$  \bfund \,: This is the $Sp(N)$ adjoint SQCD. There is an anomaly-free global $U(1)_B$ under which the symmetric field $S$ and fundamental $Q$ carry charges $1$ and $-(N+1)/N_f$ respectively. 
The (single-trace) gauge-invariant operators are given as follows:
\begin{itemize}
	\item Coulomb branch operators: $\text{Tr} \left(\Omega S\right)^{2n}$,\quad $n=1,\dots,N$
	\item Symmetric mesons: $Q_I\left(\Omega S\right)^{2n+1}\Omega Q_J,\quad n=0,\dots,N-1$
	\item Anti-symmetric mesons: $Q_I\left(\Omega S\right)^{2n}\Omega Q_J,\quad n=0,\dots,N-1$
\end{itemize}
Here we omitted the gauge indices as before and $I, J$ denote the flavor indices $1, \ldots, 2N_f$. Note that gauge indices are contracted via the $Sp(N)$ invariant skew-symmetric matrix
\begin{align}
\Omega=\left(\begin{tabular}{cc} 0 & $-I_{N}$ \\ $I_{N}$ & 0\end{tabular}\right) \ . 
\end{align}

\begin{figure}[h]
	\centering
	\includegraphics[width=8.5cm]{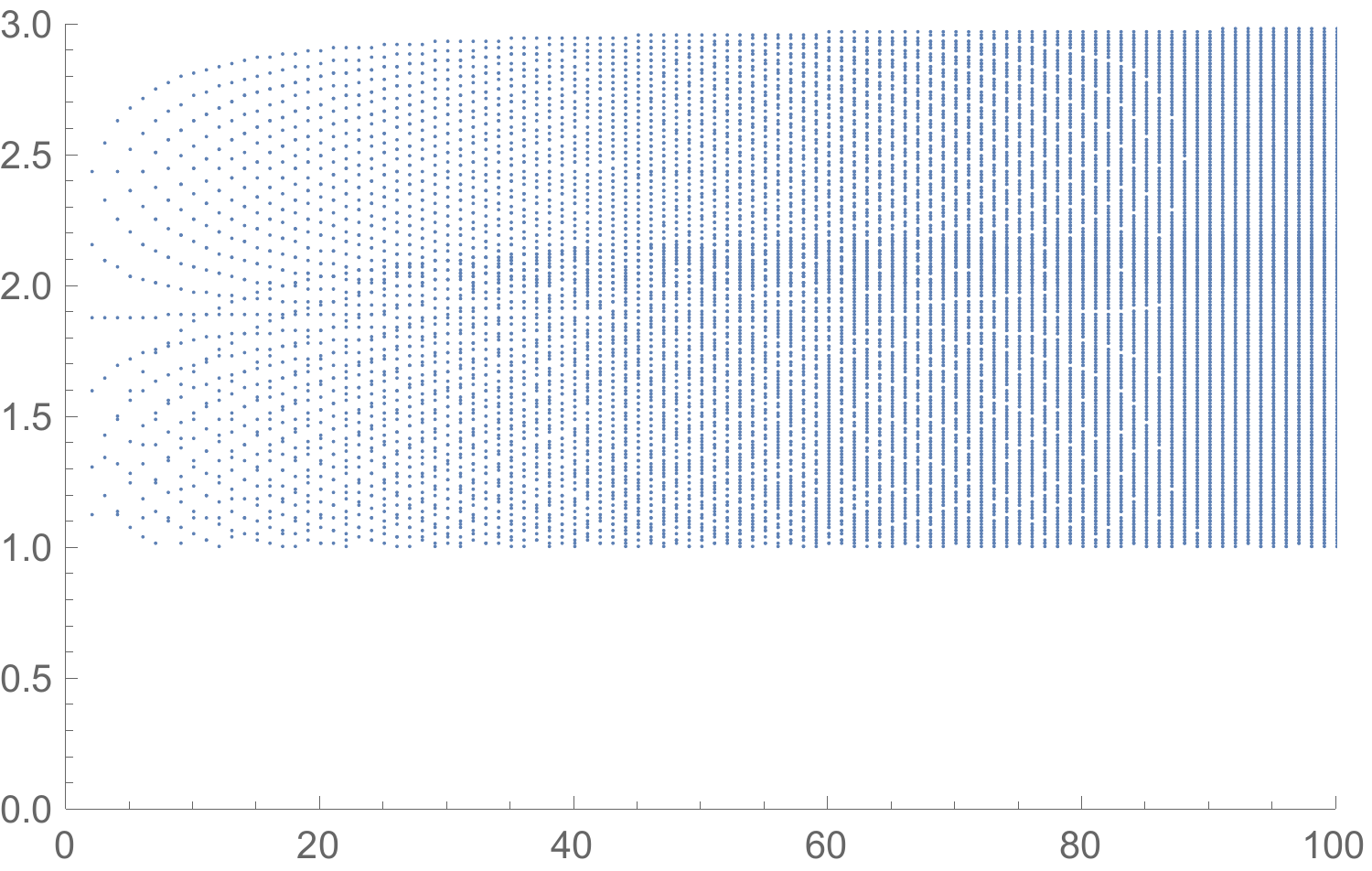}
	\caption{The (single-trace) gauge-invariant operators in $Sp(N)$ theory with 1 adjoint (symmetric) and 2 fundamentals. There is a dense band at conformal dimension $1<\Delta<3$. }
	\label{fig:spec_Sp1s2f}
\end{figure}

\begin{figure}[h]
	\centering
	\includegraphics[width=8.5cm]{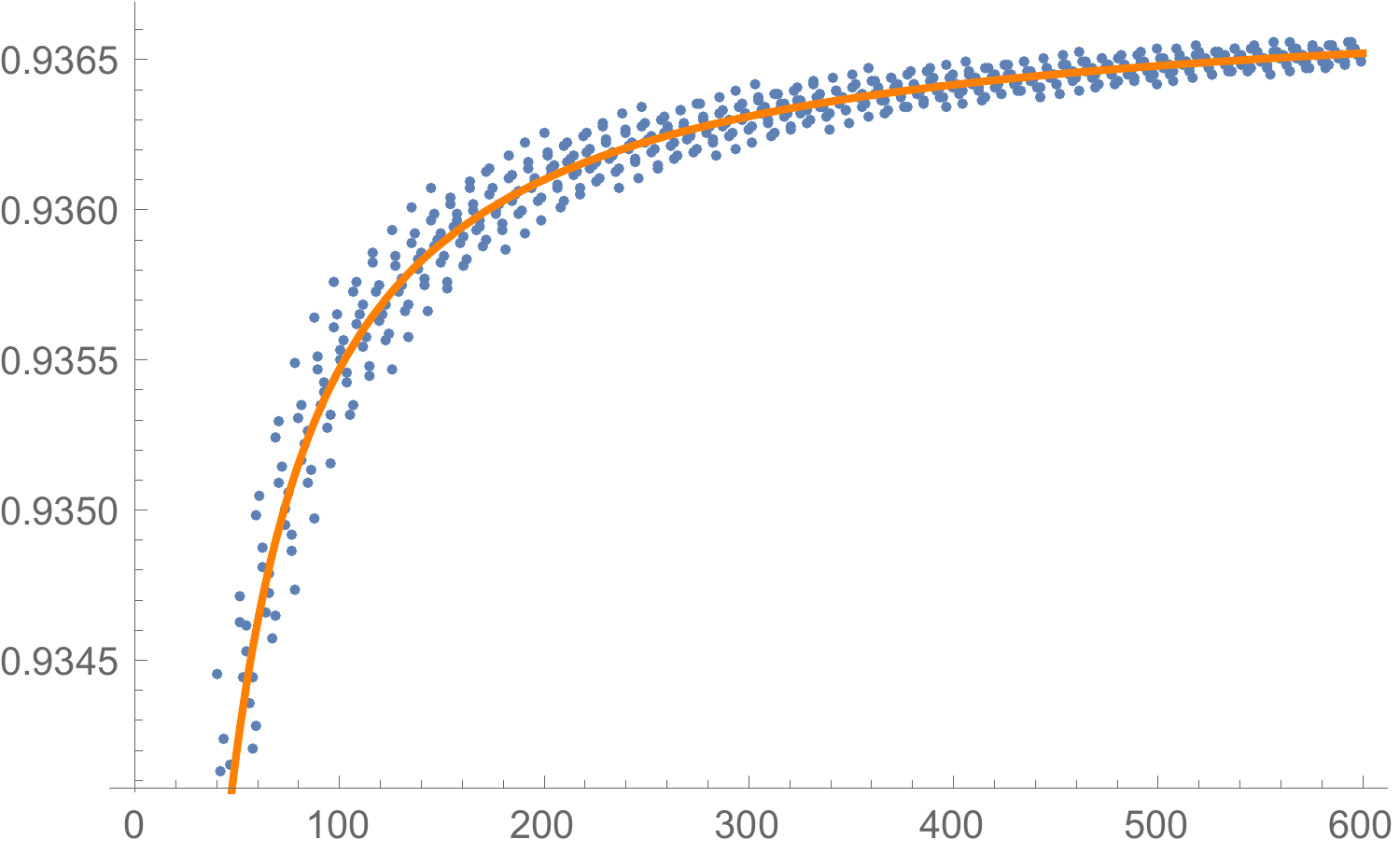}
	\caption{Plot of $a/c$ vs $N$ for the $Sp(N)$ theory with 1 adjoint (symmetric) and 2 fundamentals. The orange curve fits the plot with $a/c\sim 0.936732 -0.126216/N$.}
	\label{fig:ac_Sp1s2f}
\end{figure}

\paragraph{$N_f=1$ theory}
Let us analyze the simplest example \emph{i.e.} with 2 fundamental chiral multiplets ($N_f=1$). 
We find that the central charges and $R$-charges at large $N$ behave as 
\begin{align}
\begin{split}
&a\sim 0.942349 N-1.43282 \ ,  \\
&c\sim 1.00601 N-1.40130 \ ,  \\
&4\pi^4C_B\sim20.9961 N^3+62.5798 N^2+458.279 N-16257.2 \ ,  \\
&R_{S}\sim 0.737532/N \ ,  \\
&R_{Q}\sim 0.253570\, +0.796522/N \ .  
\end{split}
\end{align}
We see that the $R$-charge for the symmetric tensor, $S$, goes to zero at large $N$, while the $R$-charge for the fundamental remains finite at large $N$. Therefore the mesons ($Q_I S^n Q_J$) and the Coulomb branch operators ($\Tr S^{2n}$) and their filp fields (for the ones below the unitarity bound) form a dense spectrum at large $N$. We plot the spectrum of gauge-invariant operators in Figure \ref{fig:spec_Sp1s2f}.
We also plot the ratio of central charges $a/c$ vs $N$ in Figure \ref{fig:ac_Sp1s2f}. We find that $a/c<1$ and the ratio does not converge to 1 at large $N$. 

\begin{figure}[h!]
	\centering\vspace{0.3cm}
	\includegraphics[width=11cm]{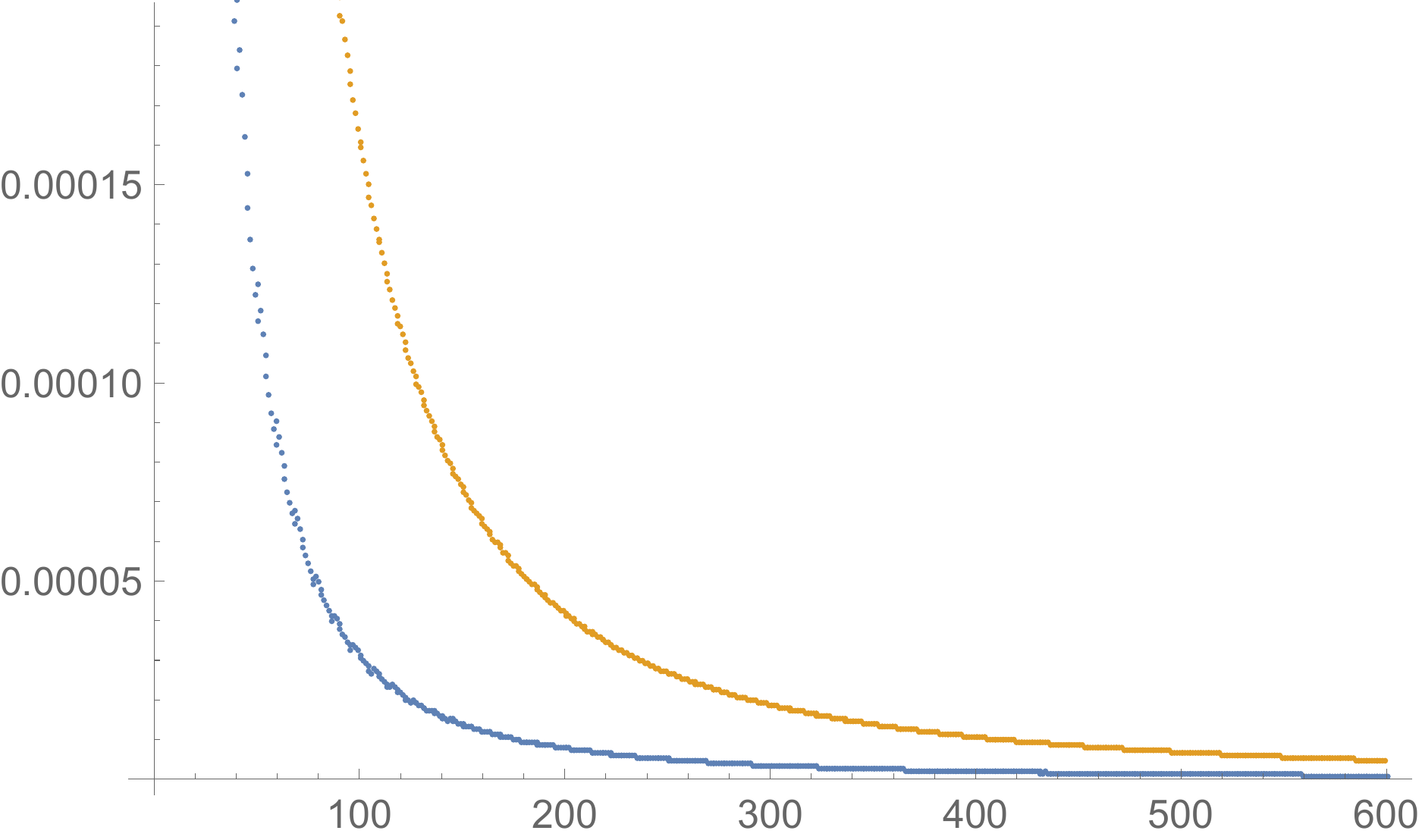}
	\caption{Checking the WGC for $Sp(N)$ theory with 1 symmetric and 2 fundamentals. Plot of $\frac{\Delta_{\text{mes}}^2}{q^2}$ (blue curve) and $\frac{9}{40}\frac{C_B}{C_T}$ (orange curve) vs $N$.}
	\label{fig:wgc_Sp1s2f}
\end{figure}

Now let us check the Weak Gravity Conjecture. We find that the lightest operator of form $Q S^n Q$, has smaller dimension-to-charge ratio than the Coulomb branch operators ${\rm Tr} S^n$. As can be seen in Figure \ref{fig:wgc_Sp1s2f}, the lightest operator of form $Q S^n Q$,  has small enough dimension-to-charge ratio for the theory to satisfy the WGC.

\begin{figure}[h]
	\centering
	\includegraphics[width=8.5cm]{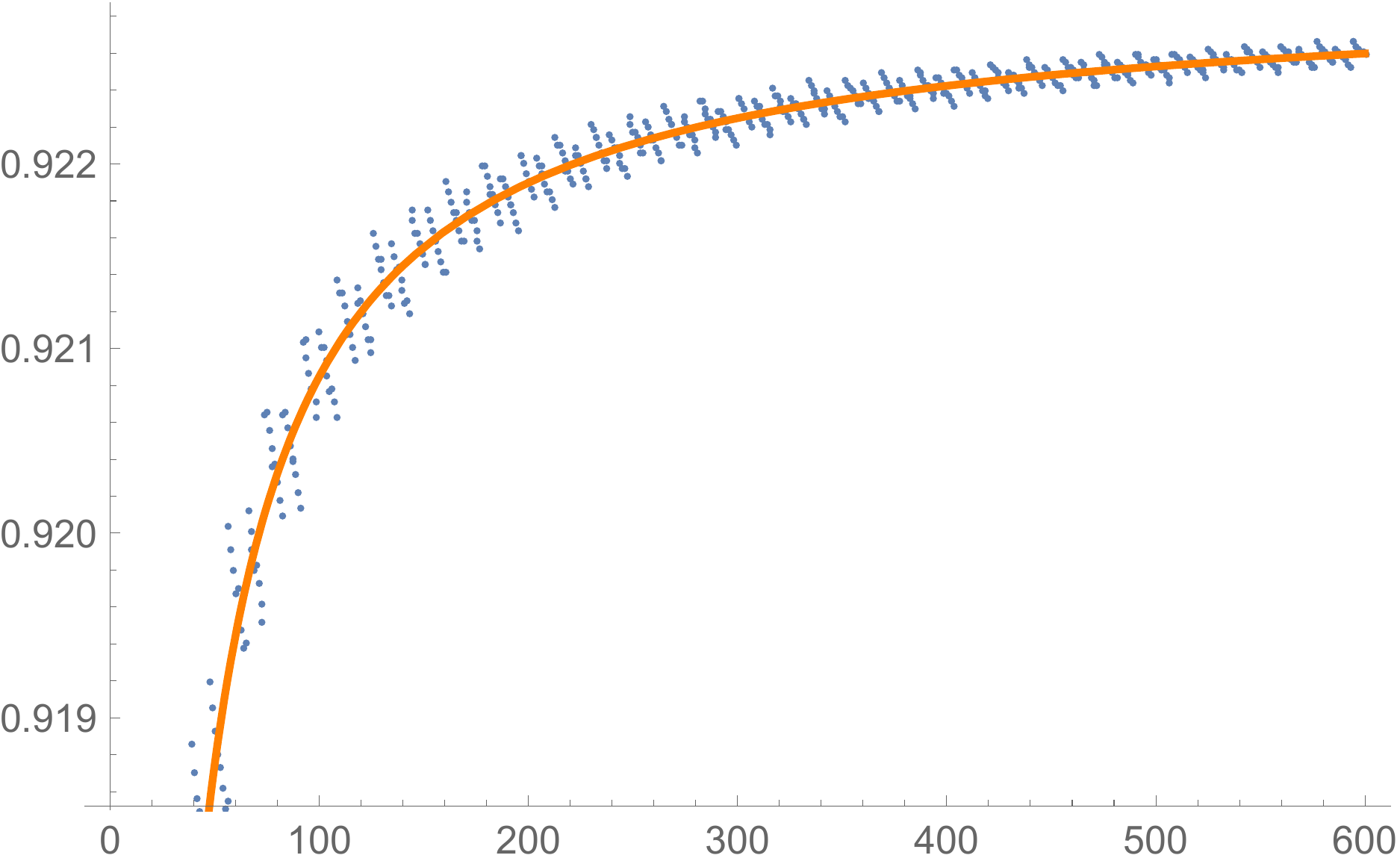}
	\caption{Plot of $a/c$ vs $N$ for the $Sp(N)$ theory with 1 adjoint (symmetric) and 4 fundamentals. The orange curve fits the plot with $a/c\sim 0.922948\, -0.209868/N$.}
	\label{fig:ac_Sp1s4f}
\end{figure}

\begin{figure}[h]
	\centering
	\includegraphics[width=8.5cm]{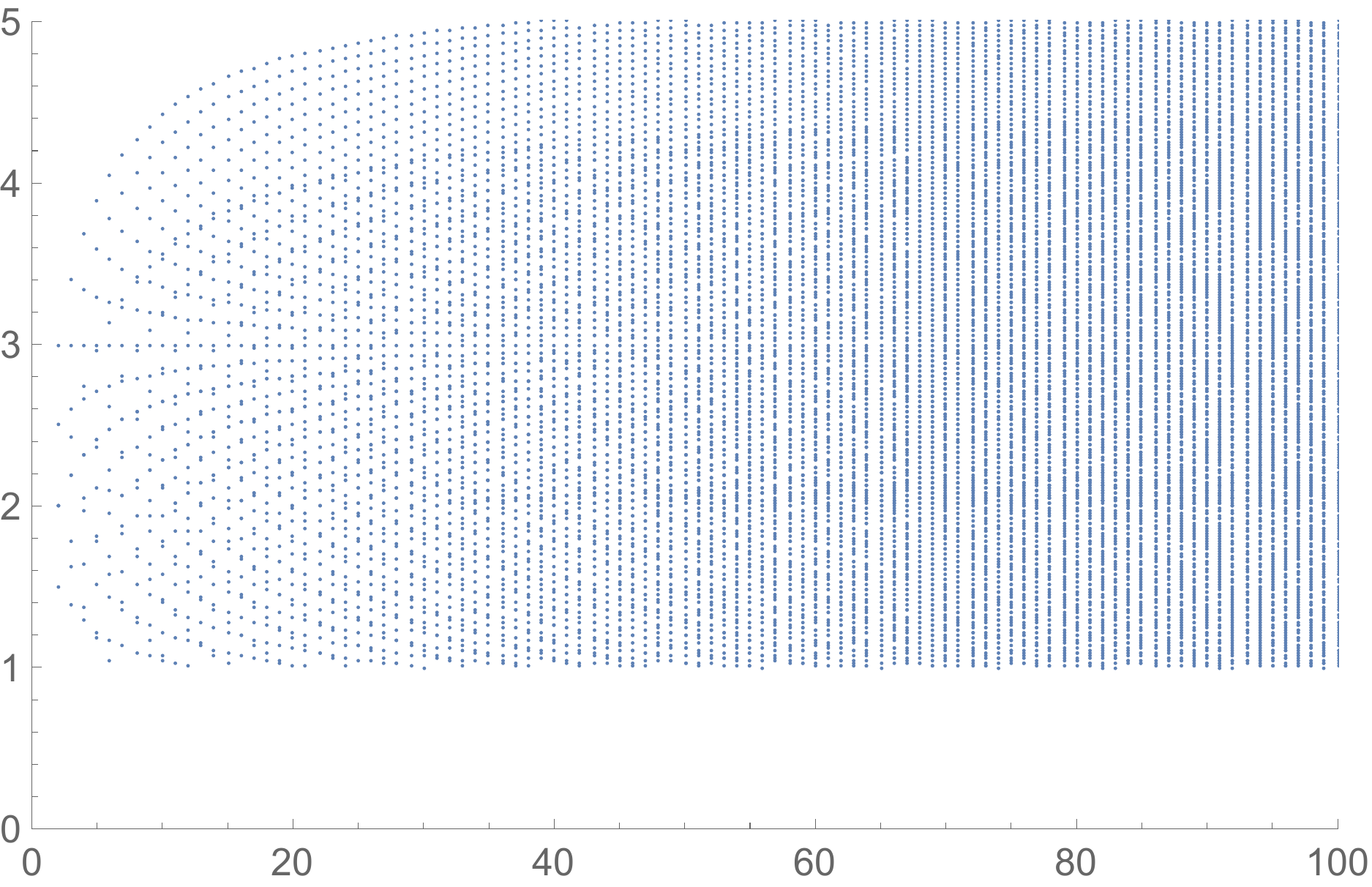}
	\caption{The (single-trace) gauge-invariant operators in $Sp(N)$ theory with 1 adjoint (symmetric) and 4 fundamentals. There is a dense band at conformal dimension $1<\Delta<5$. }
	\label{fig:spec_Sp1s4f}
\end{figure}

\begin{figure}[h!]
	\centering\vspace{0.3cm}
	\includegraphics[width=11cm]{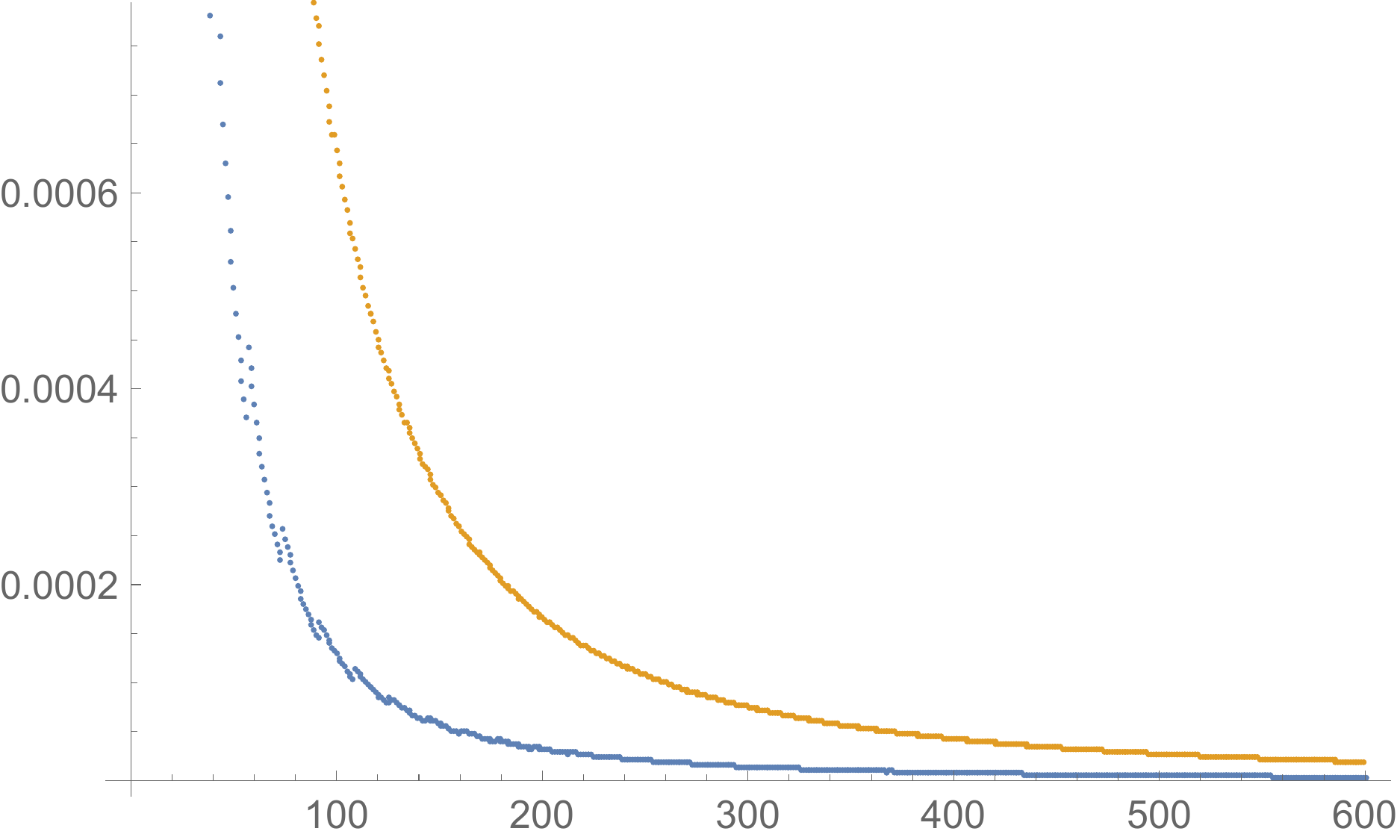}
	\caption{Checking the WGC for $Sp(N)$ theory with 1 symmetric and 4 fundamentals. Plot of $\frac{\Delta_{\text{mes}}^2}{q^2}$ (blue curve) and $\frac{9}{40}\frac{C_B}{C_T}$ (orange curve) vs $N$.}
	\label{fig:wgc_Sp1s4f}
\end{figure}

\paragraph{$N_f=2$ theory}
We now repeat the same analysis for the theory with 4 fundamentals. Their central charges, flavor central charges, and R-charges are given by
\begin{align}
\begin{split}
&a\sim 1.85520 N-4.46354 \ , \\
&c\sim 2.01015 N-4.41093 \ , \\
&4\pi^4C_B\sim10.4203 N^3+40.9721 N^2+365.603 N-11107.6 \ , \\
&R_{S}\sim 1.48286/N \ , \\
&R_{Q}\sim 0.246455\, +1.34718/N \ .  
\end{split}
\end{align}
The ratio of central charges asymptotes to $0.922948$ (as opposed to 1) as shown in Figure \ref{fig:ac_Sp1s4f}.
The R-charge of symmetric matter vanishes at large $N$ causing all the single-trace gauge-invariant operators and the flipped fields corresponding to the decoupled operators to form a dense band as shown in Figure \ref{fig:spec_Sp1s4f}. 
Finally, it can also be checked that the WGC is satisfied by the lightest meson $QS^nQ$, as depicted in Figure \ref{fig:wgc_Sp1s4f}.

\subsection{1 anti-symmetric and $2N_f$ fundamentals}
1 \banti\, + 2$N_f$  \bfund \,: This theory has a single $U(1)$ flavor symmetry under which the anti-symmetric field $A$ and fundamental $Q$ have charges 1 and $-(N-1)/N_f$ respectively.
The single-trace gauge-invariant operators are given as follows:
\begin{itemize}
	\item $\text{Tr} \left(\Omega A\right)^{n}$,\quad $n=2,\dots,N$
	\item $Q_I \left(\Omega A \right)^{n}\Omega Q_J,\quad n=0,\dots,N-1$
\end{itemize}
Here the indices $I, J$ denote the flavors $1, 2, \ldots, 2N_f$. 
The $N_f=4$ model turns out to be the simplest model that flows to an IR SCFT with a dense spectrum. 
We explicitly checked that such a spectrum appears for $N_f=4$ as well as $N_f=5$. Based on \eqref{eq:dense_conj}, one would expect the $N_f=3$ theory to be the simplest such theory, however upon an explicit numerical analysis we found that the $N_f=3$ theory flows to a free theory in the IR. During its RG-flow, all the gauge-invariant operators as listed above, decouple and the central charges of the would be interacting sector vanishes, implying that the decoupled free operators are the only IR degrees of freedom. Therefore we claim that the theory flows to a non-trivial fixed point only for $N_f \ge 4$.

\begin{figure}[h]
    \centering\vspace{0.3cm}
        \includegraphics[width=9cm]{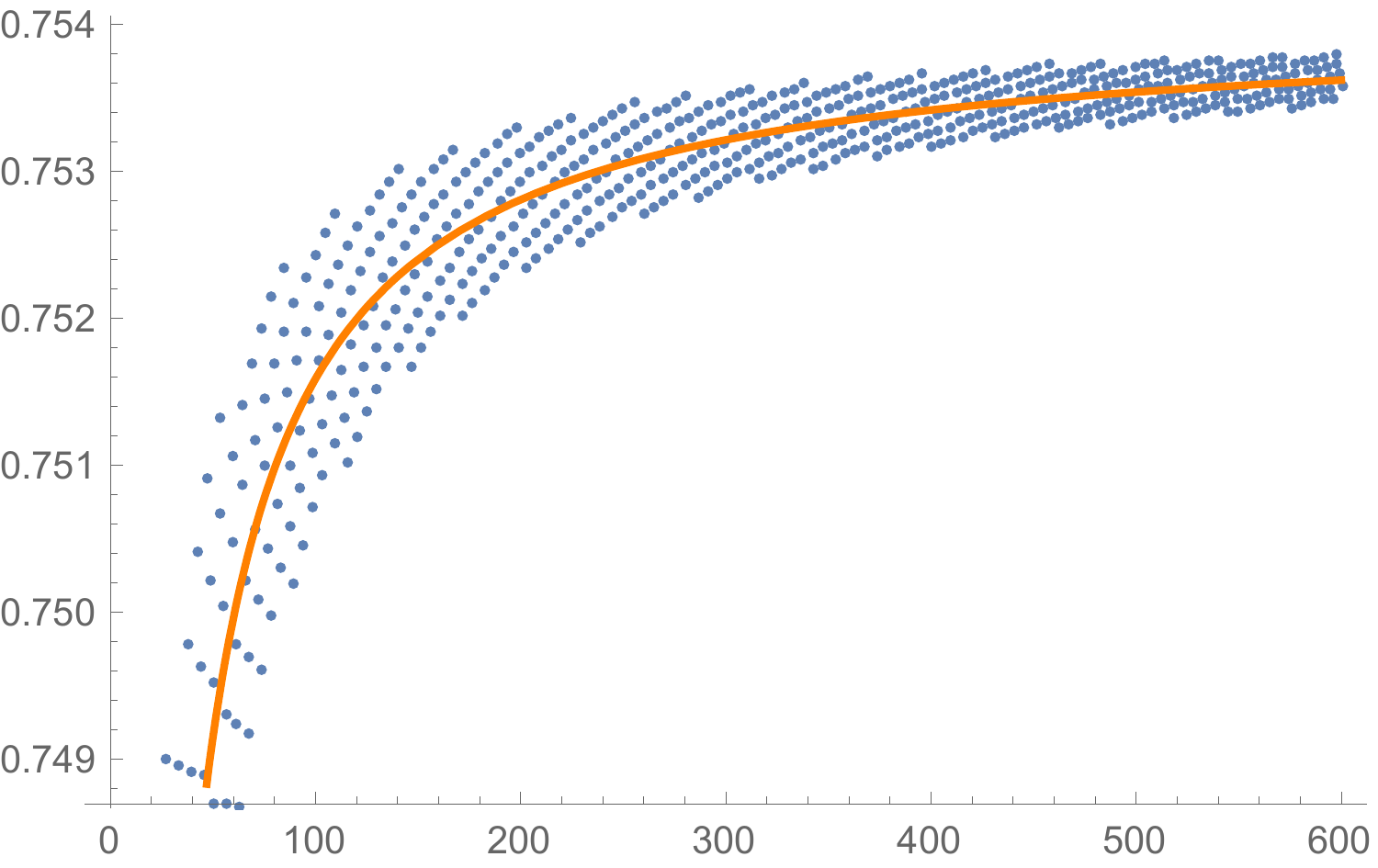}
	\caption{The plot of $a/c$ vs N of the IR theory flowed from $Sp(N)$ + 1 $\protect\banti$ +8 $\protect\bfund$ . The orange curve fits the plot with $0.754024 -0.243817/N$.}
    \label{fig:ac_Sp1a8f}
\end{figure}

\begin{figure}[h]
    \centering
		  \includegraphics[width=9cm]{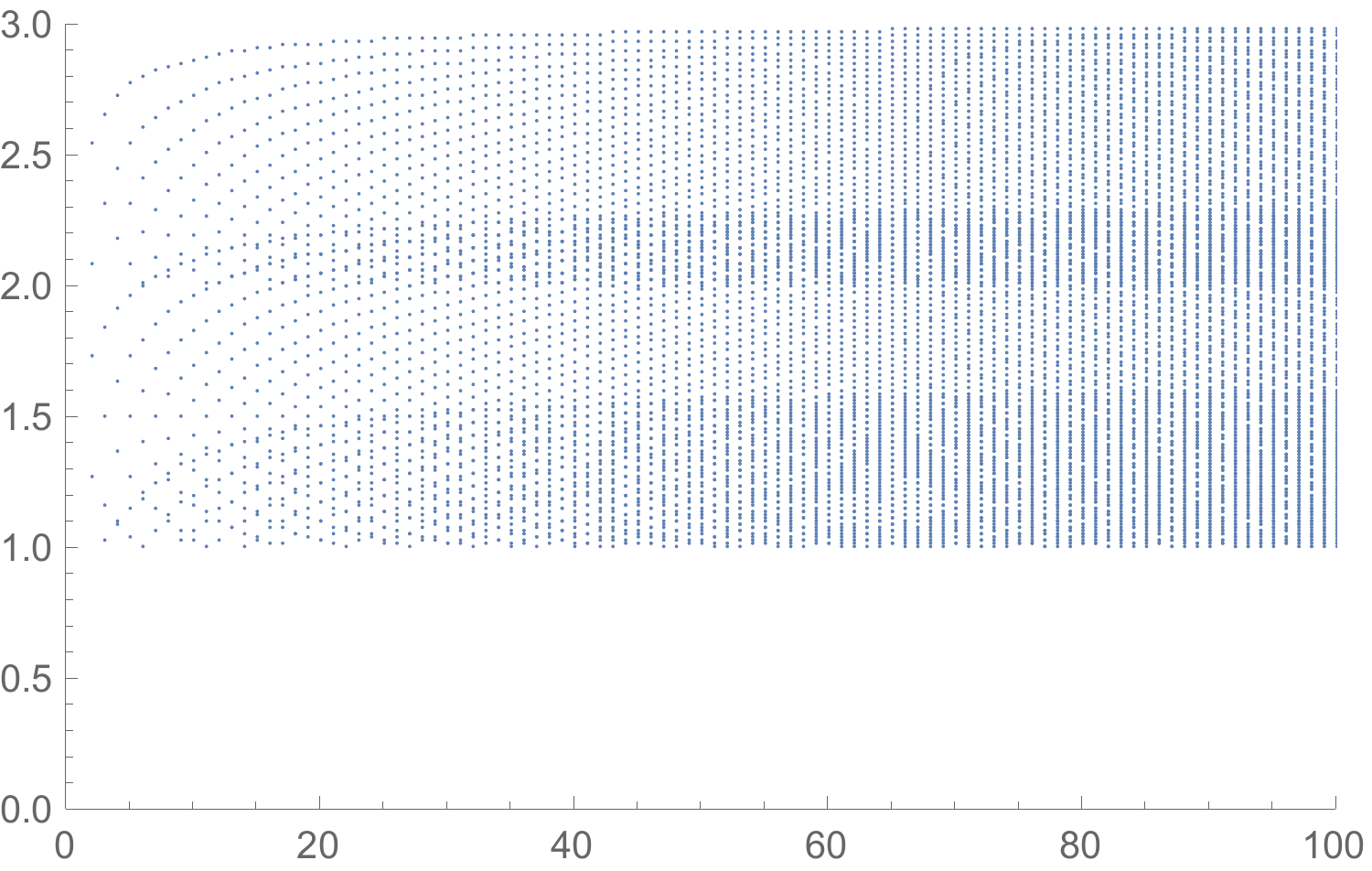}
	\caption{The spectrum of invariant operators between $1<\Delta<3$ in the IR theory of the $Sp(N)$ with 1 anti-symmetric and 8 fundamentals.}
    \label{fig:spec_Sp1a8f}
\end{figure}

\paragraph{$N_f=4$ theory}
Here we expound upon the behavior of the $N_f=4$ theory. 
In the large $N$ limit, the IR central charges and $R$-charges asymptote to
\begin{align}
\begin{split}
&a \simeq 1.03345 N-0.679128 \ ,\\
&c \simeq\,1.37060 N-0.467989 \ , \\
&4\pi^4C_F\simeq2.92641 N^3+8.82911 N^2-198.119 N+7239.57 \ ,\\
&R_{A}\simeq 1.07160/N \ , \\
&R_{Q}\simeq 0.228589 +0.872296/N \ .
\end{split}
\end{align}
The central charges grow linearly in $N$ and the spectrum of gauge-invariant operators will be dense due to the fact that $R_A \sim 1/N$ and goes to zero at large $N$. 
We plot the ratio of central charges $a/c$ in Figure \ref{fig:ac_Sp1a8f}. 
We see that $a/c<1$ and does not reach 1 for sufficiently large $N$. 
The spectrum of single-trace gauge-invariant operators is depicted in Figure \ref{fig:spec_Sp1a8f}.
We see that the operators $\Tr A^n$, $\Tr Q_I A^{n} Q_J$ and the flip fields for the decoupled operators, form a dense band with the conformal dimensions lying in the range $1 < \Delta < 3$. 

We can check this model satisfies the WGC with the spectral data. We find that the lightest meson ($Q A^n Q$ for some $n$) has a smaller dimension-to-charge ratio than the lightest Coulomb branch operator ($\Tr A^n$ for some $n$). 
\begin{figure}[h]
	\centering\vspace{0.3cm}
	\includegraphics[width=9cm]{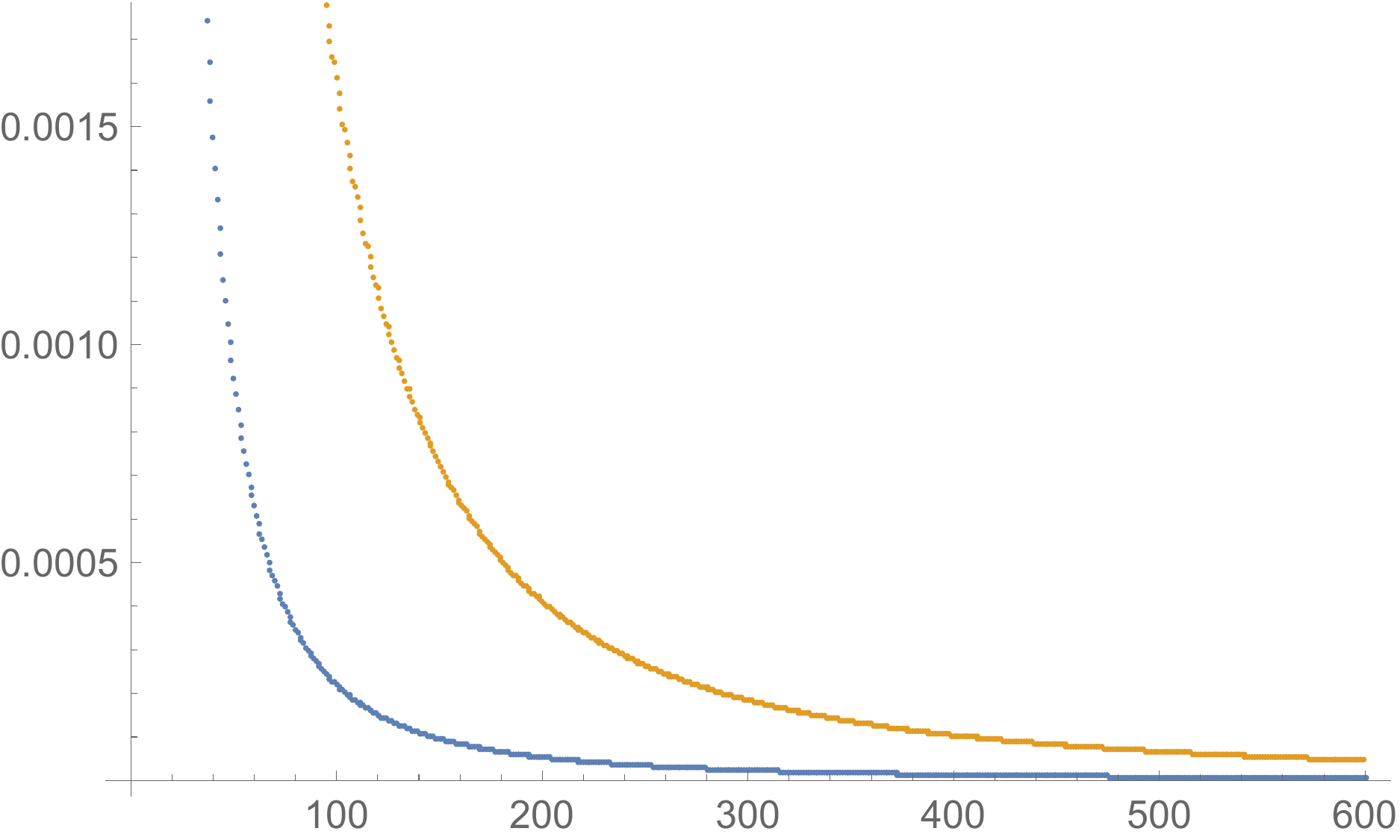}
	\caption{Checking the WGC for the $Sp(N)$ + 1 $\protect\banti$ +8 $\protect\bfund$ theory. Plot of $\frac{\Delta_{\text{mes}}^2}{q^2}$ (blue curve) and $\frac{9}{40}\frac{C_B}{C_T}$ (orange curve) vs $N$.}
	\label{fig:wgc_Sp1a8f}
\end{figure}
As shown in Figure \ref{fig:wgc_Sp1a8f}, the lightest meson has a small enough dimension, that its dual state couples to the gravity much more weakly than any other forces.

\begin{figure}[h]
    \centering\vspace{0.3cm}
        \includegraphics[width=9cm]{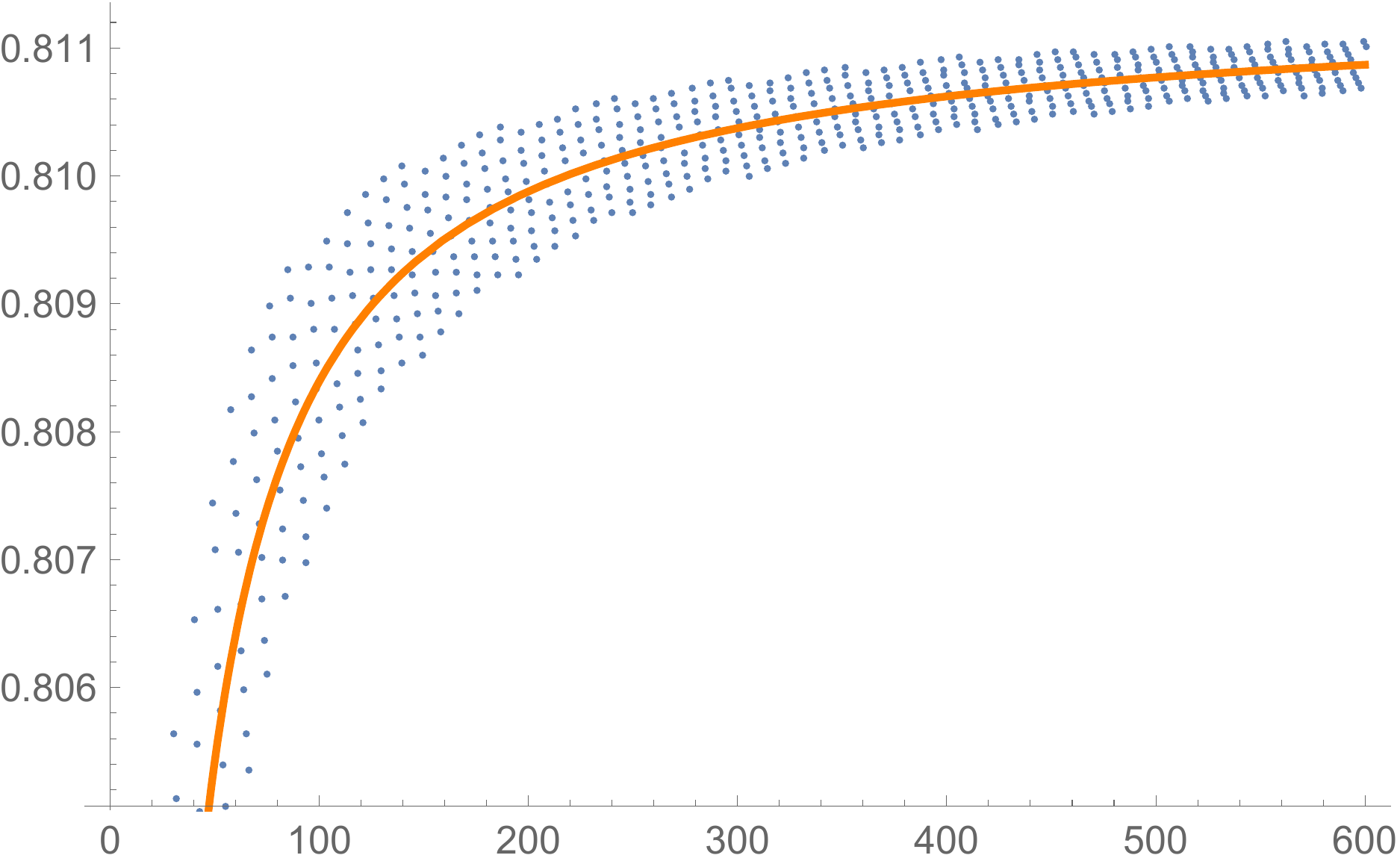}
	\caption{The plot of $a/c$ vs N of the IR theory flowed from $Sp(N)$ with 1 anti-symmetric and 10 fundamentals. The orange curve fits the plot with $0.811365\, -0.332361/N$.}
    \label{fig:ac_Sp1a10f}
\end{figure}

\begin{figure}[h]
    \centering
		  \includegraphics[width=9cm]{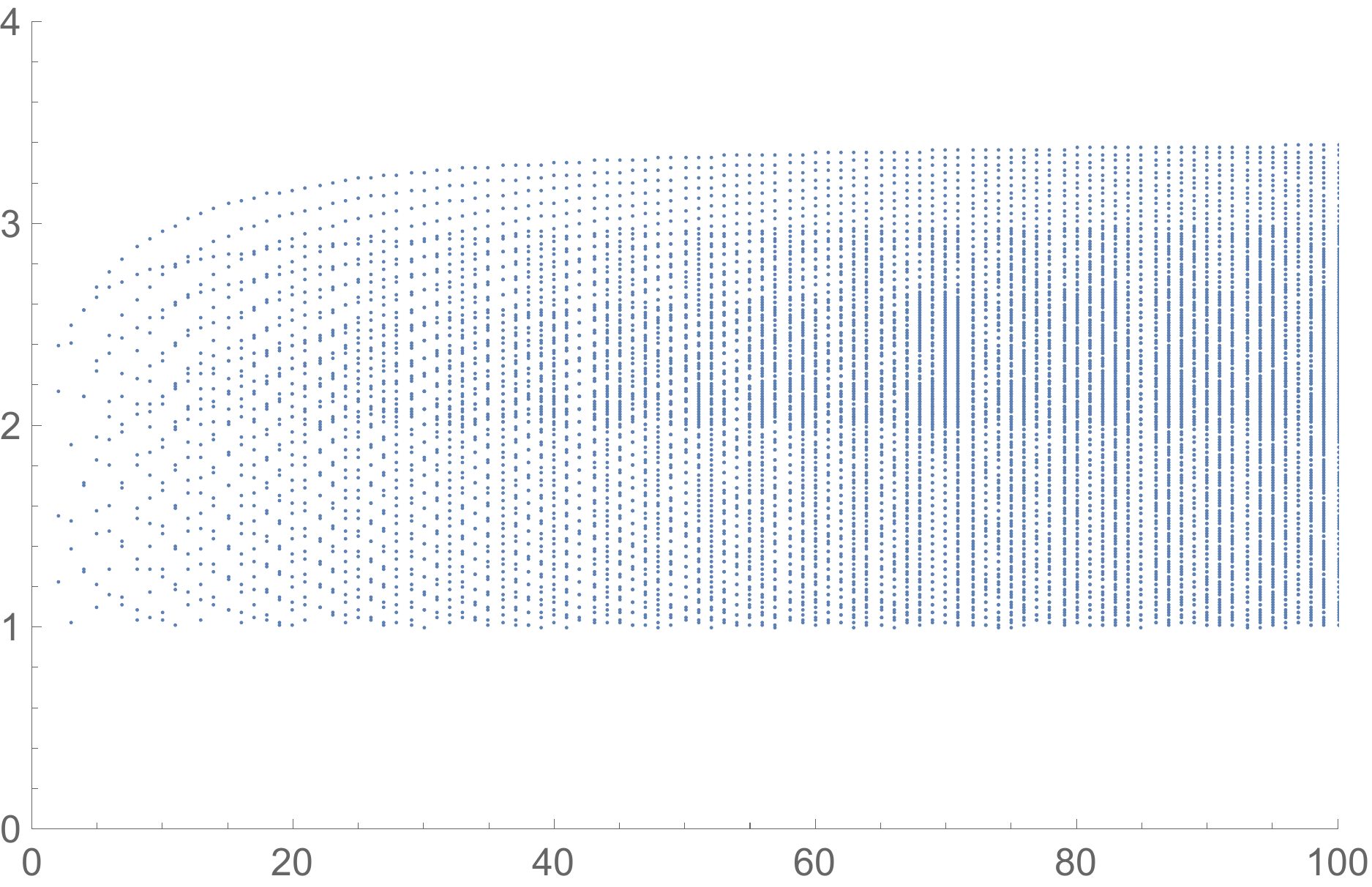}
	\caption{The spectrum of invariant operators between $1<\Delta<4$ in the IR theory of the $Sp(N)$ with 1 anti-symmetric and 10 fundamentals.}
    \label{fig:spec_Sp1a10f}
\end{figure}

\begin{figure}[h]
	\centering\vspace{0.3cm}
	\includegraphics[width=9cm]{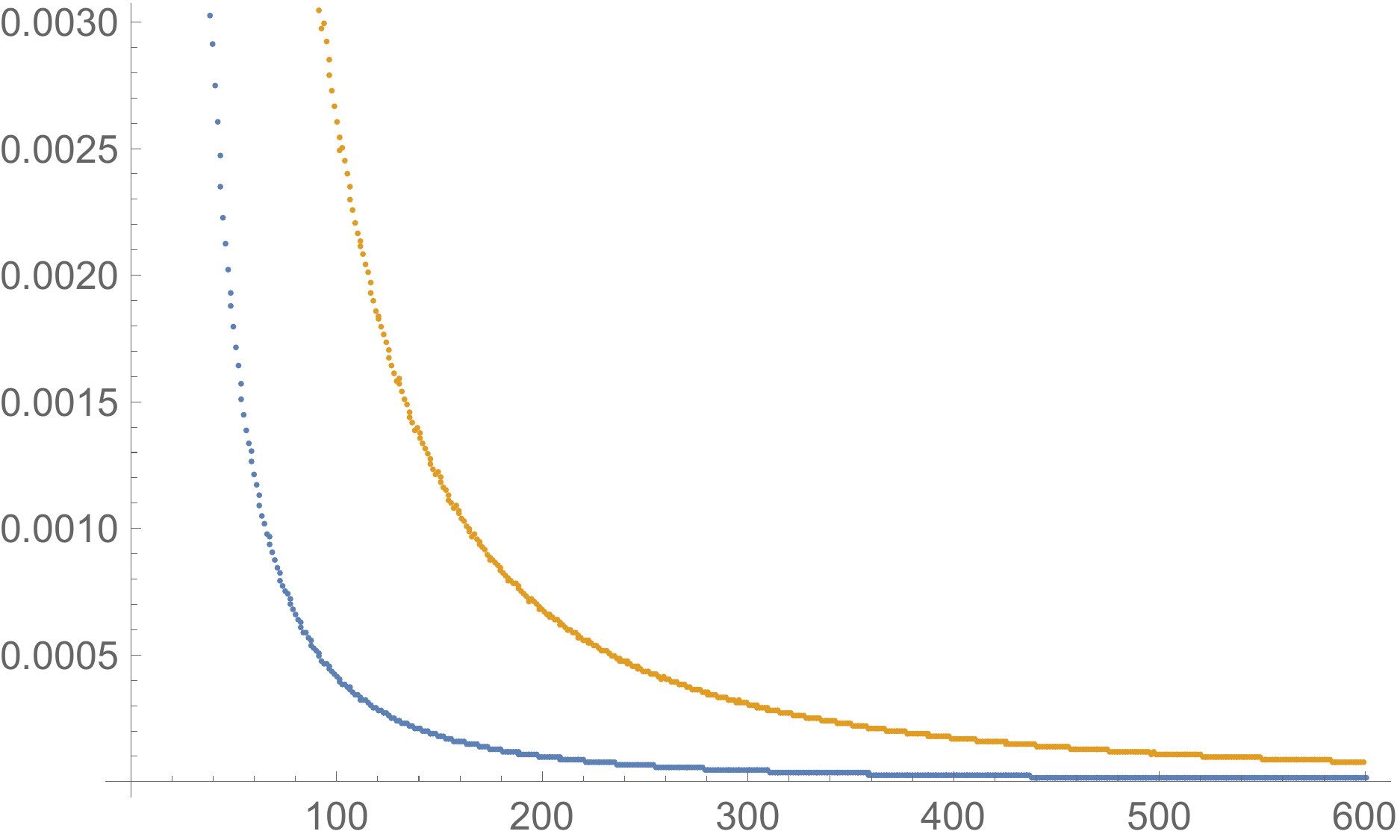}
	\caption{Checking the WGC for the $Sp(N)$ with 1 anti-symmetric and 10 fundamentals. Plot of $\frac{\Delta_{\text{mes}}^2}{q^2}$ (blue curve) and $\frac{9}{40}\frac{C_B}{C_T}$ (orange curve) vs $N$.}
	\label{fig:wgc_Sp1a10f}
\end{figure}

\paragraph{$N_f=5$ theory}
Let us now consider the $N_f =5$ theory. The IR central charges, the flavor central charge, and the R-charges are given by
\begin{align}
\begin{split}
&a \simeq 1.96119 N-3.15123 \ ,  \\
&c \simeq\,2.41728 N-2.95114 \ ,  \\
&4\pi^4C_F\simeq3.0896 N^3+11.9110 N^2+153.944 N-6697.33 \ , \\
&R_{A}\simeq 1.80211/N \ , \\
&R_{Q}\simeq 0.233477\, +1.41098/N \ .
\end{split}
\end{align}
As can easily seen by the two asymptotic central charges, the ratio $a/c$ asymptotes to a value than 1 as shown in Figure \ref{fig:ac_Sp1a10f}.
At the same time, the gap between single-trace gauge-invariant operators  decreases as $N\rightarrow \infty$. The single-trace gauge invariant operators along with the flipped fields for the decoupled operators thus form a dense band of conformal dimensions lying in the range $1<\Delta<4$. as depicted in Figure \ref{fig:spec_Sp1a10f}.
We found the lightest mesonic operator of form $Q_IA^nQ_J$ satisfies the WGC as was also the case in the $N_f=4$ theory. The corresponding plot is shown in Figure \ref{fig:wgc_Sp1a10f}.

\section{Conclusion and Discussion} \label{sec:conclusion}

In this paper, we have classified large $N$ limits of four-dimensional supersymmetric gauge theories with simple gauge groups that flow to superconformal fixed points. We have restricted the analysis to the cases with fixed flavor symmetry and no superpotential. We find a total of 35 classes of gauge theories out of which 8 theories exhibit a dense spectrum of conformal dimensions at large $N$. The central charges of the theories with a sparse spectrum grow as $N^2$ whereas for the dense theories, they grow linearly in $N$. The spectrum of single-trace operators in dense theories can have a single band or multiple bands with a gap, which is some reminiscent of the band structure in superconductors. We also checked that all the dense theories satisfy the AdS version of the Weak Gravity Conjecture (at least for large enough $N$) even though they do not have weakly-coupled gravity duals. 

Let us make a few comments regarding our dense theories and possible future directions. We notice that the dense theories share similarities with the large $N$ limit of Argyres-Douglas theories \cite{Argyres:1995jj, Argyres:1995xn, Eguchi:1996vu, Eguchi:1996ds, Cecotti:2010fi, Xie:2012hs, Wang:2015mra}. For example, $(A_1, A_{N})$, $(A_1, D_{2N})$ theory can be obtained via certain deformations of adjoint SQCDs \cite{Maruyoshi:2016tqk, Maruyoshi:2016aim, Agarwal:2016pjo, Agarwal:2017roi, Benvenuti:2017bpg}.\footnote{Of course, this is not the only way to take ``large $N$" limit for these theories. For example, one can study rank $r$ version of the $H_0=(A_1, A_2)$, $H_1=(A_1, A_3)$, $H_2=(A_1, D_4)$ theories via $r$ D3-branes probing F-theory singularities. This way of increasing the rank gives a sparse spectrum of Coulomb branch operators and its gravity dual is studied in \cite{Aharony:1998xz}, even though the string-coupling is of order 1.} This fact suggests that, like the Argyres-Douglas theories, the IR phase of a dense theory behaves more like the fixed point of rank-$N$ abelian theories than the non-abelian Coulomb phase of SQCD in the conformal window \cite{Seiberg:1994pq}. This may explain the $\CO(N)$ scaling of the degrees of freedom. 

One curious aspect for theories having a dense spectrum is that they often come with a set of decoupled operators that become free at the RG fixed points. The phenomenon of decoupling along the RG flow was brought to light soon after the discovery of $a$-maximization \cite{Kutasov:2003iy, Barnes:2004jj}. We carefully analyzed the decoupling of the operators along the RG flow in situations when there are multiple distinct possible choices of operators to decouple. We proposed a prescription to resolve this based on the $a$-theorem \cite{Komargodski:2011vj}. See also \cite{Intriligator:2005if, Amariti:2012wc} for related discussions. 

Coincidentally, the theories exhibiting a dense spectrum in the IR at large $N$ coincide with the list of theories studied in \cite{Intriligator:1995ax} (generalizing \cite{Kutasov:1995ve, Kutasov:1995np, Intriligator:1995ff, Leigh:1995qp, Aharony:1995ne, Berkooz:1995km}) when the superpotential is turned off. They study dual descriptions for the case with a polynomial superpotential turned on. For some cases, dual theories without a superpotential term have been proposed \cite{Pouliot:1995me,Pouliot:1995zc}. The superpotential deformation does not seem to change the qualitative behavior regarding the spectrum in a drastic way. This is indeed consistent with the dual descriptions described in the paper. It would be interesting to study various superpotential/flip-field deformations, as was done in \cite{Intriligator:2003mi, Barnes:2004jj, Maruyoshi:2018nod, MNS2}. Also, it will be important to consider more general gauge theories where the gauge group is not simple: such as quiver type. Since most of the known `holographic' theories with explicit supergravity dual descriptions are of quiver-type, it is important to study quiver gauge theories to assess the universality of theories with a dense spectrum. Likewise, it would be interesting to consider 3d $\CN=2$ or or 2d $\CN=(0, 2)$ theories. We expect almost the same phenomenon to occur in 3d $\CN=2$ setup given the success (and subtlety) of dimensional reduction of the 4d $\CN=1$ theory flowing to Argyres-Douglas theories \cite{Benvenuti:2017lle, Benvenuti:2017kud, Benvenuti:2017bpg, Aghaei:2017xqe, Agarwal:2018oxb}. 

One of the most pressing questions regarding our dense model would be its holographic interpretation. The $\CO(N)$ growth of the central charges/free energies reminds us of the $O(N)$ vector model and its higher-spin dual \cite{Klebanov:2002ja, Gaberdiel:2010pz}. However, our theories are far from being free and do not have higher-spin currents either. The dense spectrum at large $N$ is reminiscent of the deconstruction analysis \cite{ArkaniHamed:2001ca, ArkaniHamed:2001ie}, though we do not know if there is any connection. 
We have tested the AdS version of the Weak Gravity Conjecture (with convex hull conditions for the multiplet $U(1)$'s) and found that it still holds for large enough $N$ even though the dense theories are not dual to weakly coupled gravity. Previous works on the AdS version of the WGC include \cite{Harlow:2015lma, Heidenreich:2016aqi, Crisford:2017gsb, Conlon:2018vov, Horowitz:2019eum,  Alday:2019qrf, Cremonini:2019wdk, Nakayama:2020dle}. Our results provide strong evidence that the (at least a certain version of) WGC holds beyond the semi-classical gravity. It would be desirable to perform a more refined analysis to fully understand under which condition the WGC holds. 

One of the common aspects of gauge theories is the confinement/deconfinement phase transition. When the theory has a holographic dual description, this becomes identical to the Hawking-Page phase transition \cite{Hawking:1982dh, Witten:1998zw}. A supersymmetric (or BPS) analog of the phase structure of 4d large $N$ gauge theories has been studied in \cite{Choi:2018vbz, Cabo-Bizet:2019eaf, ArabiArdehali:2019orz, Cabo-Bizet:2020nkr} based on the recent advances in supersymmetric indices and the microscopic calculation of the AdS black hole entropy \cite{Choi:2018hmj, Benini:2018ywd, ArabiArdehali:2019tdm,Honda:2019cio, Kim:2019yrz, Cabo-Bizet:2019osg}. 
We do not expect there to be Hawking-Page or confinement/deconfinement phase transition for the `non-holographic' dense theories. However, there can still be a new type of phase in the space of chemical potentials. 

Finally, it is interesting to ask if it is possible to find a non-supersymmetric gauge theory (or any CFT) with a dense spectrum. Our analysis heavily depends on supersymmetric tools that are not readily applicable to non-SUSY theories. However, given the simplicity and genericity of our dense theories with a minimal amount of supersymmetry, it is not unreasonable to expect that such theories also exist in non-supersymmetric setups. 

\begin{acknowledgments}
We thank Sunjin Choi for collaboration at the early stages of this project.  We also thank Dongmin Gang, Hee-Cheol Kim, Seung-Joo Lee, Costis Papageorgakis and Eric Perlmutter for discussion and correspondence.
The work of P.A. is supported by the Royal Society through a Research Fellows Enhancement Award, grant no. RGF$\backslash$EA$\backslash$181049.
The work of KHL and JS is supported by the Junior Research Group Program at the APCTP through the Science and Technology Promotion Fund, Lottery Fund of the Korean Government, Gyeongsangbuk-do, and Pohang City.
The work of JS is also supported by the National Research Foundation of Korea (NRF) grant NRF-2020R1C1C1007591 and the Start-up Research Grant for new faculty provided by Korea Advanced Institute for Science and Technology (KAIST). 

\end{acknowledgments}

\bibliographystyle{jhep}
\bibliography{refs}

\end{document}